\documentstyle[a4,12pt,epsfig]{article}
\def\be{\begin{equation}}
\def\ee{\end{equation}}
\def\bq{\begin{eqnarray}}
\def\eq{\end{eqnarray}}
\def\g{\gamma}
\def\ra{\rightarrow}
\def\vp{\varphi}
\newcounter{saveeqn}
\newcounter{App} 
\newcommand{\app}{%
\stepcounter{App}%
\setcounter{saveeqn}{\value{equation}}%
\setcounter{equation}{0}%
\renewcommand{\theequation}{\Alph{App}\arabic{equation}} }
\newcommand{\appende}{%
\setcounter{equation}{\value{saveeqn}}%
\renewcommand{\theequation}{\arabic{equation}}  }
\begin{document}
\thispagestyle{empty}
\setcounter{page}{0}
\begin{flushright}
WUE-ITP-97-049\\
MPI-PhT/97-85\\
\end{flushright}
\begin{center}
{\Large \bf
QCD Sum Rules for Exclusive Decays of Heavy Mesons
}\\
\vspace{0.5em}
\large
\vspace{0.5em}
A. Khodjamirian$^{a,}$\footnote{\small on leave of absence from
the Yerevan Physics Institute, 375036 Yerevan, Armenia}
and R. R\"uckl$^{a,b}$\\
\vspace{1em}
{$^a$ \small Institut f\"ur Theoretische Physik,
Universit\"at W\"urzburg, D-97074 W\"urzburg,
Germany}\\
{$^b$ \small Max-Planck-Institut f\"ur Physik, Werner-Heisenberg-Institut,
D-80805 M\"unchen, Germany}\\
\end{center}
\begin{abstract}
Applications of QCD sum rules to exclusive decays of heavy mesons
are reviewed. In detail, 
we discuss the calculations of the $B$ and $D$ decay constants,
the heavy-to-light form factors $f^+(p^2)$ and $f^0(p^2)$, 
and the strong couplings $g_{B^*B\pi}$ and $g_{D^*D\pi}$. 
Predictions are presented for
the semileptonic weak decays 
$B \rightarrow \pi \bar{l} \nu_l$ and 
$B \rightarrow \rho \bar{l} \nu_l$,
and used to extract $V_{ub}$ from the measured decay widths.
Results are also shown on $D \rightarrow \pi \bar{l} \nu_l$ and 
$D^* \rightarrow D \pi$.  
Furthermore, we reconsider the factorization hypothesis in
nonleptonic two-body decays, and describe 
a sum rule estimate of the nonfactorizable 
contribution to the amplitude of $B \rightarrow J/\psi K$.
\end{abstract}

\vspace{0.5em}
\begin{center}
{\it To appear in Heavy Flavours, 2nd edition,\\ eds. A.J. Buras
and M. Lindner (World Scientific, Singapore)}   
\end{center}
\vspace{0.5em}

\section{Introduction}
During recent years a large amount of new experimental data
on exclusive decays of charm and bottom hadrons has become available.
In order to make optimal use of these data 
in determinations of fundamental parameters and tests of the standard model,
one needs a quantitative understanding of the impact 
of strong interactions.
This requires accurate calculations of hadronic matrix
elements of weak operators in QCD beyond perturbation theory.
Current approaches include lattice calculations, QCD sum rules, heavy
quark effective theory (HQET), chiral perturbation theory (CHPT),
and phenomenological quark models.
Each of these approaches has advantages and disadvantages.
For example, quark models are easy to use and good
for intuition. However, their relation to QCD is unclear.
On the other hand, lattice calculations are rigorous from the
point of view of QCD, but they suffer from lattice artifacts
and uncertainties connected with the necessary extrapolations 
to the physical quark masses.
Furthermore, effective theories are usually applicable only to
a restricted class of problems, and sometimes require substantial
corrections which cannot be calculated within the same framework.
For example, HQET is very powerful in treating  $b \ra c $ transitions,
but a priori less suitable for $b\to u$ transitions, while CHPT is 
designed for processes involving soft pions and kaons.

In this review, we focus on the method of
QCD sum rules \cite{SVZ,Volume,RRY}. Proceeding from the firm basis
of QCD perturbation theory this approach tries to incorporate  
nonperturbative elements of full QCD. Schematically,
hadronic matrix elements of operators composed of quark and gluon
fields are extracted from correlation functions of quark currents,
rather than estimated directly in some models.  
One makes use of operator product
expansion (OPE), the analyticity principle
(dispersion relations), and $S$-matrix unitarity. In addition, one assumes
the validity of quark-hadron duality in a rather strong form, 
sometimes called semi-local
duality. The long-distance dynamics is parameterized in terms of vacuum
condensates or light-cone wave functions. These new elements cannot
yet be calculated rigorously in QCD, but are to be determined from
experimental data in the one or other way. Nevertheless,
because of the universal nature of the nonperturbative input,
the sum rule approach retains its predictive power. 
Moreover, it is rather flexible and can therefore be
applied to a large variety of problems in hadron physics. 

The application of sum rules to heavy quark physics has always been
particularly successful. The more recent development 
of sum rule techniques for exclusive heavy meson decays again
looks very promising.
We shall substantiate this assessment for leptonic,
semileptonic as well as nonleptonic decays of $B$ and $D$ mesons.
Although the two flavours could be treated in parallel,  
we shall usually refer to $B$ mesons, for definiteness.
In most cases, it is obvious how to obtain the corresponding results 
for $D$ mesons. 

The leptonic decay process $B\to \bar{l} \nu_l$ involves 
the simplest hadronic matrix element, that is the $B$-meson decay 
constant $f_B$. We therefore have chosen the latter for our introductory
study case.
Furthermore, $f_B$ is one of the important parameters of 
mixing and CP-violation in the $B$ system, and a necessary input in 
order to  eventually extract the CKM element
$V_{ub}$ from future measurements of $B\to \bar{\tau} \nu_\tau$.
Finally, $f_B$ also enters more complicated 
QCD sum rules for other hadronic
properties of $B$ mesons, e.g., form factors.

The semileptonic decays $B\to \pi \bar{l} \nu_l$
and $B\to \rho \bar{l} \nu_l$ have already been observed experimentally. 
These exclusive modes are considered to provide interesting 
alternatives to the determination of $V_{ub}$ from inclusive $b\to u$
transitions. However, truly competitive results can only be expected if the
calculation of heavy-to-light form factors
can be improved significantly.
From the theoretical point of view, the $B \rightarrow \pi$ and
$B \rightarrow \rho$ form factors are
excellent examples for the application of light-cone sum rules.

Exclusive nonleptonic decays are an even greater challenge to theory. Although
over the years one has developed a qualitative understanding and even an
amazingly consistent quantitative description of two-body decays 
\footnote{A comprehensive study is presented by 
M. Neubert and B. Stech in this volume, ref. \cite{NS}.},
some features still lack a clear dynamical explanation. 
This concerns most of all the factorization of matrix elements
of four-quark operators into products of matrix elements of quark currents.
Since nonfactorizable amplitudes are a priori expected
to be channel-dependent, 
the apparent universality of the effective coefficients 
$a_1$ and $a_2$ constitutes a major puzzle. 
The decay mode $B\to J/\psi K $ containing two heavy quarks in the final
state may play a special role in this respect 
and therefore shed some light on 
the factorization problem. We present this example also because
it shows conceptual and technical limits 
of the sum rule approach.

The content of this review is as follows.
In section 2, we describe the derivation of the  
sum rule for $f_B$ from a two-point correlation function and summarize 
the analogous calculations for other $B$ and $D$ mesons. Also shown 
is a comparison with lattice results and experimental data.
The straightforward generalization of the above method
to three-point correlation functions is discussed in section 3 and
applied to heavy-to-light form factors.
In section 4, we explain an alternative approach
based on light-cone expansion and employ it in section 5  
to obtain sum rules
for the $B\rightarrow \pi$ form factors $f^+$ and $f^0$.
Recent results on the $B \rightarrow \rho$
form factors are also mentioned here. Section 6 is devoted to
sum rules for the strong $B^*B\pi$ and $D^*D\pi$ couplings. 
The estimates are used to normalize the single-pole approximation
for $f^+$ and to calculate the width for $D^* \rightarrow D \pi$.
In section 7, we present the sum rule predictions on 
the decay widths and distributions for $B \rightarrow \pi \bar{l}\nu $
and $B \rightarrow \rho \bar{l}\nu $, and 
extract $V_{ub}$ from the comparison with data. Results are also shown
on $D \rightarrow \pi l \bar{\nu}$. 
Section 8 deals with the heavy mass expansion of the sum rules 
for heavy-to-light form factors and couplings, and
with the asymptotic scaling behaviour.
Finally, in section 9 we discuss the 
factorization hypothesis for
$B\to J/\psi K$ and describe a first sum rule 
estimate of the nonfactorizable
piece in the $B\to J/\psi K$ amplitude.

\section{Decay constants of $B$ and $D$ mesons}

Leptonic decays of $B$ mesons are induced by the weak annihilation
process $b\bar{u}\to l \bar{\nu}_l$. 
The decay width
\be
\Gamma( B^- \to l^- \bar{\nu}_l) = \frac{G_F^2}{8\pi} |V_{ub}|^2
m_Bm_l^2 \left(1-\frac{m_l^2}{m_B^2}\right)^2 f_B^2
\label{width}
\ee
is suppressed by three small parameters: 
the CKM matrix element $V_{ub}$, 
the lepton mass $m_l$ and the decay constant $f_B$.
While the factor $m^2_l$ is enforced by helicity conservation, 
$f_B$ characterizes the size of the $B$-meson wave function at the origin. 
Usually, the decay constant is defined by the matrix element
of the relevant weak axial-vector current:
\be
\langle 0 \mid\bar{q}\gamma_\mu \gamma_5 b\mid B\rangle =if_B q_\mu~,
\label{fB}
\ee
$\bar q$ carrying the flavour of the light quark constituent in the $B$,
and $q_\mu$ being the $B$ four-momentum. 
Later we shall use the equivalent definition
in terms of the matrix element of the corresponding pseudoscalar 
current: 
\be
m_b\langle 0 \mid\bar{q}i\gamma_5 b\mid B\rangle =m_B^2f_B~,
\label{fB2}
\ee
$m_b$ being the $b$-quark mass.
Because of the strong suppression of leptonic $B$ decays
only the mode
$B\to {\overline \tau} \nu_{\tau}$
has a chance to be measured. 
However, this task will not be easy, even at future $B$ factories. 
For this and other reasons which will become clear in the course 
of this review,
accurate theoretical
predictions of $f_B$ are indispensable.

The QCD sum rule estimate of $f_B$ is based
on an analysis of the two-point correlation function
\cite{RRY,6auth,BG,AE}
\be
\Pi(q^2)= i\int d^4x e^{iqx}\langle 0\mid
T\{\bar{q}(x)i\gamma_5 b(x), \bar{b}(0)i\gamma_5 q(0)\}
\mid 0\rangle~.
\label{corr2}
\ee
By inserting a complete set of states 
with $B$-meson quantum numbers between the currents in (\ref{corr2}) 
one obtains the hadronic representation
$$
\Pi (q^2)=
\frac{\langle 0\mid \bar{q}i\gamma_5 b\mid B \rangle
\langle B\mid \bar{b}i\gamma_5q\mid 0 \rangle}{m_B^2-q^2}
$$
\be
+\sum_h \frac{ <0\mid \bar{q}i\gamma_5 b\mid h>
<h\mid \bar{b}i\gamma_5 q \mid 0 \rangle}{ m_h^2- q^2}~.
\label{hadr2}
\ee
The decay constant $f_B$ appears in the 
first term contributed by the ground state $B$-meson. 
The second term takes into account the higher resonances 
and non-resonant states. 
The result (\ref{hadr2})
can be rewritten in the form of a dispersion relation:
\be
\Pi(q^2)=
\int_{m_B^2}^\infty
\frac{\rho (s)ds}{s-q^2}~,
\label{disp2}
\ee
where the spectral density is given by
\be
\rho (s)=\delta (s-m_B^2)\frac{m_B^4f_B^2}{m_b^2}
+ \rho^{h}(s)\Theta(s-s^h_0)
\,.
\label{spect2}
\ee
Obviously, the $\delta$-function term on the r.h.s. of (\ref{spect2})
represents the $B$ meson, while $\rho^h(s)$
and $s^h_0$ are the spectral density and threshold energy squared
of the excited resonances and continuum states, respectively.
In general, the dispersion integral (\ref{disp2}) is
ultraviolet divergent. In order to make the integral  
finite, one subtracts a sufficient
number of terms of the Taylor expansion of $\Pi(q^2)$ at $q^2=0$.
In the case at hand, two subtractions are needed:
\be
\Pi_f(q^2)=
\Pi(q^2) -\Pi(0) - q^2\Pi'(0)=
q^4\int ^\infty _{m_B^2}\frac{\rho(s)ds}{s^2(s-q^2)}~.
\label{subtr}
\ee
Substitution of (\ref{spect2}) in (\ref{subtr}) yields 
\be
\Pi_f(q^2)=
\frac{m_B^4f_B^2}{m_b^2(m_B^2-q^2)}+ \int ^\infty_{s^h_0}\frac{\rho^h
(s)ds}{s-q^2} -\Pi(0) - q^2\Pi'(0).
\label{dispersion}
\ee

The subtraction terms can be removed by performing a Borel
transformation with respect to $q^2$ :
\be
{\cal B}_{M^2}\Pi(q^2)=\lim_{\stackrel{-q^2,n \to \infty}{-q^2/n=M^2}}
\frac{(-q^2)^{(n+1)}}{n!}\left( \frac{d}{dq^2}\right)^n \Pi(q^2)
\equiv \Pi(M^2)~.
\label{Borel2}
\ee
With
\be
{\cal B}_{M^2}
\left(\frac1{s-q^2}\right)^k
=\frac1{(k-1)!}\left(\frac1{M^2} \right)^{k-1}e^{-s/M^2}
\label{Borel1}
\ee
and 
\be
{\cal B}_{M^2}(-q^2)^k=0 
\label{Borel22}
\ee
for $k \ge 0$ one readily finds
\be
\Pi(M^2)=\Pi_{f}(M^2)=\frac{m_B^4f_B^2}{m_b^2}e^{-m_B^2/M^2}+
\int_{s^h_0}^\infty \rho^h(s)e^{-s/M^2}ds~.
\label{Borel12}
\ee
We see that Borel transformation removes arbitrary
polynomials in $q^2$ and suppresses the
contributions from excited and continuum states 
exponentially relative to the ground-state contribution.
The second point is actually the main motivation for this 
transformation.

At $q^2 \ll m_b^2 $, below the poles and cuts 
associated with the resonances and continuum
states, it is possible to calculate the
correlation function $\Pi(q^2)$ in terms of the quark and gluon degrees
of freedom in QCD. To this end, one expands
the $T$--product of currents in (\ref{corr2}) in a series of
local operators $\Omega_d$ constructed from quark and gluon fields
and normalized at the scale $\mu$:
\be
i\int d^4x
e^{iqx}~T\{\bar{q}(x)i\gamma_5 b(x), \bar{b}(0)i\gamma_5 q(0)\}
= \sum_d C_d(q^2,\mu) \Omega_d(\mu)
\label{exp2}
\ee
with the sum running over the canonical dimensions $d$ of the
operators. The lowest-dimensional operators with  $d=$0,3,4,5,6 are
given by 
\be
\Omega_d = 1,~ \bar{q}q ,~ G^a_{\mu\nu}G^{a\mu\nu}, ~
\bar q\sigma_{\mu\nu} \frac{\lambda ^a}2 G^{a\mu\nu}q,~
(\bar{q}\Gamma_r q)( \bar{q}\Gamma_s q)~,
\label{oper2}
\ee
respectively. Here, $\lambda^a$ are the usual $SU(3)$-colour matrices, 
and $G^a_{\mu\nu}$ is the gluon field strength tensor. 
$\Gamma_r$ denotes a certain
combination of Lorentz and colour matrices.
Note the absence of a colour-neutral, Lorentz-invariant operator with $d=2$.

The most important virtue of OPE is the possibility to
separate long- and short-distance contributions to
the correlation function
\be
\Pi(q^2) = \sum_d C_d(q^2,\mu)\langle 0 \mid \Omega_d(\mu) \mid 0 \rangle~.
\label{ope112}
\ee
While the strong interaction effects at momenta $k^2>\mu^2$ are included 
in the coefficients $C_d(q^2,\mu)$, the effects at $k^2<\mu^2$
are absorbed into the matrix elements of the operators 
$\Omega_d(\mu)$. Thus, if $\mu \gg \Lambda_{QCD}$ 
the Wilson coefficients depend only on 
short-distance dynamics, and can therefore be
calculated in perturbation
theory, while the long-distance effects are taken into 
account by the vacuum averages
$\langle 0 \mid \Omega_d \mid 0 \rangle \equiv \langle \Omega_d \rangle $.
These so-called condensates describe properties
of the full QCD vacuum and are process-independent.
At present, they can only be estimated in
some crude approximations. For this reason, they are usually determined
empirically by fitting sum rules to experimental data.
It is essential for the whole approach that at $q^2 \ll m_b^2$
the expansion (\ref{ope112}) can be cut off after a few
terms. The reason is that the higher the dimension 
of $\Omega_d$, the more suppressed by inverse powers of 
$m_b^2-q^2$ is the corresponding Wilson coefficient 
$C_d$. Most applications in the past include the set of  
operators  given in (\ref{oper2}).

\begin{figure}[ht]
\mbox{
\epsfig{file=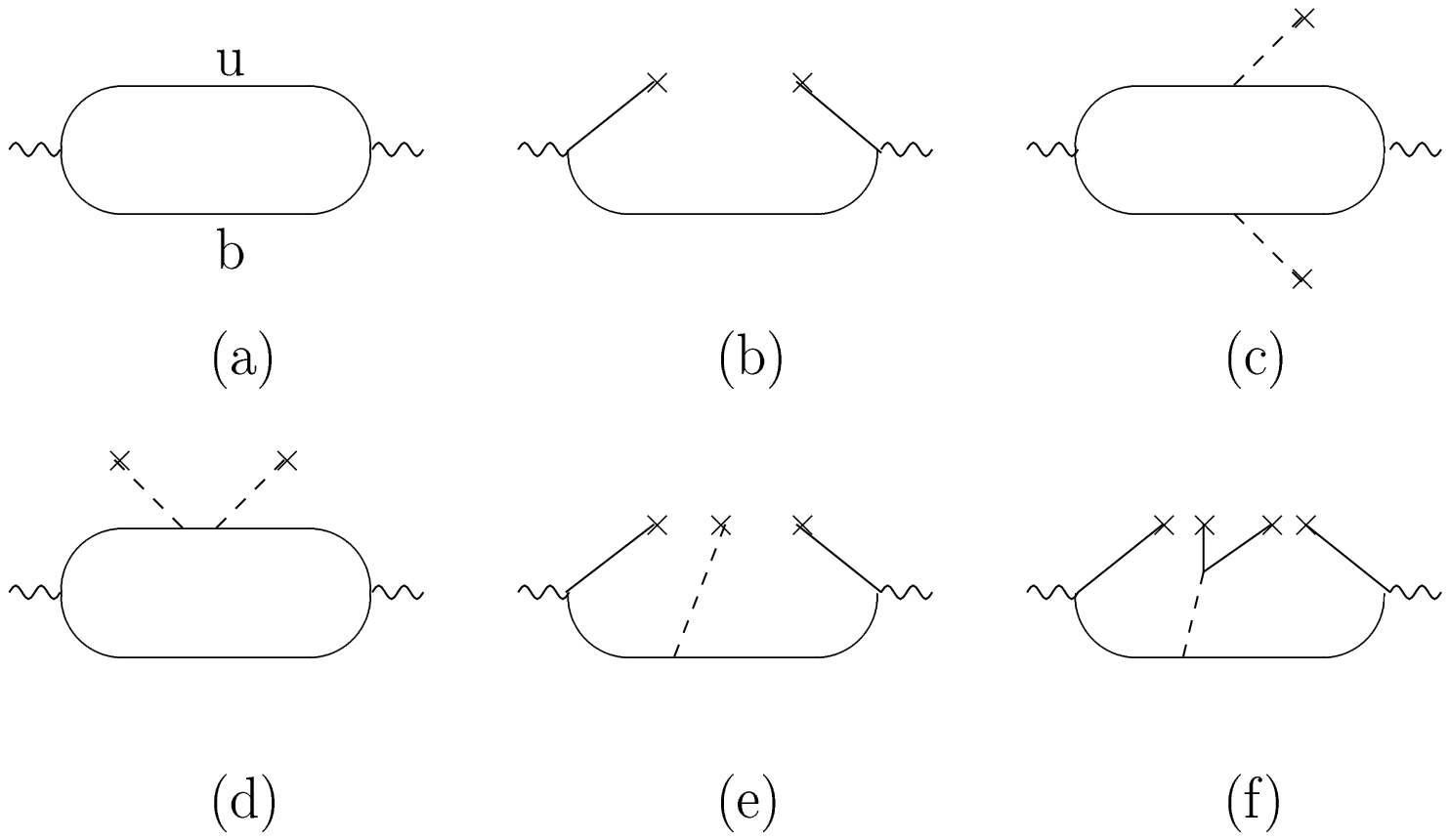,width=\textwidth,bbllx=0pt,bblly=260pt,bburx=600pt,%
bbury=570pt,clip=}
}
\caption{\it Diagrams determining the Wilson coefficients
in the OPE of the two-point correlation function (\ref{corr2}):
$C_0$ (a), $C_3$ (b), $C_4$ (c,d), $C_5$ (b,e), $C_6$ (b,f).
Solid lines denote quarks, dashed lines gluons, wavy lines
external currents. Crosses indicate vacuum fields.}
\end{figure}

The short-distance coefficients $C_d$ are calculated
from the diagrams 
depicted in Fig. 1. Specifically, the coefficient $C_0$ of the
unit operator is obtained by contracting all quark fields
in the correlation function (\ref{corr2}), and
inserting the free quark propagators
\be
\langle 0 |T\{q_i(x)\overline{q}_j(0)\}|0\rangle = i\hat{S}_{ij}^q(x)=
\delta_{ij}\int\frac{d^4k}{i(2\pi )^4}e^{-ikx}
\frac{\not\!k+m_q}{m^2_q-k^2}~,
\label{S02}
\ee
where $i,j$ are colour indices.
For light quarks we put $m_q = 0$. Diagrammatically, this approximation is
represented by Fig. 1a. It constitutes the zeroth order 
approximation for the correlation function (\ref{corr2}) in QCD
perturbation theory. The result can be written in the form of
a dispersion relation:
\be
\Pi_{Fig.1a}(q^2) = C_0(q^2)=\frac{1}{\pi}
\int_{m_b^2} ^{\infty}ds\frac{\mbox{Im} C_0(s)}{s-q^2}
\label{c021}
\ee
with
\be
\mbox{Im}C_0(s)=\frac{3}{8\pi}\frac{(s-m_b^2)^2}{s}~.
\label{c02}
\ee
The subtraction procedure necessary to make the integral (\ref{c021})
finite is postponed in anticipation of the Borel transformation
(\ref{Borel2}).

The coefficient $C_3$ of the 
$\bar{q}q$ operator in (\ref{ope112})
is obtained by treating the light-quark fields
in (\ref{corr2}) as external vacuum fields and
neglecting their momenta
as compared to the momentum of the freely propagating off-shell $b$ quark.
The corresponding diagram is shown in Fig. 1b.
Substituting (\ref{S02}) in (\ref{corr2}) one gets
\be
\Pi_{Fig.1b}(q^2)= \int d^4x e^{iqx}\langle 0\mid
\bar{q}(0)_{i\alpha} q(0)_{j\beta} \mid 0\rangle
\{\gamma_5 \hat{S}_{ij}^{b}(x)\gamma_5 \}_{\alpha\beta}~,
\label{corrr2}
\ee
where $\alpha,\beta$ are the spinor indices of the quark fields.
Because of vanishing spin and colour of the vacuum,
\be
\langle 0\mid \bar{q}(0)_{i\alpha} q(0)_{j\beta} \mid 0\rangle =
\frac1{12} \delta_{\alpha\beta}\delta_{ij} \langle \bar{q} q \rangle.
\label{neu2}
\ee
Integration of (\ref{corrr2}) over $x$ then yields
\be
\Pi_{Fig.1b}(q^2) = C_3(q^2) \langle \bar{q}q \rangle
\label{neu3}
\ee
with
\be
C_3(q^2)= m_b/(q^2-m_b^2)~.
\label{neu4}
\ee
The more complicated calculation of the coefficients
$C_{4,5,6}$ from the diagrams of Fig. 1 is discussed, for example, 
in \cite{RRY}.

\begin{figure}[ht]
\mbox{
\epsfig{file=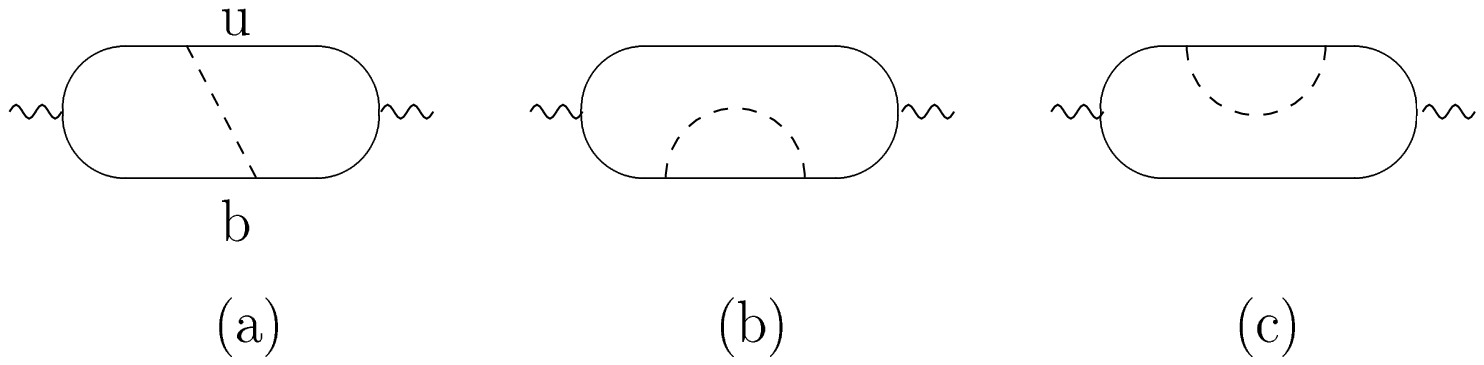,width=\textwidth,bbllx=0pt,bblly=340pt,bburx=600pt,%
bbury=480pt,clip=}
}
\caption{\it Feynman diagrams of the
$O(\alpha_s)$ correction to the Wilson coefficient $C_0(q^2)$.}
\end{figure}

The accuracy of the OPE can be further improved
by including perturbative QCD corrections to the Wilson coefficients
corresponding to hard gluon exchanges in the diagrams of Fig. 1.
Most important are the $O(\alpha_s)$ effects on the
coefficient $C_0$ \cite{RRY,BG} given by
the two-loop diagrams of Fig. 2.
Adding these corrections to the imaginary part of $C_0$ given in 
(\ref{c02}), one has
\be
\mbox{Im}C_0(s)=\frac{3}{8\pi}\frac{(s-m_b^2)^2}s
\left(1+ \frac{4\alpha_s}{3\pi}f(s,m_b^2)\right)~,
\label{im2}
\ee
where
$$
f(s,m_b^2) = \frac94 + 2\mbox{Li}_2(\frac{m_b^2}s)
+
\ln\frac{s}{m_b^2}\ln\frac{s}{s-m_b^2}
$$
\be
+ \frac32 \ln\frac{m_b^2}{s-m_b^2}+\ln\frac{s}{s-m_b^2}
+\frac{m_b^2}s \ln\frac{s-m_b^2}{m_b^2} +
\frac{m_b^2}{s-m_b^2}\ln\frac{s}{m_b^2}~,
\label{oalphas}
\ee
$\mbox{Li}_2(x)=-\int^{x}_0\frac{dt}tln(1-t)$,
and $m_b$ is the pole mass of the $b$ quark.

The complete result for the Borel-transformed 
correlation function (\ref{corr2}) given in \cite{AE} reads
$$
\Pi_{QCD}(M^2)=
\frac{3}{8\pi^2}\int_{m_b^2} ^{\infty}ds
\frac{(s-m_b^2)^2}{s}
\left(1+ \frac{4\alpha_s}{3\pi}f(s,m_b^2)\right)
\exp \left(-\frac{s}{M^2}\right)
$$
$$
+\Bigg(-m_b\langle \bar q q \rangle +
\frac1{12}\langle  \frac{\alpha_s}{\pi}
G^a_{\mu\nu}G^{a\mu\nu} \rangle
-\frac{m_b}{2M^2}\left(1-\frac{m_b^2}{2M^2}\right)
m_0^2\langle \bar q q \rangle
$$
\be
-\frac{16\pi\alpha_s}{27M^2}
\left(1-\frac{m_b^2}{4M^2}-\frac{m_b^4}{12M^4}
\right) \langle \bar q q \rangle^2
\Bigg) \exp \left(-\frac{m_b^2}{M^2}\right)~.
\label{ope12}
\ee
Here, use has been made of the conventional
parametrization for the quark-gluon condensate density:
\be
\langle  \bar q \sigma_{\mu\nu}
\frac{\lambda^a}{2}G^{a\mu\nu}q \rangle =
m_0^2 \langle  \bar q q  \rangle~,
\label{qqbarG2}
\ee
and of vacuum saturation reducing four-quark condensates
to squares of quark condensates with known
coefficients \cite{SVZ}:
\be
\langle
 \bar{q}\Gamma_rq \bar{q}\Gamma_sq\rangle=
\frac{1}{(12)^2} \{ (Tr\Gamma_r)(Tr\Gamma_s)-Tr(\Gamma_r\Gamma_s) \}
\langle  \bar{q}q\rangle^2~.
\label{vac4q2}
\ee

Now, by equating the hadronic
representation of $\Pi(M^2)$ given in (\ref{Borel12}) 
with the QCD result (\ref{ope12}) 
one obtains an interesting sum rule
which, however, does not yet allow to determine $f_B$. This is due
to the unknown spectral density $\rho^h(s)$ associated with the
excited and continuum states. In order to proceed,  
one assumes quark-hadron duality and substitutes, in (\ref{Borel12}),
the perturbative spectral density:
\be
\rho^h(s)\Theta(s-s_0^h) = \frac1{\pi} \mbox{Im} C_0(s) \Theta (s-s_0^B)~,
\label{high2}
\ee
considering $\sqrt{s_0^B}$ as an effective threshold parameter of the
order of the mass of the first excited $B$ resonance.
Subtraction of the integral in (\ref{Borel12}) 
from the corresponding term in $\Pi_{QCD}(M^2)$ then amounts to a simple 
change of the upper limit of integration, and yields 
the following sum rule for $f_B$:
$$
f_B^2m_B^4=
\frac{3m_b^2}{8\pi^2}
\int_{m_b^2} ^{s_0^B}ds
\frac{(s-m_b^2)^2}s\left( 1+ \frac{4\alpha_s}{3\pi}f(s,m_b^2)\right )
\exp\left(\frac{m_B^2-s}{M^2}\right)
$$
$$
+m_b^2\Bigg \{-m_b \langle \bar q q \rangle \left(1+
\frac{m_0^2}{2M^2}\left(1-\frac{m_b^2}{2M^2}\right)\right)
+\frac1{12}\langle  \frac{\alpha_s}{\pi}
G^a_{\mu\nu}G^{a\mu\nu} \rangle
$$
\be
-\frac{16\pi}{27}\frac{\alpha_s\langle \bar q q \rangle^2}{M^2}
\left(1-\frac{m_b^2}{4M^2}-\frac{m_b^4}{12M^4}\right)
\Bigg\}
\exp\left(\frac{m_B^2-m_b^2}{M^2}\right)~.
\label{fB1}
\ee
The uncertainty due to the crude subtraction procedure 
should not be too harmful because of the suppression 
of excited and continuum states after Borel transformation. 
In principle, since the l.h.s. of (\ref{fB1}) is a measurable 
quantity, the scale dependence of the quark masses, 
the condensates and the running coupling 
on the r.h.s. must cancel. In practice, this
can only be achieved approximately.

The $b$-quark mass appearing in the perturbative coefficient 
$C_0(q^2,\mu)$ is defined to be the pole mass, while the choice 
of $m_b$ in the leading-order coefficients of the higher-dimensional
terms in (\ref{fB1}) is arbitrary.
Following the usual procedure we take the pole mass
\be
m_b=4.7 \pm 0.1 ~\mbox{GeV}
\label{bmass}
\ee
everywhere in (\ref{fB1}).
The interval (\ref{bmass}) covers the current
estimates obtained from bottomonium sum rules \cite{mb}.

Turning to the condensates, it is important to note that 
the combinations $m_b\langle \bar{q}q\rangle $ 
and $\langle \frac{\alpha_s}{\pi} G^a_{\mu\nu}G^{a\mu\nu} \rangle$
are renormalization-group invariant.
The numerical value of the
quark condensate density at $\mu = O(1~GeV)$ is obtained
from the PCAC relation \cite{SVZ,Leutw}:
\be
\langle \bar q q  \rangle(1 ~\mbox{GeV}) = 
-\frac{f_\pi^2m_\pi^2}{2(m_u +m_d)}
\simeq-(240 ~\mbox{MeV})^3.
\label{condens}
\ee From 
that and the central value of (\ref{bmass}) 
converted to the running mass at 1 GeV, one finds
\be
m_b\langle \bar{q}q\rangle \simeq -0.084 ~\mbox{GeV}^4.
\ee
The gluon condensate density 
is determined from charmonium sum rules \cite{SVZ}:
\be
\langle \frac{\alpha_s}{\pi}
G^a_{\mu\nu}G^{a\mu\nu} \rangle \simeq 0.012 ~\mbox{GeV}^4~.
\label{glue}
\ee
For the remaining parameters we take 
\be
m_0^2(\mbox{1 GeV}) \simeq 0.8  ~\mbox{GeV}^2~,
\label{m02}
\ee
as extracted from sum rules for light baryons \cite{BI},
and 
\be
\alpha_s\langle \bar{q}q\rangle^2 = 8\cdot 10^{-5} ~\mbox{GeV}^6~,
\label{4qc}
\ee
as given in \cite{AE}.
The scale dependence 
of $m_0^2 \langle \bar{q}q\rangle$ and  $\alpha_s\langle \bar{q}q\rangle^2$
is negligible. The numerical uncertainties on the condensates
vary from 10 \% to about 50 \% or more. However, they influence the 
final result on $f_B$ rather little.

This leaves us with the choice of scale in  
the running coupling $\alpha_s$.
As the average virtuality of the quarks in the correlator (\ref{corr2})
is characterized by the Borel parameter $M^2$,
it is reasonable to take 
\be
\mu^2 = O(M^2)~.
\label{scale}
\ee

Finally, we have to determine the values of $M^2$ for which 
the sum rule (\ref{fB1}) can be trusted. 
On the one hand, $M^2$ has to be small enough such that 
the contributions from excited and continuum states are exponentially damped 
and the approximation (\ref{high2}) is satisfactory. 
On the other hand, the scale $M^2$ must be large enough 
such that higher-dimensional operators are suppressed and 
the OPE converges sufficiently fast. Consequently, there is at 
best a finite range of $M^2$ in which a given sum rule is valid. Moreover,
in this range the numerical result should be stable under variations 
of $M^2$. Whether or not these requirements can be met 
has to be investigated for each case separately. 
For the sum rule (\ref{fB1}) we have checked
that with
\be
3 ~\mbox{GeV}^2\leq M^2 \leq 5.5 ~\mbox{GeV}^2
\label{Borelint}
\ee
the necessary conditions are fulfilled:
the quark-gluon condensate contributes
less than 15 \%, the contributions from the  gluon condensate
and from the four--quark operators are negligibly small,
and the excited states and continuum contribute less than 30 \%.
Moreover, the change of $f_B$ due to variation of $M^2$
in the range (\ref{Borelint}) is indeed very small. 

The value of $f_B$ derived from the sum rule (\ref{fB1}) 
is given in Table 1. The value changes from 210 MeV to 150 MeV
when $m_b$ is varied 
in the range (\ref{bmass}) from 4.6 to 4.8 GeV,
and at the same time $s_0^B$ from 37 to 33 GeV$^2$.
The correlation between $m_b$ and $s_0^B$ 
improves the stability of the sum rule under variation of $M^2$.
As compared to the above uncertainty, the uncertainties from the 
condensates are negligible. Note that 
the effect of the $O(\alpha_s)$  correction (\ref{im2}) to $C_0$
is sizeable. Without this correction the value for $f_B$ is much lower:
\be
\overline{f}_B \equiv f_B(\alpha_s=0)
=140 \pm 30 ~\mbox{MeV}~.
\label{fB0}
\ee

Also shown in Table 1 is
the corresponding prediction on $f_D$. It follows from (\ref{fB1}) 
after replacing $m_b$ 
by the pole mass $m_c=1.3\pm 0.1 $ GeV of the $c$ quark, 
and $s_0^B$ by the effective threshold $s_0^D=6 \mp 1 $ GeV$^2$ 
in the $D$-meson channel. The allowed interval
in the Borel mass is found to be  
1 GeV$^2 < M^2 < 2$ GeV$^2$.
For later use, we again quote the result 
without the $O(\alpha_s)$ correction: 
\bq
\overline{f}_D\equiv f_D(\alpha_s=0)
=170 \pm 20 ~\mbox{MeV}.
\label{fDres}
\eq

In addition, Table 1 allows comparing QCD sum rule estimates 
with lattice results and with the available experimental data.
The mutual agreement, within the uncertainties of the two theoretical 
methods, is satisfactory. On the other hand, the data are just beginning
to challenge theory.
\begin{table}[htb]
\caption{\label{tab2}
{\it Decay constants of
$B$ and $D$ mesons in MeV.}}
\begin{center}
\begin{tabular}{|c||c|c|c|c|c|}
\hline
&&&&&\\
Method & Ref.& $f_B$ & $f_{B_s}$ & $f_D$& $f_{D_s}$ \\
&&&&&\\
\hline
&&&&&\\
&$\stackrel{\mbox{this}}{\mbox{review}}$& 180 $\pm$ 30 & ~~~--&
190 $\pm$ 20&~~~-- \\
QCD sum rules &&&&&\\
&\cite{Dom}~$^{a)}$ & 175& 210 & $180 \pm 10$ & $220 \pm 10$ \\
&&&&&\\
\hline
&&&&&\\
& \cite{Flynn}& 175$\pm$ 25 & 200$\pm$ 25 & 205$\pm$ 15 &235 $\pm$ 15\\
&&&&&\\
Lattice &\cite{APE}&180$\pm$ 32 & 205$\pm$ 35 &221$\pm$ 17&237$\pm$ 16\\
&&&&&\\
&\cite{Wittig}&$172 ^{+27}_{-31}$ &$196^{+30}_{-35}$ &$191^{+19}_{-28} $
&$206^{+18}_{-28}$\\
&&&&&\\
\hline
&&&&&\\
&\cite{m3}&~~~--&~~~--
&~~~-- &$< 310$\\
Experiment&&&&&\\
&\cite{fBDexp}~$^{b)}$ &~~~-- &~~~-- &~~~-- &241 $\pm$ 21$\pm$ 30\\
&&&&&\\
\hline
\end{tabular}
\\
\end{center}
\hspace*{1.5cm}
$^{a)}$ update of results of \cite{RRY,AE,fB} taking 
        $m_b=4.67$ GeV, $s_0^B=35$ GeV$^2$, \\
\hspace*{1.9cm}
        and $m_c=1.3$ GeV, $s_0^D=5.5$ GeV$^2$.\\
\hspace*{1.5cm}
$^{b)}$ world average  \\
\end{table}

It should be mentioned that the $B$--meson decay constant has been derived 
also from a two-point sum rule in the
heavy quark limit \cite{Shuryak} using HQET and
including $1/m_Q$ corrections  \cite{Neubert,BBBD}. In this approach,
the scale in $\alpha_s$ is lower than (\ref{scale}), of the order of 
the reduced mass in the $B$-meson bound state.
This leads to a somewhat larger value for $f_B$, close to the upper end
of the range obtained from the sum rule (\ref{fB1}).
For more details, one may consult the original papers quoted above.

Finally, for the calculation of the $B^*B\pi$ and $D^*D\pi$ couplings
in section 6, we will need the $B^*$ and $D^*$ decay constants.
Since they can be estimated from similar two-point sum rules as
$f_B$ and $f_D$, we briefly summarize the result here.
The decay constant of the $B^*$ is defined by the matrix element
\bq
\langle 0 \mid \bar{q} \gamma_\mu b \mid B^* \rangle = m_{B^*} f_{B^*} 
\epsilon_\mu ~,
\label{fBstar2}
\eq
$\epsilon_\mu$ being the $B^*$ polarization vector. 
From the correlation function of the 
vector currents $\bar{q}\gamma_\mu b$ and $\bar{b} \gamma_\mu q$
similar to (\ref{corr2}), and with the same approximations as in 
(\ref{fB1}),
one derives the following sum rule:
$$
f_{B^*}^2m_{B*}^2=
\frac{1}{8 \pi^2}\int_{m_b^2} ^{s_0^{B^*}} ds
\frac{(s-m_b^2)^2}s
\left(2+\frac{m_b^2}{s}\right)\left(1+ 
\frac{4\alpha_s}{3\pi}f^*(s,m_b^2) \right)\exp\left(
\frac{m_{B^*}^2-s}{M^2}\right)
$$
$$
+\Bigg\{ -m_b \langle \bar{q} q \rangle
\left( 1-\frac{m_0^2 m_b^2}{4M^4} \right)
+C_4(M^2)\langle \frac{\alpha_s}{\pi} G^a_{\mu\nu}G^{a\mu\nu}\rangle
$$
\be
+C_6(M^2)\alpha_s\langle\bar{q}q\rangle ^2 \Bigg\}
\exp\left(\frac{m_{B^*}^2-m_b^2}{M^2}\right) ~.
\label{fBstarSR}
\ee
The explicit expressions for the coefficient $f^*$ of the
$O(\alpha_s)$ correction, as well as the coefficients
$C_4$ and $C_6$ of the numerically insignificant contributions
from the gluon and four-quark condensates are given in
\cite{RRY}. The corresponding sum rule for $f_{D^*}$ can be
directly inferred from (\ref{fBstarSR}).
For the parameters one may take the same values as in the case of the
pseudoscalars.
For later use we give below the leading-order results in which
the $O(\alpha_s)$ loop-corrections to the coefficient
of the unity operator are not included:
\be
\bar{f}_{B^*} = f_{B^*} (\alpha_s = 0) = 160 \pm 30 ~\mbox{MeV}~,
\label{fBstar0}
\ee
\be
\bar{f}_{D^*} = f_{D^*} (\alpha_s = 0) = 240 \pm 20 ~\mbox{MeV}~.
\label{fDstar0}
\ee

\section{Three-point sum rules for heavy meson form factors}

We next turn to the more complicated matrix elements
of hadronic transitions induced by weak currents.
An important example is the $B\to \pi$ transition.
The relevant matrix element
is parametrized by two independent form factors: 
\be
\langle\pi(q)|\bar{u} \gamma_\mu b |B(p+q)\rangle=
2f^+(p^2)q_\mu +
\left(f^+(p^2)+f^-(p^2)\right)p_\mu
\label{form3}
\ee
with $p+q$, $q$ and $p$ being the $B$ and $\pi$ momenta, and the
momentum transfer, respectively.

Generalizing the method employed in section 2
one starts from the following vacuum correlation function of three
currents:
\be
T_{\alpha \mu}(p,q)=
-\int d^4xd^4y e^{ipx+iqy}
\langle 0|T\Big\{\bar{d}(y)\gamma_{\alpha} \gamma_5u(y),
\bar{u}(x)\gamma_{\mu}b(x),\bar{b}(0)i\gamma_5 d(0)\Big\}|0 \rangle~,
\label{corr3}
\ee
where the $B$ and $\pi$ meson states of (\ref{form3})  are replaced 
by the corresponding generating
currents. Insertion of the
complete sets of hadronic states carrying
$B$ and $\pi$  quantum numbers, respectively, 
yields the double dispersion relation
\bq
T_{\alpha\mu}(p,q)=
\frac{m_{B}^2 f_{B}f_{\pi}q_\alpha\left( 2f^+(p^2)q_\mu 
+
(f^+(p^2)+f^-(p^2))p_\mu  \right)
}{ m_b(m_B^2-(p+q)^2)
(m_{\pi}^2-q^2)}
\nonumber
\\
+ \int\int_{\Sigma_{12}}
ds_1ds_2 \frac{\rho^h_{\alpha\mu}
(s_1,s_2,p^2)}{(s_1-(p+q)^2)(s_2-q^2)}
\nonumber
\\
+P_1(q^2,p^2)\int_{\Sigma_1} 
ds_1\frac{\rho ^h _{1\alpha\mu}(s_1,p^2)}{s_1-(p+q)^2}
+P_2((p+q)^2,p^2)\int_{\Sigma_2} ds_2\frac{\rho ^h _{2\alpha\mu}
(s_2,p^2)}{s_2-q^2}~.
\label{t13}
\eq
In the above, the matrix elements (\ref{fB2}), (\ref{form3}), and
\be
\langle 0 \mid \bar{d}\gamma_\alpha\gamma_5 u \mid \pi(q) \rangle=
if_\pi q_\alpha ~
\label{fpi3}
\ee
have been used. While  
the first term on the r.h.s. of (\ref{t13}) is the 
the $B$ and $\pi$ ground state contribution, the integral
over the double spectral density $\rho^h_{\alpha\mu}$
takes into account the contributions of higher resonances and
continuum states in the  $B$ and $\pi$ channels.
$\Sigma_{12}$ denotes the integration region in the $(s_1,s_2)$ plane. 
$\Theta$-functions defining the actual thresholds
are implicitly contained  in $\rho^h_{\alpha\mu}$.  
The additional single dispersion integrals multiplied
by polynomials $P_{1,2}$
arise from subtractions, similar to the polynomial
in (\ref{dispersion}). They vanish after Borel transformation. 

\begin{figure}[ht]
\mbox{
\epsfig{file=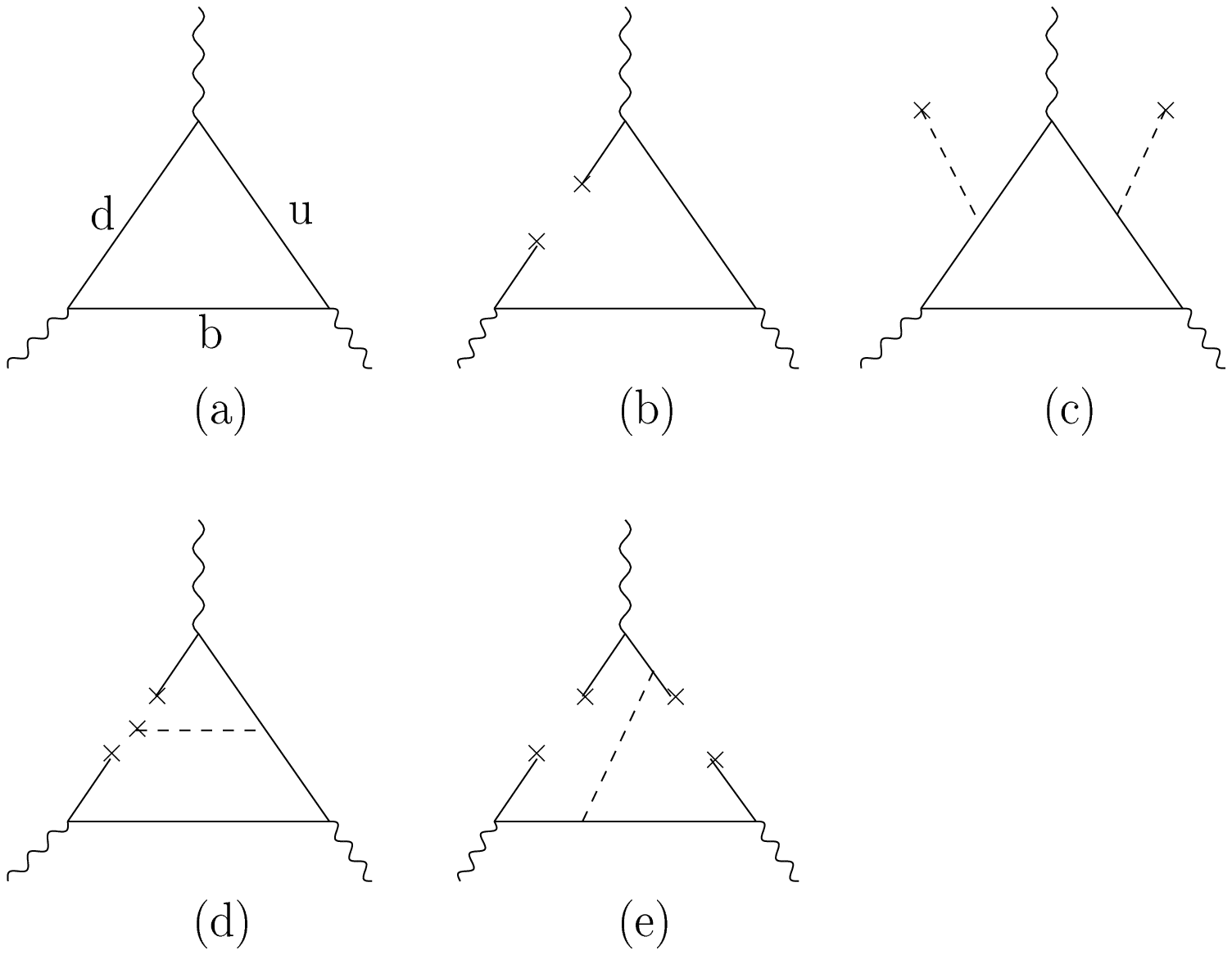,width=\textwidth,bbllx=0pt,bblly=209pt,bburx=600pt,%
bbury=600pt,clip=}
}
\caption{\it Diagrams determining the Wilson coefficients 
of the OPE of the three-point correlation function
(\ref{corr3}). The symbols are as in Fig. 1.}
\end{figure}

At $(p+q)^2 \ll m^2_b$, $q^2 \ll 0$,
and $p^2 \leq m_b^2 - 2 m_b \chi$, 
$\chi$ being a $m_b$-independent scale of order $\Lambda_{QCD}$,  
the correlation function $T_{\alpha \mu}(p,q)$ 
can be approximated by the first few terms in 
the OPE of the $T$-product of currents 
in (\ref{corr3}), in analogy to (\ref{ope112}): 
\be
T^{\alpha \mu}(p,q)
= \sum_d C^{\alpha\mu}_d(p,q,\mu)
\langle  \Omega_d(\mu) \rangle.
\label{ope113}
\ee
The restriction on $p^2$ is necessary
in order to stay sufficiently far away from
the physical states in the $\bar{b}u$ channel, 
most notably the $B^*$.
For $d \leq 6$ the short-distance coefficients $C^{\alpha\mu}_d$
can be calculated from the diagrams shown in Fig. 3.
The calculation follows the
procedure outlined in the previous section. However,
the kinematics is more complicated 
because of the presence of two independent external four-momenta.
The explicit expressions 
can be found in  \cite{AEK,BBD1,Ball}.
Decomposing $C^{\alpha\mu}_d$
as well as the spectral densities $\rho^h_{\alpha\mu}$
in the independent tensor structures, 
\be
C_d^{\alpha\mu}(p,q) = C_d((p+q)^2,q^2,p^2)q^\alpha(2q+p)^\mu+...~,
\label{expans3}
\ee 
\be
\rho^h_{\alpha\mu}(s_1,s_2,p^2)= \rho^h(s_1,s_2,p^2)q_\alpha (2q+p)_\mu +...~.
\label{spectral}
\ee
and equating the corresponding invariant functions
in (\ref{t13}) and (\ref{ope113}), one obtains 
$$
\frac{m_{B}^2 f_{B}f_{\pi}f^+(p^2)}{ m_b(m_B^2-(p+q)^2)
(m_{\pi}^2-q^2)}
+ \int\int_{\Sigma_{12}}  d s_1ds_2 \frac{\rho^h(s_1,s_2,p^2)}
{(s_1-(p+q)^2)(s_2-q^2)} + ...
$$
\be
=\sum_{d} C_d((p+q)^2,q^2,p^2,\mu)\langle  \Omega_d(\mu) \rangle~,
\label{rel3}
\ee
where the ellipses denote the subtraction terms.
Since light quarks are taken massless, we put $m_\pi=0$ for 
consistency.

Analogously to (\ref{high2}),
the double spectral density
$\rho^h $ is approximated by
\be
\rho^{h}(s_1,s_2,p^2)=
\frac{1}{\pi^2}\mbox{Im}_{s_1} \mbox{Im}_{s_2} C_0(s_1,s_2,p^2)
\Theta (s_1-s_0^B)\Theta (s_2-s_0^\pi)~,
\label{high3}
\ee
where $C_0$ is the perturbative coefficient calculated from Fig. 3a,
and $s_0^\pi$ is the effective threshold parameter
in the $\pi$-meson  channel. After
Borel transformation (\ref{Borel2})
with respect to the variables $(p+q)^2$ and $q^2$ the subtraction terms
disappear from (\ref{rel3}) and the contributions from excited
and continuum states become exponentially suppressed.
The final sum rule for $f^+$ takes the following schematical form
\cite{AEK,BBD1,Ball,3point}:
$$
f^+(p^2)
= \frac{m_b}{m_{B}^2f_Bf_\pi}
\exp\left(\frac{m_B^2}{M_1^2} +\frac{m_\pi^2}{M_2^2}\right)
$$
$$
\times \Bigg \{\frac{1}{\pi^2}\int \int_{\Sigma_{12}(s_0^B,s_0^\pi)}
ds_1 ds_2 \mbox{Im}_{s_1}\mbox{Im}_{s_2}C_{0}(s_1,s_2,p^2,\mu)
\exp\left(-\frac{s_1}{M_1^2} -\frac{s_2}{M_2^2}\right)
$$
\be
+\sum_{d =3}^6C_{d}(M_1^2,M_2^2,p^2,\mu)
\langle {\Omega}_d(\mu) \rangle
\Bigg\}~
\label{sr3}
\ee
with $M_1^2$ and $M_2^2$ being the Borel parameters in the $B$ and $\pi$
channels, respectively. The appearance of the threshold parameters
$s_0^B$ and $s_0^{\pi}$ reflects the continuum subtraction.
Similarly as for the decay constants, there may
be important perturbative QCD
corrections to the short-distance coefficients,
in particular, to $C_0$. To our knowledge, the two-loop diagrams
corresponding to hard-gluon exchanges in Fig. 3a
have not yet been calculated.

The numerical estimate \cite{BBD2} for $f^+(p^2)$ obtained 
from (\ref{sr3})  
is plotted in Fig. 4 in comparison with other 
predictions. The values for $m_b$, $s_0^B$, and the 
vacuum condensates are already stated in the previous section; for
$f_B$ the leading-order estimate (\ref{fB0}) is used for consistency.
Furthermore, the pion decay constant 
is $f_\pi= 132$ MeV, while the threshold 
parameter $s_0^\pi = 0.75$ GeV$^2$ is inferred from the 
two-point sum rule \cite{SVZ} for the
correlation function  $\langle 0\mid T\{\bar{d}\gamma_\alpha\gamma_5 u(x),
\bar{u}\gamma_\beta\gamma_5 d(0)\}\mid 0\rangle $. 
\begin{figure}[htb]
\centerline{
\epsfig{bbllx=100pt,bblly=209pt,bburx=507pt,%
bbury=490pt,file=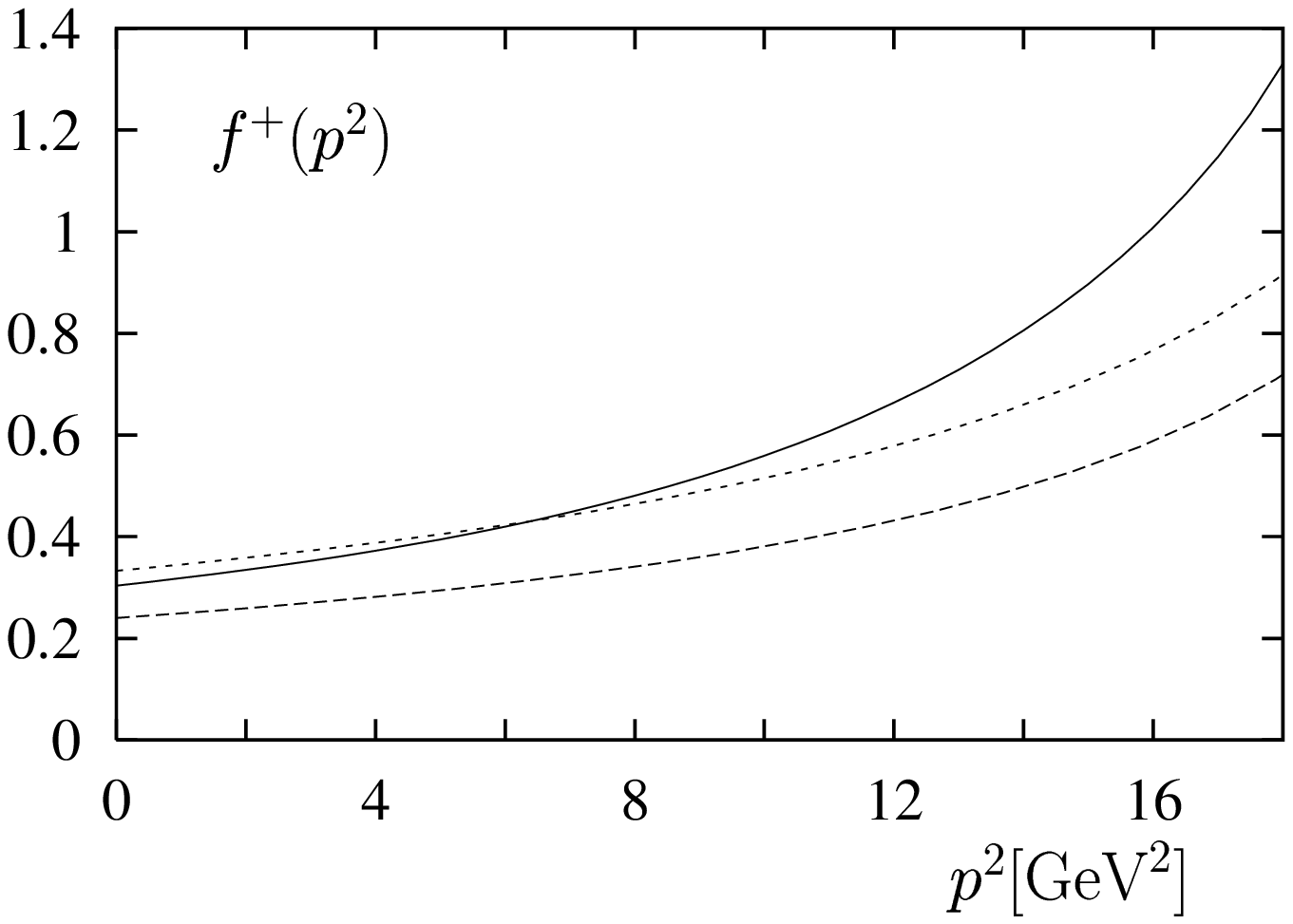,scale=0.9,%
clip=}
}
\caption{\it $B \to \pi$ form factor $f^+$ 
calculated in different approaches: 
three-point sum rule \cite{BBD2} (dashed),
quark model \cite{BSW} (dotted), and
light-cone sum rule \cite{BKR,KRW} (solid).}
\end{figure}

Similar three-point sum rules 
have been applied to a variety of hadronic
problems including
the pion form factor \cite{pion}, 
radiative charmonium transitions  \cite{RRY,AK,BRcc},
and rare $B$ decays \cite{penguin3point,Ballpenguin,COL96}.
However, there are theoretical difficulties which put some doubts 
on the reliability of this method.
Most disturbing in 
the case of heavy-to-light form factors
is the breakdown of the OPE in the heavy mass 
limit. More specifically, the coefficients of the subleading quark and 
quark-gluon condensate terms grow faster with $m_b$ 
than the coefficient of the leading unity operator, that is the 
perturbative contribution. 
In contrast,
the heavy mass limit of the two-point sum rules for 
decay constants
discussed in the previous section is completely well-behaved.  
A more detailed discussion on this issue 
can be found in \cite{BallBraun,ABS}.

Although three-point sum rules still 
remain a useful calculational tool for selected problems,
the above difficulties call for more consistent and reliable
sum rule methods. One important development is explained 
in the next section. 


\section{Light-cone expansion}

For processes where a light meson such as a $\pi$, $K$, or $\rho$
is involved, there is an interesting alternative  
to the short-distance OPE of vacuum-vacuum correlation 
functions in terms of condensates,
namely the expansion 
of vacuum-meson correlators near the light-cone 
in terms of meson wave functions \cite{exclusive,BL,CZ}.
The latter are functions of the light-cone momentum fractions
carried by the constituents of a given meson.  
Similarly as the deep-inelastic structure functions,
the wave functions can be classified
by the twist of the corresponding composite operators.
While the light mesons are taken on mass shell,
the heavy meson channels are treated in the 
usual way: choice of a generating current,
contraction of the heavy quark fields,
dispersion relation, Borel transformation and continuum subtraction.
The light-cone variant of QCD sum rules suggested in \cite{BBK,BF,CZ1}
allows
to incorporate additional information about the Euclidean asymptotics
of correlation functions in QCD for arbitrary external momenta.
Moreover, it avoids the problems of the three-point sum rules 
mentioned at the end of the previous section.

Here, we explain the general idea using 
the $B \rightarrow \pi$  transition element (\ref{form3}) as an example.
The detailed derivation of the light-cone sum rules
for $f^+$ and $f^-$ \cite{BKR,BBKR,KRW} is postponed 
to the next section. The starting point is the vacuum-pion
correlation function 
\be
F_\mu (p,q)=
i \int d^4xe^{ipx}\langle \pi(q)\mid T\{\bar{u}(x)\gamma_\mu b(x),
\bar{b}(0)i\gamma_5 d(0)\}\mid 0\rangle
\label{1a4}
\ee
$$
= F(p^2,(p+q)^2) q_\mu + \tilde{F}(p^2,(p+q)^2) p_\mu~.
$$
Since the pion is on-shell, $q^2 = m_\pi^2$ vanishes in the
chiral limit adopted throughout this discussion.
For the momenta in the $\bar{b}d$ and $\bar{u}b$ channels 
we respectively require  $(p+q)^2 \ll m_b^2$ and 
$p^2 \leq m_b^2 -2 m_b \chi$, as before.
Contracting the $b$-quark fields in (\ref{1a4}) 
and inserting the free $b$-quark propagator (\ref{S02}), 
one gets
\bq
F^{(a)}_\mu(p,q)&=&i\int \frac{
d^4x\,d^4k}{(2\pi )^4(m_b^2-k^2)}
e^{i(p-k)x}\left(m_b
\langle \pi (q)|\bar{u}(x)\gamma_\mu\gamma_5d(0)|0\rangle \right.
\nonumber
\\
&&{}+\left.
k^\nu \langle\pi(q) |\bar{u}(x)\gamma_\mu\gamma_\nu\gamma_5
d(0)|0\rangle\right)~.
\label{234}
\eq
This contribution is diagrammatically depicted in Fig.~5a.

\begin{figure}[ht]
\mbox{
\epsfig{file=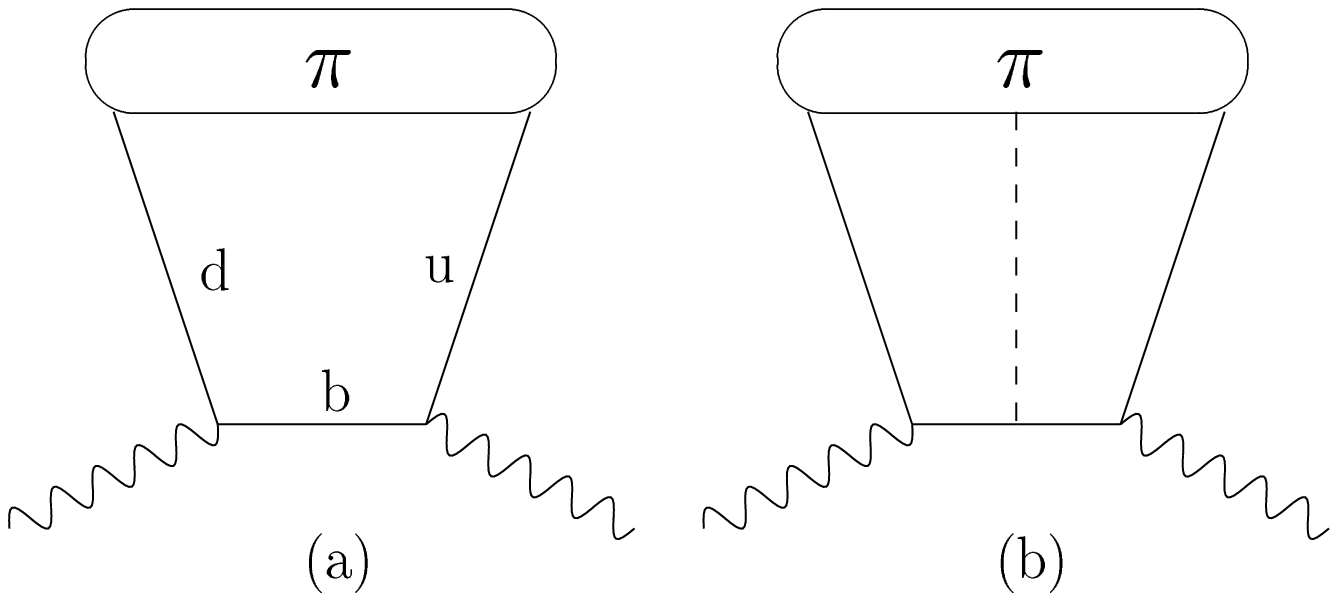,width=\textwidth,bbllx=0pt,bblly=290pt,bburx=600pt,%
bbury=520pt,clip=}
}
\caption{\it Diagrammatic representation of the correlation
function (\ref{1a4}): terms involving two-particle (a) 
and three-particle (b) wave functions.
Solid lines represent quarks, dashed lines gluons,
wavy lines external currents, and
ovals light-cone wave functions of the pion.}                       
             
\end{figure}

Let us first focus on the matrix element
\be
\langle \pi (q)|\bar{u}(x)\gamma_\mu\gamma_5 d(0)|0\rangle ,
\label{matrix4}
\ee
and expand the bilocal quark-antiquark
operator around $x=0$: 
\be
\bar{u}(x)\gamma_\mu\gamma_5d(0)=\sum_r\frac{1}{r!}
\bar{u}(0)(\stackrel{\leftarrow}{D}\cdot x)^r\gamma_\mu\gamma_5 d(0)~.
\label{expan4}
\ee
The matrix elements of the local operators
can be written in the form  
\be
\langle\pi(q) |\bar{u}\stackrel{\leftarrow}{D}_{\alpha_1}
\stackrel{\leftarrow}{D}_{\alpha_2}...
\stackrel{\leftarrow}{D}_{\alpha_r}\g_\mu\g_5 d |0\rangle        
 =(i)^r q_\mu q_{\alpha_1} q_{\alpha_2}...q_{\alpha_r}M_r + ...~,
\label{local4}
\ee
where the ellipses stand for 
additional terms containing various combinations of the  
metric tensor $g_{\alpha_i\alpha_k}$.
Note that the lowest twist \footnote{
Twist is defined as the difference between
the canonical dimension and the spin of a traceless
and totally symmetric local operator.} in (\ref{local4})
is equal to 2.
Substituting (\ref{expan4}) and (\ref{local4}) in the first
term of (\ref{234}), integrating over $x$ and $k$,
and comparing the result with (\ref{1a4}), one obtains
\bq
F(p^2,(p+q)^2)=i\frac{m_b}{m_b^2-p^2}
\sum_{r=0}^\infty \xi ^rM_r~
\label{expans4}
\eq
with 
\bq
\xi =\frac{2(p \cdot q)}{m_b^2-p^2}= \frac{(p+q)^2-p^2}{m_b^2-p^2}~.
\label{ksi4}
\eq
Now, one immediately encounters the following problem. If
the ratio $\xi$
is finite one must keep an infinite series of matrix elements
in (\ref{expans4}). All of them give
contributions of the same order in the heavy quark propagator
$1/(m_b^2-p^2)$, differing only by powers
of the dimensionless parameter $ \xi$.
In other words,
the expansion (\ref{expan4})
is useful only in the case  $ \xi \rightarrow 0$,
i.e., when $p^2 \simeq (p+q)^2$,
or equivalently when the momentum of the light meson vanishes.
Under this condition, the series in
(\ref{expans4}) can be truncated after a few terms 
involving only a manageable
number of unknown matrix elements $M_r$. However, generally, when 
$p^2 \neq (p+q)^2$ one has to sum up the infinite series of matrix elements
of local operators in some way.

One possible solution is provided by expanding 
the bilocal operator (\ref{expan4}) near the light-cone.
For the matrix element (\ref{matrix4}) the
first term of this expansion is given by 
\bq
\langle\pi(q)|\bar{u}(x)\g_\mu\g_5d(0)|0\rangle_{x^2=0}=
-iq_\mu f_\pi\int_0^1du\,e^{iuqx}\varphi_\pi (u)  ~.
\label{pionwf4}
\eq
The function  $\varphi_\pi(u)$ is known as the twist 2
light-cone wave function of the pion.
It is normalized to unity, as can be seen by putting
$x=0$ in (\ref{pionwf4}). As already mentioned, $\varphi_\pi$ 
represents the distribution in the fraction
of the light-cone momentum $q_0 +q_3 $ of the pion carried by
a constituent quark. The path-ordered gluon operator
\be
Pexp \left\{ig_s \int ^1_0d\alpha~ x_\mu A^\mu (\alpha x)\right \}
\label{factor}
\ee
necessary for gauge invariance of the matrix element (\ref{matrix4}),
is unity in the light-cone gauge, $x_\mu A^\mu=0$,
assumed here.  Substitution of (\ref{pionwf4}) in (\ref{234})
and integration over $x$ and $k$ yields 
\be
F(p^2,(p+q)^2)=m_bf_\pi\int_0^1\frac{du ~\varphi_\pi(u) }{m_b^2-(p+uq)^2} ~.
\label{Fzeroth4}
\ee
We see that the infinite series of matrix elements of local operators
encountered before in (\ref{expans4})
is effectively replaced by a wave function.
If one expands (\ref{Fzeroth4}) in $q$,
\be
F(p^2,(p+q)^2)=m_bf_\pi\sum_{r=0}^{\infty}
\frac{(2p\cdot q)^r}{(m_b^2-p^2)^{r+1}}
\int_0^1du ~u^r\varphi_\pi(u)~,
\label{Fexp}
\ee
and compares the above expression  with (\ref{expans4}),
one can directly read off the relation between the matrix elements $M_r$
defined in (\ref{local4})
and the moments of $\varphi_\pi(u)$:
\be
M_r= -if_\pi
\int_0^1du ~u^r\varphi_\pi(u).
\label{local}
\ee
It should be noted that the pion wave function $\varphi_\pi(u)$ 
is a universal function. It is the same quantity 
which enters, for example, the $ \pi^0 \g ^* \g^* $ form factor
\cite{exclusive}.
This universality is essential for the whole
approach, similarly as the universality of the
vacuum condensates for the short-distance sum rules.

Including the next-to-leading terms in $x^2$, the light-cone
expansion  of the matrix element (\ref{matrix4})
reads
\begin{eqnarray}
\langle\pi(q)|\bar{u}(x)\gamma_\mu\gamma_5d(0)|0\rangle =
-iq_\mu f_\pi\int_0^1du\,e^{iuqx}
\left(\varphi_\pi (u)+x^2g_1(u)\right)
\nonumber \\
+
f_\pi\left( x_\mu -\frac{x^2q_\mu}{qx}\right)\int_0^1
du\,e^{iuqx}g_2(u),
\label{phi}
\end{eqnarray}
where  $g_1$
and $g_2$ are twist~4 wave functions.
Proceeding to the second term
in (\ref{234}) and using the relation  
\be
\gamma_\mu\gamma_\nu\gamma_5= g_{\mu\nu}\gamma_5-
i\sigma_{\mu\nu}\gamma_5~,
\label{gamma1}
\ee
one encounters two further matrix elements:
\be
\langle\pi(q)\mid \bar{u}(x)i\gamma_5d(0)\mid 0\rangle=
f_\pi \mu_\pi \int_0^1du~e^{iuqx}\varphi_{p}(u) 
\label{phip}
\ee
and
\be
\langle\pi(q)\mid\bar{u}(x)\sigma_{\mu\nu}\gamma_5d(0)\mid 0\rangle=
i(q_\mu x_\nu -q_\nu x_\mu )\frac{f_\pi \mu_\pi}{6}
\int_0^1 du ~e^{iuqx}\varphi_{\sigma }(u) \, 
\label{phisigma}
\ee
with $\mu_\pi=m_\pi^2/(m_u + m_d)$. 
Only the leading terms in the expansion are considered here. 
They have twist 3 and are parameterized by the
wave functions $\varphi_p$ and $\varphi_\sigma$.

Beyond twist 2, one should also take into account
the higher-order terms \cite{BB} resulting from the contraction 
of the $b$-quark fields in the correlator (\ref{1a4}):
\begin{eqnarray}
\langle 0 |T\{b(x)\bar{b}(0)\}|0\rangle &=& i\hat{S}^b(x)
-ig_s\int\frac{d^4k}{(2\pi )^4}e^{-ikx}
\int_0^1dv\left[ \frac12\frac{\not\!k+m_b}{(m^2_b-k^2)^2}
G^{\mu\nu}(vx)\sigma_{\mu\nu}\right.
\nonumber\\
&& \mbox{}
+\left.\frac1{m_b^2-k^2}vx_\mu G^{\mu\nu}(vx)\g_\nu
\right],
\label{32}
\end{eqnarray}
where $\hat{S}^b$
is the free propagator given in (\ref{S02}), 
and $G_{\mu\nu}=G_{\mu\nu}^c\frac{\lambda^c}2$ with
$\mbox{\rm tr}(\lambda^a\lambda^b)=2\delta^{ab}$.
Insertion of the gluonic term in the 
correlation function (\ref{1a4}) yields the contribution
represented by the diagram Fig. 5b:
\bq
F_\mu^{(b)}
(p,q)
&=&i g_s \int\frac{d^4k\,d^4x\,dv}{(2\pi)^4(m_b^2-k^2)}
e^{i(p-k)x}\langle \pi |\bar{u}(x)\gamma_\mu
\Bigg( vx_\rho G^{\rho\lambda}(vx)\gamma_\lambda
\nonumber
\\
&&{}+\frac12\frac{\not\!k+m_b}{m_b^2-k^2}G^{\rho\lambda}(vx)
\sigma_{\rho\lambda}\Bigg)\gamma_5d(0)|0\rangle~.
\label{33}
\eq
Use of the identities
\bq
\gamma_\mu\sigma_{\rho\lambda}&=&
i(g_{\mu\rho}\gamma_\lambda-g_{\mu\lambda}\gamma_\rho)
+\varepsilon_{\mu\rho\lambda\delta}\gamma^\delta\g_5
\label{ggg}
\eq
and
\bq
\gamma_\mu\gamma_\nu\sigma_{\rho\lambda}&=&
(\sigma_{\mu\lambda}g_{\nu\rho}-\sigma_{\mu\rho}g_{\nu\lambda})+
i(g_{\mu\lambda}g_{\nu\rho}-g_{\mu\rho}g_{\nu\lambda})
\nonumber
\\
&&\mbox{}-
\varepsilon_{\mu\nu\rho\lambda}\gamma_5-i\varepsilon_{\nu\rho\lambda\alpha}
g^{\alpha\beta}
\sigma_{\mu\beta}\gamma_5~
\label{34}
\eq
leads to the following matrix elements
of quark-antiquark-gluon operators which 
bring three-particle wave functions into the game: 
\bq
\lefteqn{
\langle\pi |\bar{u}(x)
\sigma_{\mu\nu}\gamma_5g_s G_{\alpha\beta}(vx)d(0)|0\rangle =}
\nonumber\\
&=&if_{3\pi}[(q_\alpha q_\mu g_{\beta\nu}-q_\beta q_\mu g_{\alpha\nu})
-(q_\alpha q_\nu g_{\beta\mu}-q_\beta q_\nu g_{\alpha\mu})]
\int{\cal D}\alpha_i\,\varphi_{3\pi}(\alpha_i)e^{iqx(\alpha_1+v\alpha_3)}\,,
\label{29}
\eq
\bq
\lefteqn{
\langle\pi |\bar{u}(x)\gamma_\mu\gamma_5 g_sG_{\alpha\beta}(vx)d(0)|0\rangle =}
\nonumber
\\
&=&f_\pi\left[ q_\beta\left( g_{\alpha\mu}-\frac{x_\alpha q_\mu}{qx}\right) -
q_\alpha\left(g_{\beta\mu}-\frac{x_\beta q_\mu}{qx}\right)\right]
\int{\cal D}\alpha_i\varphi_\perp (\alpha_i)e^{iqx(\alpha_1+v\alpha_3)}
\hspace{1.5cm}{}
\nonumber
\\
&& {}+f_\pi\frac{q_\mu}{qx}(q_\alpha x_\beta -q_\beta x_\alpha )
\int{\cal D}\alpha_i\,\varphi_\parallel (\alpha_i)e^{iqx
(\alpha_1+v\alpha_3)}\,,
\label{30}
\eq
\bq
\lefteqn{
\langle\pi |\bar{u}(x)\gamma_\mu g_s\tilde{G}_{\alpha\beta}(vx)d(0)|0\rangle=}
\nonumber
\\
&=&if_\pi\left[ q_\beta\left( g_{\alpha\mu}-\frac{x_\alpha q_\mu}{qx}\right) -
q_\alpha\left( g_{\beta\mu}-\frac{x_\beta q_\mu}{qx}\right)\right]
\int{\cal D}\alpha_i\,\tilde{\varphi}_\perp (\alpha_i)e^{iqx
(\alpha_1+v\alpha_3)}
\hspace{1.4cm}{}
\nonumber
\\
&&{}+if_\pi\frac{q_\mu}{qx}(q_\alpha x_\beta -q_\beta x_\alpha )
\int{\cal D}\alpha_i\,\tilde{\varphi}_\parallel (\alpha_i)
e^{iqx(\alpha_1+v\alpha_3)}
\label{31}
\eq
with $\tilde{G}_{\alpha\beta}= \frac12
\epsilon _{\alpha\beta \sigma\tau}G^{\sigma\tau}$, and
${\cal D}\alpha_i= d\alpha_1 d\alpha_2 d\alpha_3 \delta(1-\alpha_1-\alpha_2
-\alpha_3)$. 
The wave function 
$\varphi_{3\pi}(\alpha_i)=\varphi_{3\pi}(\alpha_1,\alpha_2,\alpha_3)$ 
has twist 3, while  
$\varphi_\perp$, $\varphi_\parallel$,
$\tilde{\varphi}_\perp$ and $\tilde{\varphi}_\parallel$ are all of twist 4.
Gluons emitted from the light quark lines in Fig. 5a  are effectively
taken care of by the twist 3 and 4 wave functions as was shown in
\cite{BF,Gorsky,BF2}. 
Components of the pion wave function with two extra gluons,
or with an additional $\bar q q$ pair
are neglected.

Finally, a comment is in order concerning the  
perturbative corrections to the correlation
function (\ref{1a4}) from hard gluon exchanges. 
The first-order diagrams are shown in Fig. 6. 
Returning to the twist 2 approximation (\ref{Fzeroth4}) of the invariant
function F, one can more generally write it  
as a convolution of a hard scattering 
amplitude $T$ with the twist 2 wave function:
\begin{eqnarray}\label{represent}
F(p^2,(p+q)^2)=f_\pi\int^1_0 du \varphi_\pi (u,\mu) T(p^2,(p+q)^2,u,\mu)~,
\end{eqnarray}
where $T$ is given by 
\begin{eqnarray}
T(p^2,(p+q)^2,u,\mu)=T_0(p^2,(p+q)^2,u)+\frac{\alpha_s C_F}{4\pi}
T_1(p^2,(p+q)^2,u,\mu)
+ O(\alpha_s^2)~.
\label{T}
\end{eqnarray}
As well known, the renormalization procedure and the factorization of the 
collinear logarithms generated by gluon radiation 
induce scale dependences. In (\ref{represent}) we have
chosen a common scale $\mu$, for simplicity.

\begin{figure}[htb]
\centerline{
\epsfig{file=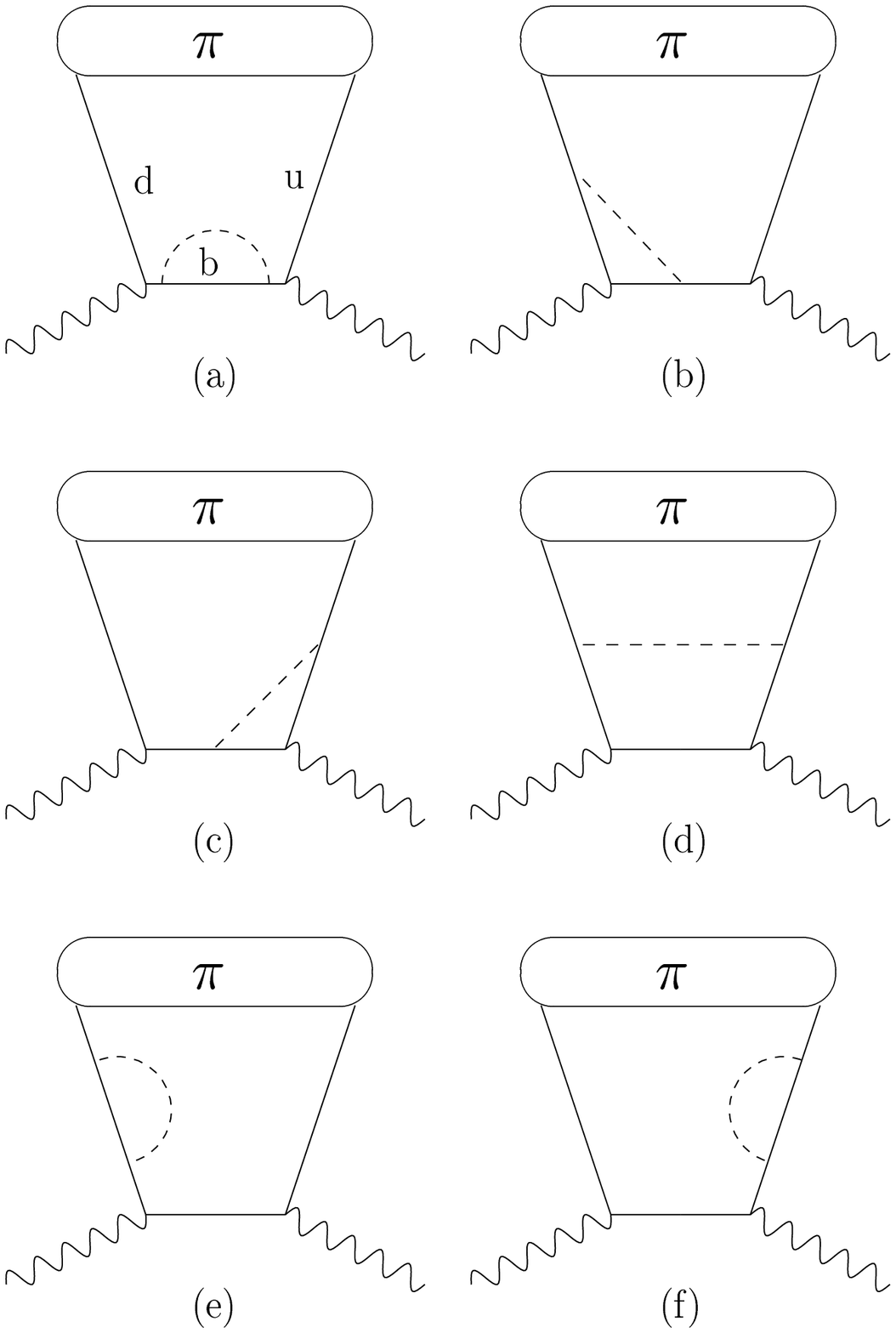,width=12cm,bbllx=0pt,bblly=100pt,bburx=600pt,%
bbury=720pt,clip=}
}
\caption{\it Diagrams of the $O(\alpha_s)$ corrections to the correlation
function (\ref{1a4}).}
\end{figure}

In leading-order approximation (LO), the lowest-order amplitude  
\begin{equation}
 T_0(p^2,(p+q)^2,u)=\frac{m_b}{m_b^2-p^2(1-u)-(p+q)^2u}
\label{Born}
\end{equation}
is convoluted with the wave function obeying 
the Brodsky-Lepage evolution equation \cite{exclusive}:
\begin{equation}
d\varphi_\pi (u,\mu)/d\ln \mu = \int\limits_{0}^{1} dw
V(u,w)\varphi_\pi(w,\mu)
\label{BLL}
\end{equation}
with 
\begin{eqnarray}
V(u,w)  = \frac{\alpha_s(\mu) C_{F}}{\pi} \left[ \frac{1-u}{1-w}
\left( 1 + \frac{1}{u-w} \right) \Theta(u-w)
+ \frac{u}{w} \left( 1 + \frac{1}{w-u} \right)
 \Theta(w-u) \right]_{+}~,
\label{BL}
\end{eqnarray}
and
\begin{eqnarray}
 R(u,w)_{+} & = & R(u,w) - \delta(u-w) \int\limits_{0}^{1} R(v,w) dv ~.
\label{plus}
\end{eqnarray}
This equation effectively sums up the leading logarithms
to all orders. 
It can be solved by expanding $\varphi_\pi(u,\mu)$ in
terms of Gegenbauer polynomials in which case the coefficients
are multiplicatively renormalizable \cite{exclusive}.

The next-to-leading order approximation (NLO)
is given by the hard amplitude (\ref{T}) 
including the first-order term $T_1$ \cite{KRWY,Bagan}, together with
the wave function $\varphi_\pi(u,\mu)$ solving the evolution equation in
next-to-leading order \cite{kad}. 
To date, the NLO program is completed
only for the twist 2 contribution to the invariant amplitude $F$. 
The relevant wave function parameters and correction terms 
are given in Appendices 1 and 2.


\section{Light-cone sum rules for heavy-to-light form factors}

In order to determine the $B\to \pi$ form factors $f^+(p^2)$ and $f^-(p^2)$
from the
correlation function (\ref{1a4}), we employ a QCD sum rule with
respect to the $B$-meson channel following essentially the same
steps as in the derivation of  the sum rule for $f_B$ in section 2.
The hadronic representation of (\ref{1a4}) is obtained by
inserting a complete set
of intermediate states with $B$-meson quantum numbers:
$$
F_\mu (p,q)=
\frac{\langle \pi \mid \bar{u}\gamma_\mu b\mid B\rangle
\langle B\mid \bar{b}i\gamma_5d\mid 0\rangle}{m_B^2-(p+q)^2}
$$
\be
\label{2a}
+\sum_h \frac{ \langle \pi \mid \bar{u}\gamma_\mu b\mid h\rangle
\langle h\mid \bar{b}i\gamma_5 d\mid 0\rangle}{m_h^2-(p+q)^2}~.
\ee
Using the matrix elements (\ref{fB2}) and (\ref{form3}),
and representing the sum over
excited and continuum states by a dispersion integral
with the spectral density $\rho^h_\mu = \rho^h q_\mu + \tilde{\rho}^hp_\mu$,
one obtains the following relations for the invariant amplitudes
$F$ and $\tilde{F}$:
\be
F(p^2,(p+q)^2)=
\frac{2m_B^2f_Bf^+(p^2)}{m_b(m_B^2-(p+q)^2)}
+\int\limits_{s_0^h}^{\infty}
\frac{\rho^h (p^2,s)ds}{s-(p+q)^2}~,
\label{dispa}
\ee
\be
\widetilde{F}(p^2,(p+q)^2)=
\frac{m_B^2 f_B(f^+(p^2)+f^-(p^2))}{m_b(m_B^2-(p+q)^2)}
+\int\limits_{s_0^h}^{\infty} ds \frac{\tilde{\rho}^h(p^2,s)}{s-(p+q)^2}~.
\label{hadr4}
\ee
Similarly as in 
(\ref{high2}), the integrals over 
$\rho^h$ and $\tilde{\rho}^h$ are again approximated by integrals over
the corresponding spectral densities 
calculated from the light-cone 
expansion of (\ref{1a4}) using
\be
\rho^h(p^2,s) \Theta(s - s_0^h)
 = \frac1\pi \mbox{Im} F_{QCD}(p^2, s) \Theta( s-s_0^B) ~,
\label{rhoh5}
\ee
and the analogous relation for $\tilde{\rho}_h$ and 
$\mbox{Im}\tilde{F}_{QCD}$.
The calculation of 
$F_{QCD}$ and $\tilde{F}_{QCD}$ 
has already been outlined in the preceding section.
With (\ref{rhoh5}) it is now straightforward to subtract
the contribution of the excited and continuum states 
from the corresponding integrals 
on the l.h.s. of (\ref{dispa}) and (\ref{hadr4}).
After performing 
the Borel transformation in $(p+q)^2$, one finally arrives at the  
sum rules
\begin{eqnarray}
f_Bf^+(p^2)=\frac{m_b}{2\pi m^2_B}\int\limits_{m_b^2}^{s_0^B} 
\mbox{Im} F_{QCD}(p^2, s)
\exp\left(\frac{m^2_B-s}{M^2}\right )ds~,
\label{sr}
\end{eqnarray}
\begin{eqnarray}
f_B(f^+(p^2)+f^-(p^2))=\frac{m_b}{\pi m^2_B}\int\limits_{m_b^2}^{s_0^B} 
\mbox{Im} \tilde{F}_{QCD}(p^2, s)
\exp\left(\frac{m^2_B-s}{M^2}\right)ds~.
\label{sr5}
\end{eqnarray}
From the above, the form factor $f^-$ follows directly by subtraction,
while the scalar form factor $f^0$ is given by  
\be
f^0(p^2) = f^+(p^2) + \frac{p^2}{m_B^2-m_\pi^2} f^-(p^2) ~.
\label{f0}
\ee

The remaining task is to complete the calculation of
$F_{QCD}$ and $\tilde{F}_{QCD}$, and to determine the
imaginary parts. We proceed along the lines 
explained in section 4. The contribution of two-particle
wave functions is derived from the  diagram Fig. 5a  
inserting the matrix elements (\ref{phi}), (\ref{phip}), and
(\ref{phisigma}) in (\ref{234}). This results in
\bq
F^{(a)}_{QCD}(p^2,(p+q)^2)=f_\pi\int\limits_{0}^{1}
\frac{du}{m_b^2-(p+uq)^2}\Bigg\{
m_b\varphi_\pi (u)
\nonumber
\\
+
\mu_\pi\Bigg[ u\varphi_p(u)+
\frac16 \left(
2+\frac{p^2+m_b^2}{m_b^2-(p+uq)^2}\right)\varphi_\sigma (u)
\Bigg]
\nonumber
\\
+m_b\left[\frac{2ug_2(u)}{m_b^2-(p+uq)^2}-
\frac{8m_b^2}{(m_b^2-(p+uq)^2)^2}\left(g_1(u) -\int^u_0dv 
g_2(v)\right)\right]\Bigg\}~,
\label{28}
\eq

\bq
\widetilde{F}^{(a)}_{QCD}(p^2,(p+q)^2)=f_\pi 
\int\limits_{0}^{1} \frac{du}
{m_b^2-(p+uq)^2}\Bigg\{
\mu_\pi\varphi_{p}(u)
+\frac{\mu_\pi \varphi_{\sigma }(u) }{6u}
\nonumber
\\
\times\Bigg[ 1-
\frac{m_b^2-p^2}{m_b^2-(p+uq)^2}\Bigg]
+ \frac{2m_b g_2(u)}{m_b^2-(p+uq)^2} \Bigg\} ~.
\label{Ftilde}
\eq
Furthermore, the three-particle contribution
is obtained from Fig. 5b using (\ref{33}) and
the matrix elements (\ref{29}) to (\ref{31}):
\bq
F^{(b)}_{QCD}(p^2,(p+q)^2)
&=& \int\limits_{0}^{1} dv\int{\cal D}\alpha_i\left\{
\frac{4f_{3\pi}\vp_{3\pi}(\alpha_i)v(pq)}
{[m_b^2-(p+(\alpha_1+v\alpha_3)q)^2]^2}\right.
\nonumber
\\
&&{}+ m_bf_\pi
\left.\frac{2\varphi_\perp (\alpha_i)-\varphi_\parallel (\alpha_i)+
2\tilde{\varphi}_\perp (\alpha_i)-\tilde{\varphi}_\parallel (\alpha_i)}
{[m_b^2-(p+(\alpha_1+v\alpha_3)q)^2]^2}\right\} ~,
\label{35}
\eq

\be
\tilde{F}^{(b)}_{QCD}(p^2,(p+q)^2)=0~.
\ee
It is interesting to note that there are no contributions 
from twist 2 and three-particle wave functions
to $\tilde{F}_{QCD}$, and hence to $f^+ + f^-$.
Including all operators  up to twist 4 one has in total  
\be
F_{QCD}(p^2,(p+q)^2)
= F^{(a)}_{QCD}(p^2,(p+q)^2) + F^{(b)}_{QCD}(p^2,(p+q)^2) ~,
\label{answer}
\ee
\be
\tilde{F}_{QCD}(p^2,(p+q)^2) = \tilde{F}^{(a)}_{QCD}(p^2,(p+q)^2).
\label{answer5}
\ee
We see that every power of $x^2$
in the light-cone expansion of the integrand in (\ref{1a4})
leads to an additional power of the denominator
$m_b^2-(p+uq)^2$ in $F_{QCD}$ and $\tilde{F}_{QCD}$.
This justifies the neglect of higher-twist operators,
provided $(p+q)^2 \ll m_b^2$ and 
$p^2 < m_b^2$ by a few GeV$^2$. More definitely,
numerical studies indicate that contributions 
beyond twist 4 can safely be neglected.

The imaginary parts of the two-particle functions 
$F^{(a)}_{QCD}$ and $\tilde{F}^{(a)}_{QCD}$ are relatively easy to find.
Terms proportional to $1/[m_b^2-(p+uq)^2]$ can be directly converted
into disperson integrals with respect to $(p+q)^2$ by changing the variable
$u$ into $s=(m_b^2-p^2)/u+p^2$.
Subtraction of the continuum simply
shifts the lower limit of intergration in (\ref{28}) and in 
(\ref{Ftilde}) from 0 to
$\Delta = (m_b^2-p^2)/(s^B_0-p^2)$.
Terms proportional to higher powers of $1/[m_b^2-(p+uq)^2]$ 
need to be partially integrated. After continuum subtraction, 
this leads to additional surface terms which are not completely 
negligible.
A similar procedure is possible in the case of the
three-particle function $F^{(b)}_{QCD}$. However, 
since this contribution is anyway only a small correction,    
we neglect the corresponding surface term, for simplicity. 

In summary, (\ref{sr}), (\ref{28}), (\ref{35}), and (\ref{answer})     
yield the following sum rule for 
$f^+$, or more precisely, for the product $f_B f^+$ \cite{BKR,BBKR}:
$$
f_Bf^+( p^2)=\frac{f_{\pi} m_b^2}{2m_B^2}
\exp \left( \frac{m_B^2}{M^2}\right)
\Bigg\{
\int\limits_{\Delta}^{1}\frac{du}{u}
exp\left[-\frac{m_b^2-p^2(1-u)}{uM^2} \right]
$$
$$
\times\Bigg( \varphi_\pi(u)
+ \frac{\mu_\pi}{m_b}
\Bigg[u\varphi_{p}(u) + \frac{
\varphi_{\sigma }(u)}{3}\left(1 + \frac{m_b^2+p^2}{2uM^2}\right) \Bigg]
$$
$$
- \frac{4m_b^2 g_1(u)}{ u^2M^4 }
+\frac{2}{uM^2} \int\limits_{0}^{u} g_2(v)dv
\left(1+ \frac{m_b^2+p^2}{uM^2} \right ) \Bigg)+ t^+(s_0^B,p^2,M^2)
$$
\be
+f^+_G(p^2,M^2)
+ \frac{\alpha_sC_F}{4\pi}\delta^+(p^2,M^2) \Bigg\} .
\label{fplus}
\ee
Here, $t^+$ 
denotes the surface term just mentioned, 
$f^+_G$ the contribution from the 
three-particle wave functions, and $\delta^+$ the $O(\alpha_s)$
correction to the leading twist 2 term \cite{KRWY,Bagan}.  
Explicit expressions can be found in Appendix 2.
Similarly, the sum rule for $f^+ + f^-$ \cite{KRW} is obtained
from (\ref{sr5}) and (\ref{answer5}): 
$$
f_B(f^+(p^2) + f^-(p^2))=\frac{f_\pi \mu_\pi m_b}{m_B^2 }\exp\left(
\frac{m_B^2}{M^2}\right)\Bigg\{
\int\limits_{\Delta}^{1}
\frac{du}{u}
\exp \left[ - \frac{m_b^2-p^2(1-u)}{u M^2}\right]
$$
\be
\times\Bigg[
\varphi_{p}(u)
+ \frac{\varphi_\sigma(u)}{6u}
\left(1-
\frac{m_b^2-p^2}{uM^2}\right)
+ \frac{2m_bg_2(u)}{\mu_\pi uM^2}\Bigg]+ t^{\pm}(s_0^B,p^2,M^2)\Bigg\}
\label{fplusminus}
\ee
with the surface term $t^{\pm}$ being given in Appendix 2.
Perturbative QCD corrections to this sum rule are still missing.

\begin{figure}[htb]
\centerline{
\epsfig{bbllx=100pt,bblly=209pt,bburx=507pt,%
bbury=490pt,file=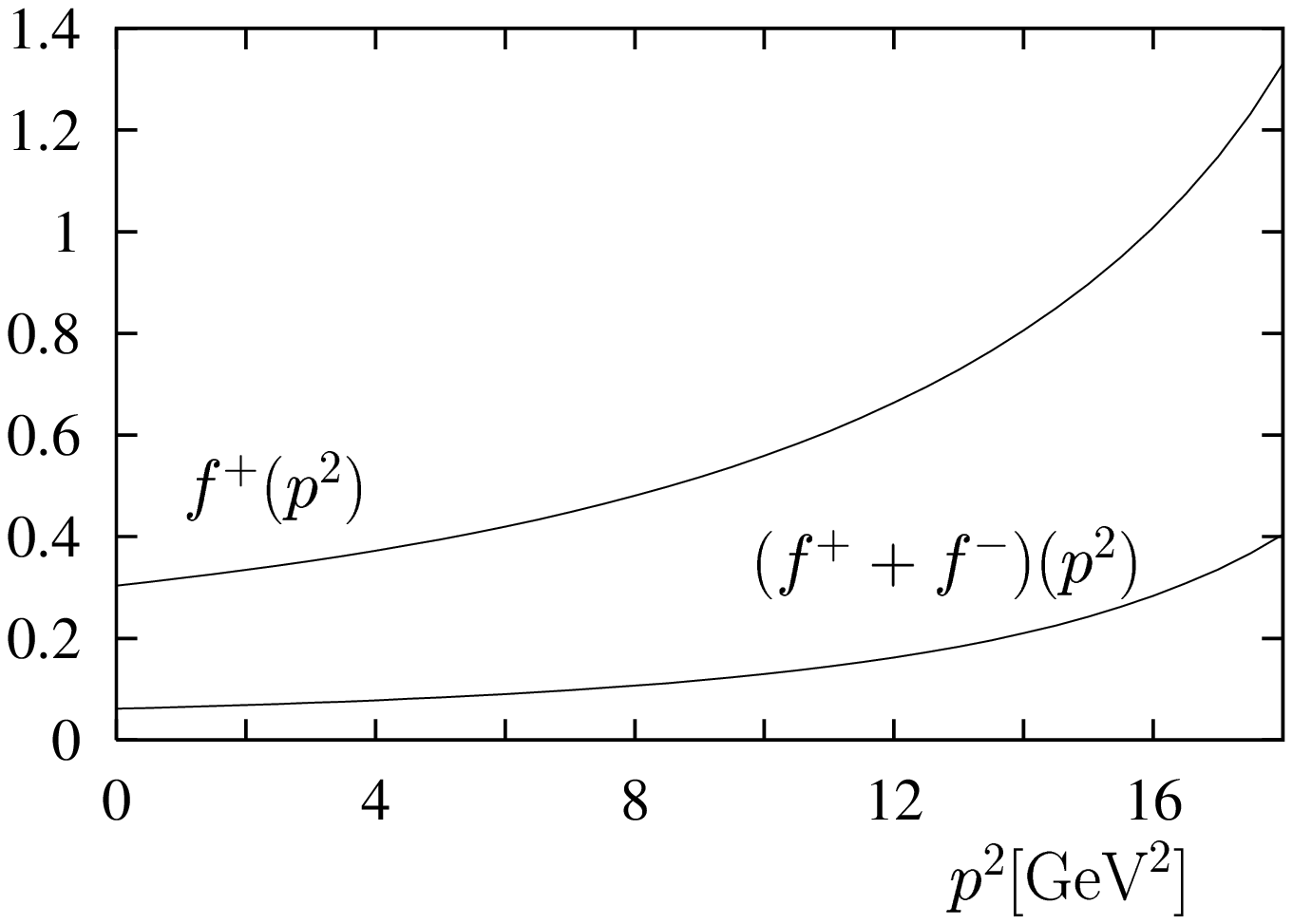,scale=0.9,%
clip=}
}
\caption{\it $B \to \pi$ form factors
obtained from light-cone sum rules in LO.}
\end{figure}

In the following, we shall illustrate the above results numerically.
For this purpose, it suffices 
to work in LO. Strictly speaking,
in the absence of a complete NLO analysis this is also more consistent.
Furthermore, for $m_b$ and $f_B$ we choose 
the values given in (\ref{bmass}) and (\ref{fB0}), respectively,
for the threshold parameter we take $s_0^B = 35 \pm 2$ GeV$^2$.
The pion wave functions are listed in Appendix 1.  
With this input we have checked that for 
$M^2 = 10 \pm 2$ GeV$^2$ the twist 4 corrections are very small,
and the contributions from excited and continuum states
do not exceed $30\%$. Up to $p^2 \simeq 18$ GeV$^2$, the sum rules are also
quite stable with respect to
a variation of $M^2$. However, at larger momentum transfer 
the sum rules become unstable, and the twist 4 contribution grows rapidly.
This is not surprising because the light-cone expansion and the sum rule method
are expected to break down as $p^2$ approaches $m_b^2$. If not stated otherwise,
the numerical results given below are obtained from the central values of the
input parameters, which we refer to as our nominal choice.

The momentum dependence of the form factors 
$f^+$ and $f^+ + f^-$ can be seen in Fig. 7.
Specifically, at zero momentum transfer we predict
\bq
f^+(0)=0.30,
\nonumber
\\
f^+(0) + f^-(0)= 0.06.
\label{fpl0}
\eq
If the known $O(\alpha_s)$ corrections are included
in the sum rules for $f_Bf^+(p^2)$ and $f_B$, one gets instead
\be 
f^+(0)=0.27.
\label{Bfpnlo}
\ee

\begin{figure}[htb]
\centerline{
\epsfig{file=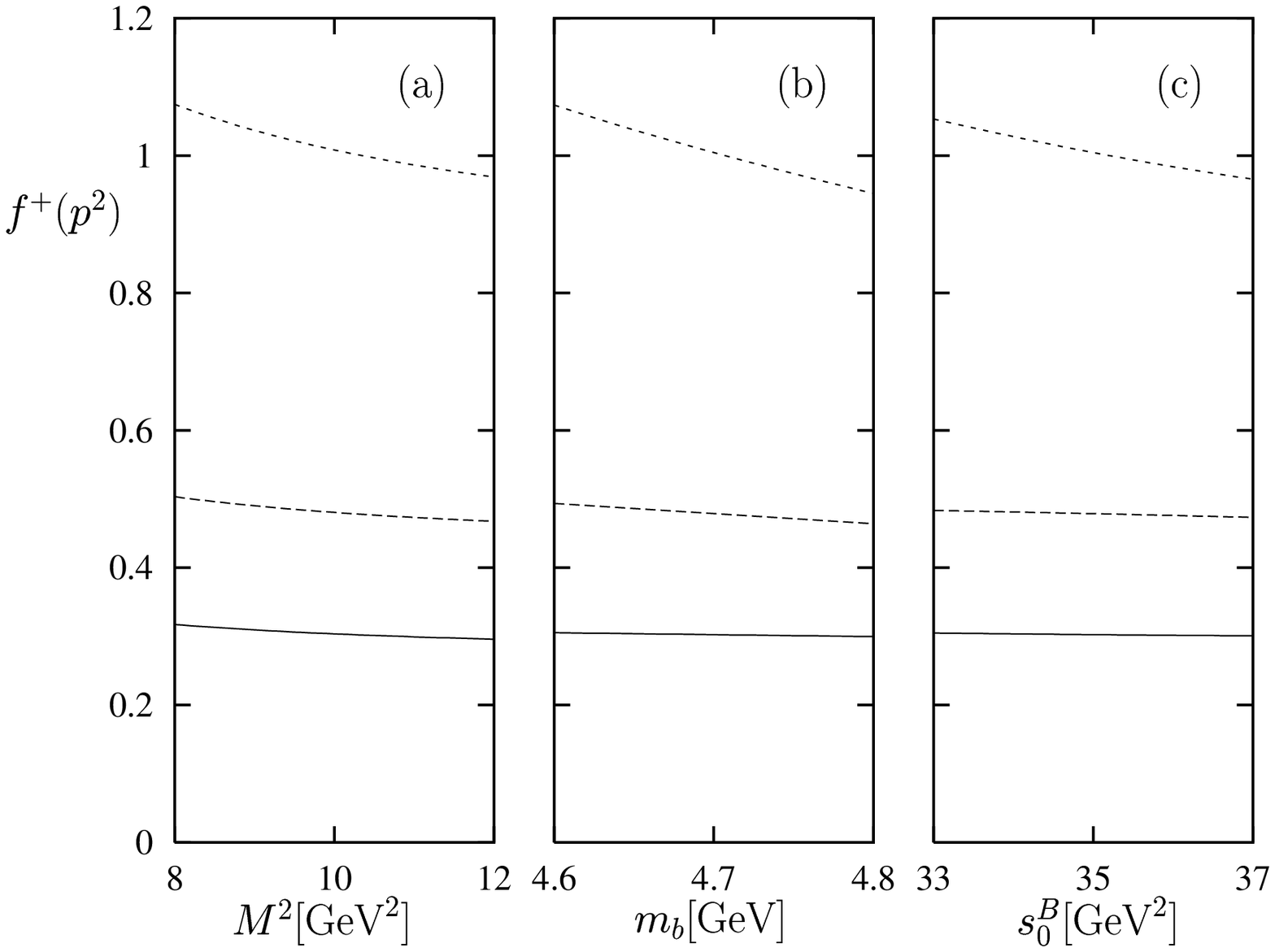,width=\textwidth,bbllx=38pt,bblly=207pt,bburx=580pt,%
bbury=660pt,clip=}
}
\caption{\it Variation
of the sum rule prediction on $f^+(p^2)$ 
with the Borel parameter (a), the $b$-quark mass (b),
and the threshold parameter $s_0^B$ (c).
Considered are three typical values of momentum transfer:
$p^2=0$ (solid),
$p^2=8$ GeV$^2$ (long-dashed), and $p^2=16$ GeV$^2$ (short-dashed).}
\end{figure}

Of course,
a most important question concerns the reliability of these 
predictions. Below we comment on the main sources of uncertainties.
Allowing the parameters to vary within the ranges given above and
in Appendix 1 we use the deviations of $f^+$ from the value obtained
with the nominal choice of parameters as an uncertainty estimate.

(a) Borel mass parameter

\noindent The dependence of $f^+$ on $M^2$ is illustrated in Fig. 8a.
In the allowed interval of $M^2$, $f^+$ varies by only
$\pm$(3 to 5)~\%, depending on $p^2$.

(b) $b$-quark mass and continuum threshold 

\noindent Fig. 8b shows the variation of $f^+$ with $m_b$
keeping all other parameters except $f_B$ fixed. 
The value of $f_B$ is taken from the sum rule ({\ref{fB1})  
dropping the $O(\alpha_s)$ corrections. 
The analogous test for $s_0^B$ is performed in Fig. 8c.  
If $s_0^B$ and $m_b$ are varied simultaneously such that 
one achieves maximum stability of the sum rule for $f_B$, 
the change in $f^+$ is negligible at 
small $p^2$ rising to about $\pm$ 3~\% at large $p^2$.

\begin{figure}[p]
\centerline{
\epsfig{file=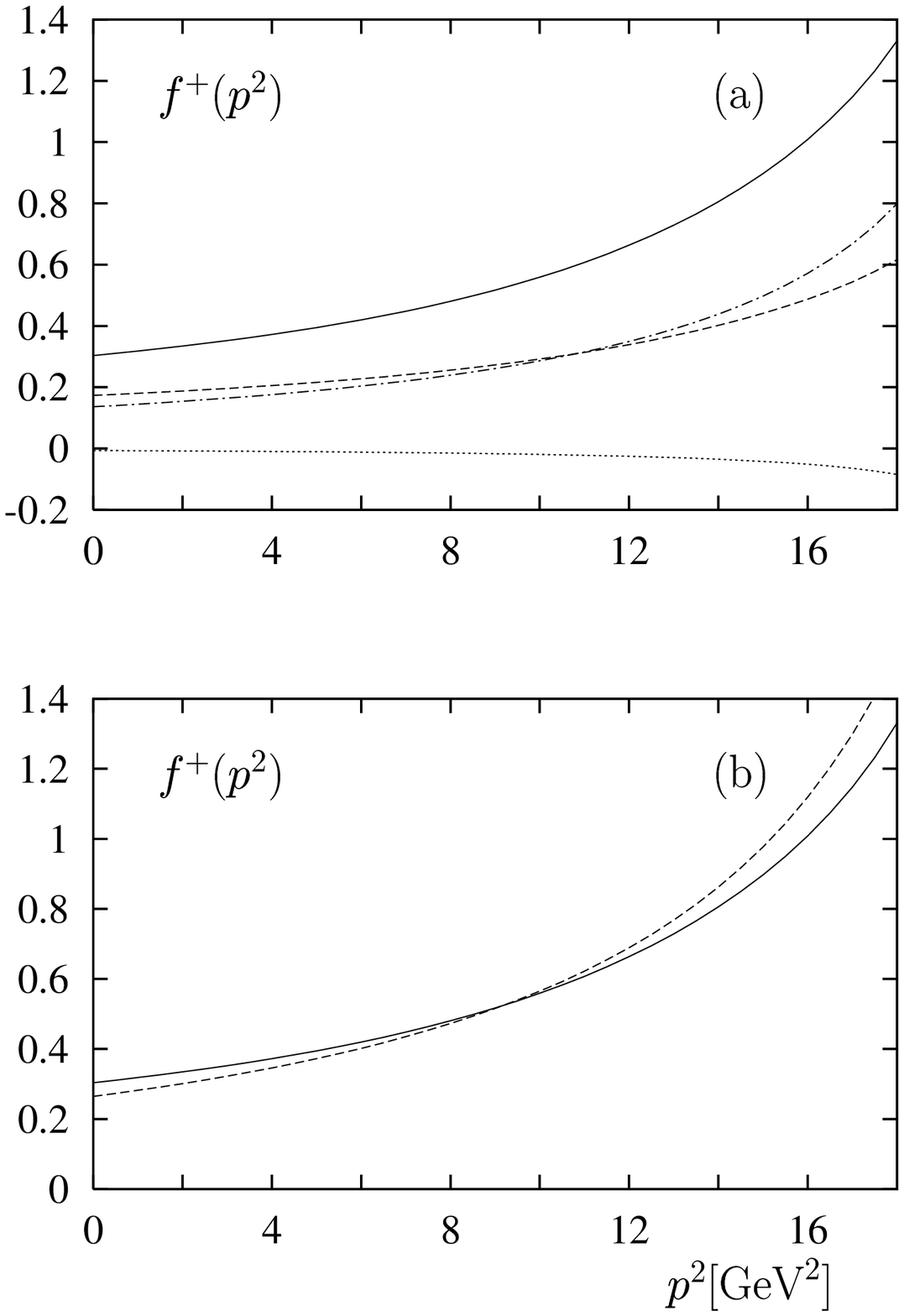,scale =0.6,width=\textwidth,bbllx=0pt,bblly=80pt,
bburx=600pt,%
bbury=800pt,clip=}
}
\caption{\it  $B \to \pi$ form factor $f^+$: (a) individual contributions
of twist 2 (dashed), 3 (dash-dotted), and 4 (dotted), 
and the total sum (solid); 
(b) wave functions with (solid) and without (dashed) 
nonasymptotic corrections.}
\end{figure}

(c) higher-twist contributions

\noindent No reliable estimates exist for wave functions beyond twist 4.
Therefore, we use the magnitude of the twist 4 contribution to $f^+$
as an indicator for the uncertainty due to the neglect
higher-twist terms. 
From Fig. 9a we see that the impact of the twist 4 components is 
comfortably small, less than 2~\% at low
$p^2$ and about 5~\% at large $p^2$.
This suggests a conservative estimate
of $\pm$ 5~\%  due to unknown higher-twist.

(d) light-cone wave functions

\noindent
The asymptotic wave functions and
the scale dependence of the nonasymptotic coefficients
are given in perturbative QCD. However, the values of these
coefficients at a certain input scale $\mu_0$ have to be
determined empirically. They are presently only known with 
considerable uncertainties. 
In order to clarify  the sensitivity of $f^+$ to nonasymptotic effects,
we put the latter to zero and compare the result in Fig. 9b 
with the nominal prediction. The change amounts to about -10~\% 
at small $p^2$ and +10~\% at large $p^2$, while
the intermediate region around $p^2\simeq 10$ GeV$^2$ is very little affected.
Since this exercise is rather extreme, a more careful estimate should
give a smaller number. Note that the sum rule
being dominated by convolutions of smooth amplitudes with
normalized wave functions,
is actually expected 
to be relatively insensitive to the precise shape of the latter.

\begin{figure}[htb]
\centerline{
\epsfig{bbllx=100pt,bblly=209pt,bburx=507pt,%
bbury=490pt,file=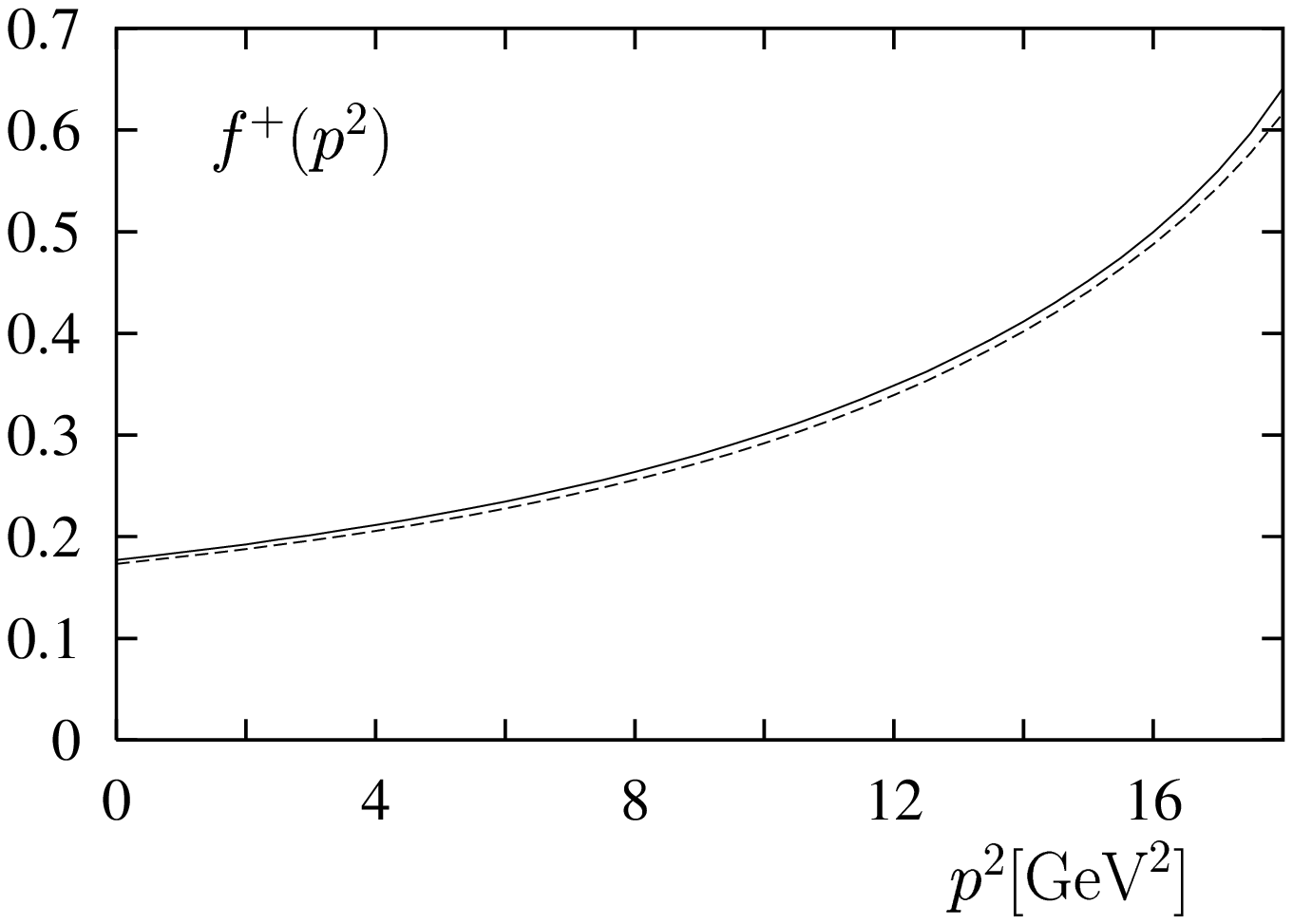,scale=0.9,%
clip=}
}
\caption{\it $B\to \pi$ form factor $f^+(p^2)$ in leading twist 2
approximation: LO (dashed) in comparison to NLO (solid).} 
\end{figure}

(e) perturbative corrections

\noindent The $O(\alpha_s)$ corrections
to the leading twist 2 term in the sum rule (\ref{fplus}) for $f_Bf^+$ 
and to the sum rule (\ref{fB1}) for 
$f_B$, both being about 30~\%, cancel in the ratio.
The net effect is the unusually small correction to $f^+$
shown in Fig. 10. 
This result \cite{KRWY,Bagan} eliminates one main 
uncertainty. Unfortunately, perturbative corrections to the higher-twist
terms have not yet been calculated. Considering that
the twist 3 terms contribute about 50~\% to the sum rule for $f_B f^+$, 
and assuming again radiative corrections of about 30~\% but no 
cancellation, one has to face a remaining uncertainty of about 15~\%.
This, however, may be an overestimate. More importantly,
with some major effort this deficiency can be cured.

In summary, the present total uncertainty in $f^+(p^2)$ is estimated to
be about 25~\%, if the uncertainties (a) to (e) with the 
exception of (d) are added up linearly, 
and about 17~\% if they are added in quadrature
as is often done in the literature. 
Once the perturbative QCD correction to the twist 3 term is calculated,
this uncertainty, which mainly concerns the normalization,
can be reduced to 10~\%. In addition, there is a 
shape-dependent uncertainty from (d) of another 10~\% at low and high $p^2$.
However, in the integrated width the latter averages out almost completely.
The uncertainties from (a) to (d) on $f^+ + f^-$ are of comparable size,
while the effect of radiative corrections is still unknown.

The sum rules  (\ref{fplus}) and
(\ref{fplusminus}) for the $ B\rightarrow \pi$ form factors
are formally converted into sum rules for
the $D\rightarrow \pi$  form factors by replacing
$b$ with $c$ and $B$ with $D$.
The input parameters are taken over from the calculation of
$f_D$ in section 2.
In addition, one has to rescale the wave functions
from $\mu_b \simeq 2.4$ GeV to $\mu_c \simeq 1.3$ GeV 
as specified in Appendix 1.  
The allowed Borel mass window is 
$ 3 ~\mbox{GeV}^2 < M^2 < 5 ~\mbox{GeV}^2$. With this choice, the $D \to \pi$
form factor at $p^2 = 0$ is predicted to be \cite{BBKR,KRW} 
\bq f^{+}(0) = 0.68.
\label{Dfplo}
\eq
$O(\alpha_s)$ corrections are not included here. 
The momentum dependence of $f^+$ in the range
$0 \le p^2 \le m_c^2 - O(1 ~\mbox{GeV}^2)$ is shown in the next
section, together with an extrapolation to higher $p^2$.

Other important applications of light-cone sum rules
include the estimate \cite{BKR} of the $B\to K$ form factor 
which determines the factorizable part of the $B\to J/\psi K$ 
amplitude, the prediction \cite{ABS} 
of the matrix element of the electromagnetic penguin operator for 
$B \to K^* \gamma$, and the more recent
calculation \cite{BallBraun} of the 
$B\to \rho$ form factors. In the latter work, essentially 
the same procedure is applied as outlined above. However, the relevant weak
currents and light-cone wave functions are very different. 
For illustration, one has to deal with the following 
matrix elements \cite{BallBraun,ABS,CZ}:
\be
\langle\rho^+(q,\lambda)\mid \bar{u}(0)\sigma_{\mu\nu}d(x)\mid 0
\rangle=
-if^{\perp}_\rho (\epsilon(\lambda)_\mu q_ \nu-\epsilon(\lambda)_\nu q_ \mu)
\int\limits_0^1du~e^{iuqx}\phi_\perp(u)~,
\label{phirho1}
\ee

\begin{eqnarray}
\langle\rho^+(q,\lambda)\mid \bar{u}(0)\gamma_{\mu}d(x)\mid 0\rangle 
& = & q_\mu\frac{\epsilon(\lambda)x}{qx}f_\rho m_\rho
\int\limits_0^1du~e^{iuqx}\phi_\parallel(u)  \nonumber\\
&  & +\left(\epsilon(\lambda)_\mu- q_\mu\frac{\epsilon(\lambda)x}{qx}\right)
f_\rho m_\rho \int\limits_0^1du~e^{iuqx}g_\perp^{(v)}(u)~,  
\label{phirho2}
\end{eqnarray}

\be
\langle\rho^+(q,\lambda)\mid \bar{u}(0)\gamma_{\mu}\gamma_5d(x)\mid 0
\rangle= \frac14\epsilon_{\mu\nu\rho\sigma}
\epsilon(\lambda)^\nu q^ \rho x^\sigma
f_\rho m_\rho\int\limits_0^1du~e^{iuqx}g_\perp^{(a)}(u)~,
\label{phirho3}
\ee
Here, $\phi_\perp$ and $\phi_\parallel$ are twist 2 wave functions
of transversely and longitudinally polarized $\rho$ mesons, 
respectively, while $g^{(v)}_\perp$ 
and $g^{(a)}_\perp$ are associated with both twist 2 and twist 3 operators.
Higher-twist components and three-particle
wave functions are still missing, as are perturbative corrections.

There are four independent $ B\to \rho$ form factors:
\begin{eqnarray}
\langle \rho(\lambda) | \bar{u}\gamma_\mu(1-\gamma_5)b| B \rangle & = &
-i (m_B + m_\rho) A_1(t) \epsilon(\lambda)_\mu +
\frac{iA_2(t)}{m_B
+ m_\rho} (\epsilon(\lambda) p_B) (p_B+p_\rho)_\mu \nonumber\\
& & + \frac{iA_3(t)}{m_B + m_\rho} (\epsilon(\lambda) p_B)
(p_B-p_\rho)_\mu \\
& & + \frac{2V(t)}{m_B + m_\rho}
\epsilon_\mu^{\phantom{\mu}\alpha\beta\gamma}\epsilon(\lambda)_\alpha
p_{B\beta} p_{\rho\gamma} \nonumber
\label{Brho}
\end{eqnarray}
with $t=(p_B-p_\rho)^2$.  Only $V(t)$, $A_1(t)$ and $A_2(t)$
contribute to the semileptonic decay $B\to\rho \bar{l} \nu_l$ for $l=e,\mu$.
These form factors are shown in Fig. 11. They all
are predicted to rise with momentum transfer $t$,
contrary to some earlier
claims based on three-point sum rules.
This contradiction was studied in detail in
\cite{BallBraun} with the conclusion that
the three-point sum rules are not reliable 
for reasons indicated in section 4.

\begin{figure}[p]
\centerline{
\epsfig{file=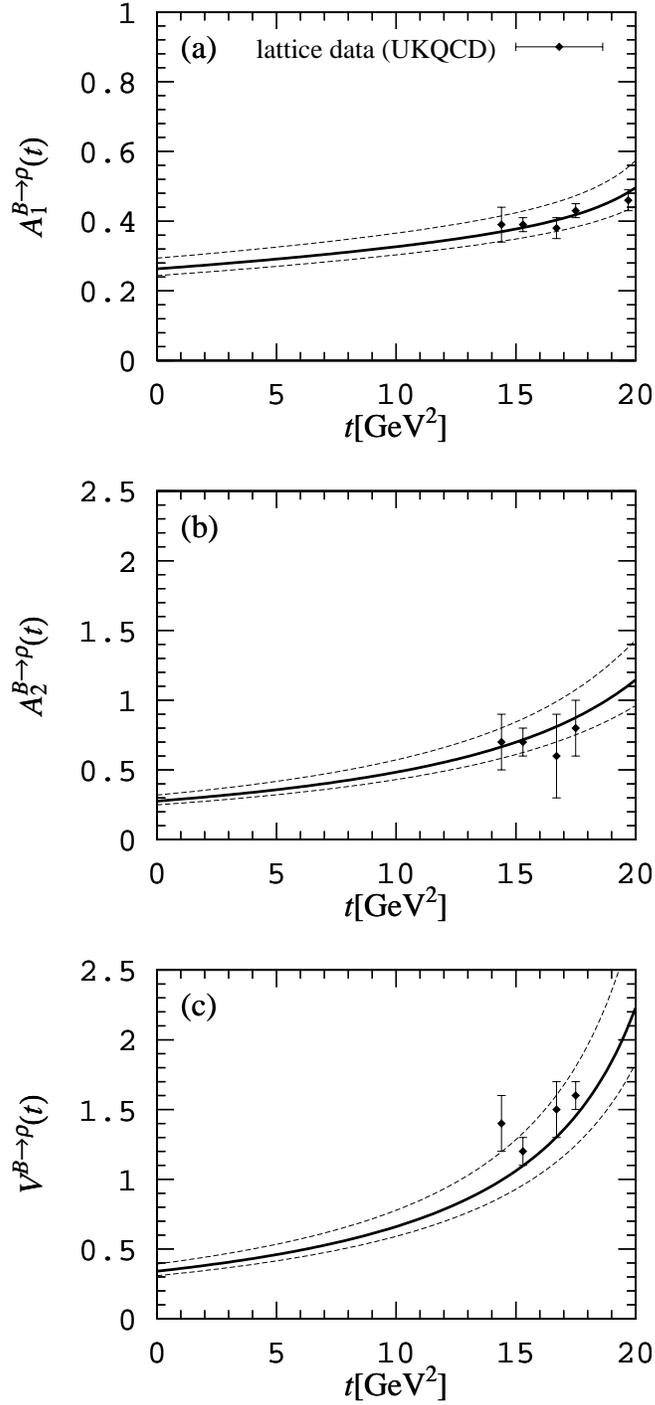,width=9cm,bbllx=50pt,bblly=50pt,bburx=300pt,%
bbury=600pt,clip=}
}
\caption{\it The $B \to \rho$ form factors: 
predictions from light-cone sum rules (solid)
in comparison to lattice results \cite{Flynn}.
The dashed curves indicate the uncertainties in the sum rule
results (from \cite{BallBraun}).}
\end{figure}


\section{$B^*B\pi$ and $D^*D\pi$ couplings}

The strong couplings of $B$ and $D$ mesons to pions have
been studied with different variants of sum rules and in a variety of 
quark models.
We shall use the following definition of the $B^*B\pi$ 
coupling constant:
\begin{equation}
\langle \bar{B}^{*0}\pi^-\mid B^-\rangle =
-g_{B^*B\pi}(q \cdot\epsilon ).
\label{16}
\end{equation}
The couplings of the different charge states 
are related by isospin symmetry:
\bq
g_{B^*B\pi} \equiv g_{\bar{B}^{*0}B^-\pi^+}=
-\sqrt{2}g_{\bar{B}^{*0}\bar{B}^{0}\pi^0}=\sqrt{2}g_{B^{*-}B^-\pi^0}
=-g_{B^{*-}\bar{B}^0\pi^-}~.
\label{gd}
\eq

In \cite{BBKR}, a light-cone sum rule for
$g_{B^*B\pi}$ has been suggested which is derived from
the same correlation function (\ref{1a4}) as the
corresponding $B\to \pi$ form factors and which
depends on the same nonperturbative input.
The key idea is to write a double dispersion integral for the
invariant function $F(p^2, (p+q)^2)$.
Inserting in (\ref{1a4})
complete sets of intermediate hadronic states
carrying $B$ and $B^*$ quantum numbers, respectively, and using 
the matrix elements (\ref{fB2}), (\ref{fBstar2}) 
and (\ref{16}), one obtains
\bq
F(p^2,(p+q)^2)=\frac{m_B^2m_{B^*}f_Bf_{B^*}g_{B^*B\pi}}{m_b
(p^2-m_{B^*}^2)((p+q)^2-m_B^2)}
+\int_{\Sigma}\frac{\rho^h(s_1,s_2)ds_1ds_2}{(s_1-p^2)(s_2-(p+q)^2)}
\label{2216}
\eq
$$
+ ~(subtractions)~.
$$
Again, the first term is the ground-state contribution
and contains the $B^*B\pi$ coupling,
while the spectral function $\rho^h(s_1,s_2)$
represents higher resonances and continuum states in the
$B^*$ and $B$ channels. The integration region in the $(s_1,s_2)$ -- plane
is denoted by $\Sigma$. The subtraction terms are polynomials in
$p^2$ and/or $(p+q)^2$ which vanish by Borel transformation of  
(\ref{2216}) with respect to
both variables $p^2$ and $(p+q)^2$. The transformed hadronic representation
of $F$ is given by
\bq
F(M_1^2,M_2^2)&\equiv&
{\cal B}_{M_1^2}{\cal B}_{M_2^2}F(p^2,(p+q)^2)=
\frac{m_B^2m_{B^*}f_Bf_{B^*}g_{B^*B\pi} }{m_b}
\exp\left[-\frac{m_{B^*}^2}{M_1^2}-\frac{m_{B}^2}{M_2^2}\right]
\nonumber
\\
&+&\int_{\Sigma_{12}} \exp\left[-\frac{s_1}{M_1^2}-\frac{s_2}{M_2^2}\right]
\rho^h(s_1,s_2)ds_1ds_2~,
\label{2226}
\eq
where  $M_1^2$ and $M_2^2$  are the Borel parameters associated with
$p^2$ and $(p+q)^2$, respectively.
The same transformation is also applied to
(\ref{answer}) yielding $F_{QCD}(M_1^2,M_2^2)$. Then, the sum rule for
$g_{B^*B\pi}$ results from the equality of $F$ and $F_{QCD}$,
and continuum subtraction.
This last step deserves some more explanations.

Let us consider the twist 2 expression for $F$ given in (\ref{Fzeroth4}).
In order to write it in the form of a double dispersion relation,
\be
F_{QCD}( p^2, (p+q)^2)=\int^\infty_{m_b^2} \frac{ds_1}{s_1-p^2}
\int^\infty_{m_b^2} \frac{ds_2}{s_2-(p+q)^2}\rho^{QCD}(s_1,s_2)~,
\label{repr6}
\ee
we change $u$ to $s =(m_b^2-p^2)/u +p^2$, and get
\be
F_{QCD}(p^2, (p+q)^2)=m_bf_\pi\int^\infty_{m_b^2} \frac{ds~
\varphi_\pi(s)}{(s-p^2) (s-(p+q)^2)}  ~.
\label{integrs6}
\ee
In general, the wave function $\varphi_\pi(u)$ can be expressed as
a power series in $(1-u)$:
\be
\vp_\pi(u) = \sum_k a_k(1-u)^k =
\sum_ka_k\left(\frac {s-m_b^2}{s-p^2} \right)^k~.
\label{sum6}
\ee
Substituting (\ref{sum6}) in (\ref{integrs6})
and introducing formally two variables $s_1$ and $s_2$ instead of $s$,
one reproduces the double integral (\ref{repr6}) with
\be
\rho^{QCD}(s_1,s_2)=m_bf_\pi\sum_k \frac{(-1)^ka_k}{\Gamma(k+1)}
(s_1-m_b^2)^k\delta^{(k)}(s_1-s_2)~.
\label{dens6}
\ee
Now we Borel-transform (\ref{repr6}),
$$
F_{QCD}(M_1^2,M_2^2) = m_bf_\pi\sum_k
\int^\infty_{m_b^2}ds_1
\int^\infty_{m_b^2} ds_2~
\frac{(-1)^ka_k}{\Gamma(k+1)}
(s_1-m_b^2)^k
$$
\be
\times
~\delta^{(k)}(s_1-s_2)
\exp\left[-\frac{s_1}{M_1^2}-\frac{s_2}{M_2^2}\right]~,
\label{check6}
\ee
introduce again new variables $s=s_1+s_2$ and $v=s_1/s$, and
use the $\delta$-function to integrate
over $v$. The result is
\bq
F_{QCD}(M_1^2,M_2^2) = m_bf_\pi\sum_k
\frac{a_k}{2^{k+1}k!}\int_{2m_b^2}^\infty\!\! ds
\left( \frac {d}{ dv} \right)^k\!\!\Bigg[\left(v-\frac{m_b^2}{s}\right)^k
\nonumber
\\
\times
\exp\left(-\frac{svM_2^2+s(1-v)M_1^2}{M_1^2M_2^2}\right)\Bigg]_{v=1/2}~.
\label{form16}
\eq
For $M_1^2=M_2^2=2M^2$ the $v$-dependence of the exponent disappears, and
the differentiation of the bracket gives a factor $k!$. As a consequence,
(\ref{form16}) reduces to
\be
F_{QCD}(M^2, M^2)= m_bf_\pi\sum_k \frac{a_k}{2^{k+1}}
\int_{2m_b^2}^\infty ds\,
e^{-\frac{s}{2M^2}}
= m_bf_\pi\varphi_\pi(u_0)M^2 e^{-\frac{m_b^2}{M^2}} ~
\label{form6}
\ee
with $u_0=1/2$. For arbitrary values of $M_1^2$ and $M_2^2$ a similar
expression is obtained, with 
$
u_0= M_1^2/(M_1^2+M_2^2)
$
and 
$
M^2= M_1^2M_2^2/(M_1^2+M_2^2)
$.

With these manipulations, and the replacement of the integral over
$\rho^h$ in (\ref{2226}) by a corresponding integral over
$\rho^{QCD}$, it is straightforward to subtract 
the contributions from
excited and continuum states from (\ref{check6}).
The remaining integral
is restricted to the region below a given boundary in $(s_1,s_2)$.
For the latter one may take
\be
(s_1)^a +(s_2)^a \leq (s_0)^a.
\label{boundary}
\ee
For $a=1$, the duality region is a triangle
in the $(s_1,s_2)$ -- plane, while for $a\rightarrow \infty$ it is a
square. Since the spectral density (\ref{dens6}) vanishes everywhere except
at $s_1=s_2$, it is actually irrelevant which form of the boundary
one chooses provided the length of the duality interval at $s_1=s_2$
is the same. For example, the triangle with $s_0 = 2s_0^B$ is equivalent
to the square with $s_0 = s_0^B$.  Here, we take $s_0^B$ as
the effective threshold in both the $B$ and $B^*$ channels.
Repeating the steps following (\ref{check6})
one obtains an
expression similar to (\ref{form16}), but with the upper limit of
integration in $s$ lowered to $2s_0^B$ and with the addition of surface
terms. The latter disappear for $M_1^2=M_2^2$, in which case
the subtracted invariant function replacing (\ref{form6}) is 
simply given by
$$
F_{subtr}(M^2, M^2)= m_bf_\pi\sum_k \left(\frac{a_k}{2^{k+1}}\right)
\int_{2m_b^2}^{2s_0^B} ds\,
e^{-\frac{s}{2M^2}}
$$
\be
= m_bf_\pi\varphi_\pi(u_0)M^2
\left[e^{-\frac{m_b^2}{M^2}}-
e^{-\frac{s_0^B}{M^2}}\right]~.
\label{formh6}
\ee
However, in general, the proportionality of 
$F_{QCD}(M_1^2, M_2^2)$ to the wave function
$\vp_\pi$ at $u_0=M_1^2/(M_1^2+M_2^2)$ found in (\ref{form6}) 
is destroyed by continuum subtraction.
For this reason we adopt the particular choice $M_1^2=M_2^2$ ($u_0=1/2$)
in what follows.

So far we have solved the problem only for the twist 2 component
of the sum rule. Unfortunately, the subtraction procedure explained above
does not work for terms in $F_{QCD}$ which
contain higher powers of $1/(m_b^2-(p+uq)^2)$ (see (\ref{28})).
The main problem is that the corresponding double spectral densities are
not concentrated near $s_1=s_2$, making
the continuum subtraction rather complicated in these cases.
For further discussion we refer the reader to \cite{BB94}.
On the other hand, this difficulty only concerns higher-twist terms which
contribute relatively little to the sum rule. Hence, to a good approximation
one may disregard the continuum subtraction in these terms.

Applying these recipes to $F_{QCD}$ as given in (\ref{answer})
one ends up with the following light-cone sum rule
for the $B^*B\pi$ coupling:
$$
f_Bf_{B^*}g_{B^*B\pi}=\frac{m_b^2f_\pi}{m_B^2m_{B^*}}
e^{\frac{m_{B}^2+m_{B^*}^2}{2M^2}}
\Bigg\{M^2 \Big[e^{-\frac{m_b^2}{M^2}} - e^{-\frac{s_0^B}{M^2}}\Big]
$$
$$
\times
~\Big[\varphi_\pi(u_0)
+ \frac{\mu_\pi}{m_b} \left( u_0\varphi_p(u_0)
+\frac13\varphi_\sigma (u_0)+\frac16u_0\frac{d\varphi_\sigma}{du}(u_0)\right)
+ \frac{2f_{3\pi}}{m_bf_\pi}I^G_3(u_0)\Big]
$$
\be
+e^{-\frac{m_b^2}{M^2}}
\Big[\frac{\mu_\pi m_b}3 \vp_\sigma (u_0)
+2u_0g_2(u_0)-\frac{4m_b^2}{M^2}
\left( g_1(u_0)-\int_0^{u_0}g_2(v)dv\right) + I^G_4(u_0)\Big]\Bigg\}_{u_0=1/2}~,
\label{fin}
\ee
where
\be
I^G_3(u_0)=
\int_0^{u_0}d\alpha_1 \left[\frac{\varphi_{3\pi}(\alpha_1,1-u_0,u_0-\alpha_1)}
{u_0-\alpha_1}
-\int_{u_0-\alpha_1}^{1-\alpha_1}d\alpha_3
\frac{\varphi_{3\pi}(\alpha_1,1-\alpha_1-\alpha_3,\alpha_3)}{\alpha_3^2}\right]
\label{fg3}
\ee
and
\be
I^G_4(u_0)=
\int_0^{u_0}d\alpha_1\int_{u_0-\alpha_1}^{1-\alpha_1}
\frac{d\alpha_3}{\alpha_3}[2\varphi_\perp (\alpha_i)-\varphi_\parallel
 (\alpha_i)+
2\tilde{\varphi}_\perp (\alpha_i)-\tilde{\varphi}_\parallel (\alpha_i)] 
\label{fg4}
\ee
are the contributions from the three-particle amplitude (\ref{35}).
As a further simplification, $G$-parity implies
$g_2(u_0)= \frac{d\varphi_\sigma}{du}(u_0)=0$ at $u_0 = 1/2$, 
whence these terms can be dropped in the above sum rule.

The wave function $\varphi_\pi(u_0)$ at the symmetry point $u_0=1/2$
also enters the sum rules for other
hadronic couplings involving the pion. We take the
value
\be
\varphi_\pi(u=\frac12) = 1.2\pm0.2
\label{phi12}
\ee
obtained from the light-cone sum rule for the
pion-nucleon coupling \cite{BF}. This value is consistent
with the choice of the coefficients $a_{2,4}$ given in (\ref{aa}).
For the remaining parameters
we use the same values as in the calculation of the form factor
$f^+$ in section 5. With this choice the sum rule (\ref{fin})
yields
\be
f_Bf_{B^{*}}g_{B^*B\pi }=  0.64 \pm 0.06 \,\mbox{\rm GeV}^2 ~,
\label{combinB}
\ee
or
\bq
g_{B^*B\pi}=29\pm 3~,
\label{41}
\eq
where $f_B$ and $f_{B^*}$ 
have been substituted from (\ref{fB0}) and (\ref{fBstar0}), respectively.
The uncertainties quoted above are to be interpreted as the range
of values corresponding to the allowed window in the Borel mass,
$6 \le M^2 \le 12$ GeV$^2$,
in which the excited and continuum states contribute
less than 30~\% and the twist 4 corrections do not exceed 10~\%.
Also these predictions can and should be improved by calculating
the radiative gluon corrections.

The sum rule (\ref{fin}) for $g_{B^*B\pi}$ is translated into  
a sum rule for $g_{D^*D\pi}= g_{D^{*+}D^0\pi^-}$
by formally changing $b$ to $c$, $\bar{B}$ to $D$, and $\bar B^*$ to $D^*$.
With (\ref{phi12}) and the input parameters from the
calculation of the $D\to \pi$ form factor,
one finds
\be
f_Df_{D^{*}}g_{D^*D\pi }~= ~0.51 \pm 0.05~ GeV^2  ~,
\label{combinD}
\ee
and using the LO estimates for $f_D$ and $f_{D^*}$ from
(\ref{fDres}) and (\ref{fDstar0}), respectively,
\bq
g_{D^*D\pi}= 12.5\pm 1.0~.
\label{constD*Dpi}
\eq
The variation of the numerical results with 
$M^2$ in the allowed interval $2 \le M^2 \le 4$ ~GeV$^2$
is again quoted as an uncertainty.

From (\ref{constD*Dpi}) one can calculate the width for the decay
$D^* \to D \pi$. The predicted value,
\be
\Gamma( D^{*+} \rightarrow D^0 \pi^+)~ =
~ \frac{g_{D^*D\pi}^2}{24\pi m_{D^*}^2}|~\vec q ~|^3
~ = ~32 \pm 5 \,\mbox{\rm keV}~,  
\label{Gamma6}
\ee
lies well below the current experimental upper limit,
\be
\Gamma( D^{*+} \rightarrow D^0 \pi^+)~ < ~ 89 ~\mbox{\rm keV}~,
\label{exp}
\ee
which is derived from the upper limit
$\Gamma_{tot}(D^{*+}) < 131$ keV and from the branching ratio 
$BR(D^{*+} \to D^0 \pi^+) = (68.3 \pm 1.4)~\%$ \cite{PDG}.
Predictions for other charge combinations are
readily obtained from (\ref{Gamma6}) and isospin relations 
analogous to (\ref{gd}). Accounting also for the
differences in phase space, one expects
\be
\Gamma( D^{*+} \rightarrow D^0 \pi^+)~ =
~2.2\,\Gamma (D^{*+} \rightarrow D^+ \pi^0)=~
1.44\,\Gamma (D^{*0} \rightarrow D^{0} \pi^0)~.
\label{widths}
\ee

In Table 2, we summarize the numerical results discussed above
and compare them with other estimates. One observes significant
differences.
Some of the predictions 
are rather close to the experimental upper limit, some even violate it.
It would be very interesting to have more precise data.

{\footnotesize
\begin{table}
\caption{\it Theoretical estimates of the strong
$B^*B\pi$ and $D^*D\pi$ couplings (from \cite{BBKR}).}
\bigskip
\begin{center}
\begin{tabular}{lllll}
\hline
\hline
\\
Reference & $\hat{g}$ &
$g_{B^*B\pi}$ & $g_{D^*D\pi}$ &$\Gamma(D^{*+}\ra D^0\pi^+)$ (keV)\\
\\
\hline
\\
\cite{BBKR} & 0.32 $\pm$ 0.02 & 29 $\pm$ 3 & 12.5 $\pm$ 1.0 & 32
 $\pm$ 5\\
\\
\cite{BBKR}$^a$ & -- & 28 $\pm$ 6& 11 $\pm$ 2 & --\\
\\
\cite{Ovch89}$^a$ &--& 32 $\pm$ 6& -- & --\\
\\
\cite{GY}$^a$ &0.2 $\div$ 0.7 &-- & -- & --\\
\\
\cite{Colang}$^a$ &0.39 $\pm$ 0.16& 20 $\pm$ 4&9 $\pm$ 1& --\\
\\
\cite{Colang}$^{a *}$ &0.21 $\pm$ 0.06& 15 $\pm$ 4 &7 $\pm$ 1& 10
 $\pm$ 3\\
\\
\cite{NW}$^b$ & 0.7&--& --& --\\
\\
\cite{IW}$^{b}$ &--& 64&--&--\\
\\
\cite{Yan}$^b$ & 0.75 $\div$ 1.0&--& --& 100 $\div$ 180\\
\\
\cite{CG}$^c$ & 0.6 $\div $ 0.7 & -- & -- & 61 $\div$ 78\\
\\
\cite{Ametal}$^c$ & 0.4 $\div$ 0.7 & -- & -- & --\\
\\
\cite{BardeenHill}$^d$ & 0.3 & -- & -- & --\\
\\
\cite{Eichtetal}$^e$ & --& -- & 16.2&53.4\\
\\
\cite{DoXu}$^f$ & --& -- &19.5 $\pm$ 1.0&76 $\pm$ 7\\
\\
\cite{Miller}$^g$ & --& -- & 16.2 $\pm$ 0.3&53.3 $\pm$ 2.0\\
\\
\cite{Kam}$^h$ & --& -- &8.9&16\\
\\
\cite{KN}$^i$ & --& -- &8.2& 13.8\\
\\
\cite{PDG}$^k$&--& -- &$<$ 21 & $<$ 89\\
\\
\hline
\hline
\end{tabular}
\end{center}
\hspace*{2.4cm}
$^a$ QCD sum rules in external axial field or soft pion limit. \\
\hspace*{2.4cm}
$^*$ including perturbative correction to the heavy meson decay constants.  \\
\hspace*{2.4cm}
$^b$ Quark model + chiral HQET.  \\
\hspace*{2.4cm}
$^c$ Chiral HQET with experimental constraints on $D^*$ decays. \\
\hspace*{2.4cm}
$^d$ Extended NJL model + chiral HQET . \\
\hspace*{2.4cm}
$^e$ Quark Model + scaling relation. \\
\hspace*{2.4cm}
$^f$ Relativistic quark model. \\
\hspace*{2.4cm}
$^g$ Bag model. \\
\hspace*{2.4cm}
$^h$ SU(4) symmetry. \\
\hspace*{2.4cm}
$^i$ Reggeon quark-gluon string model.  \\
\hspace*{2.4cm}
$^k$ Experimental limits\\
\end{table} }

\begin{figure}[htb]
\centerline{
\epsfig{bbllx=100pt,bblly=209pt,bburx=507pt,%
bbury=490pt,file=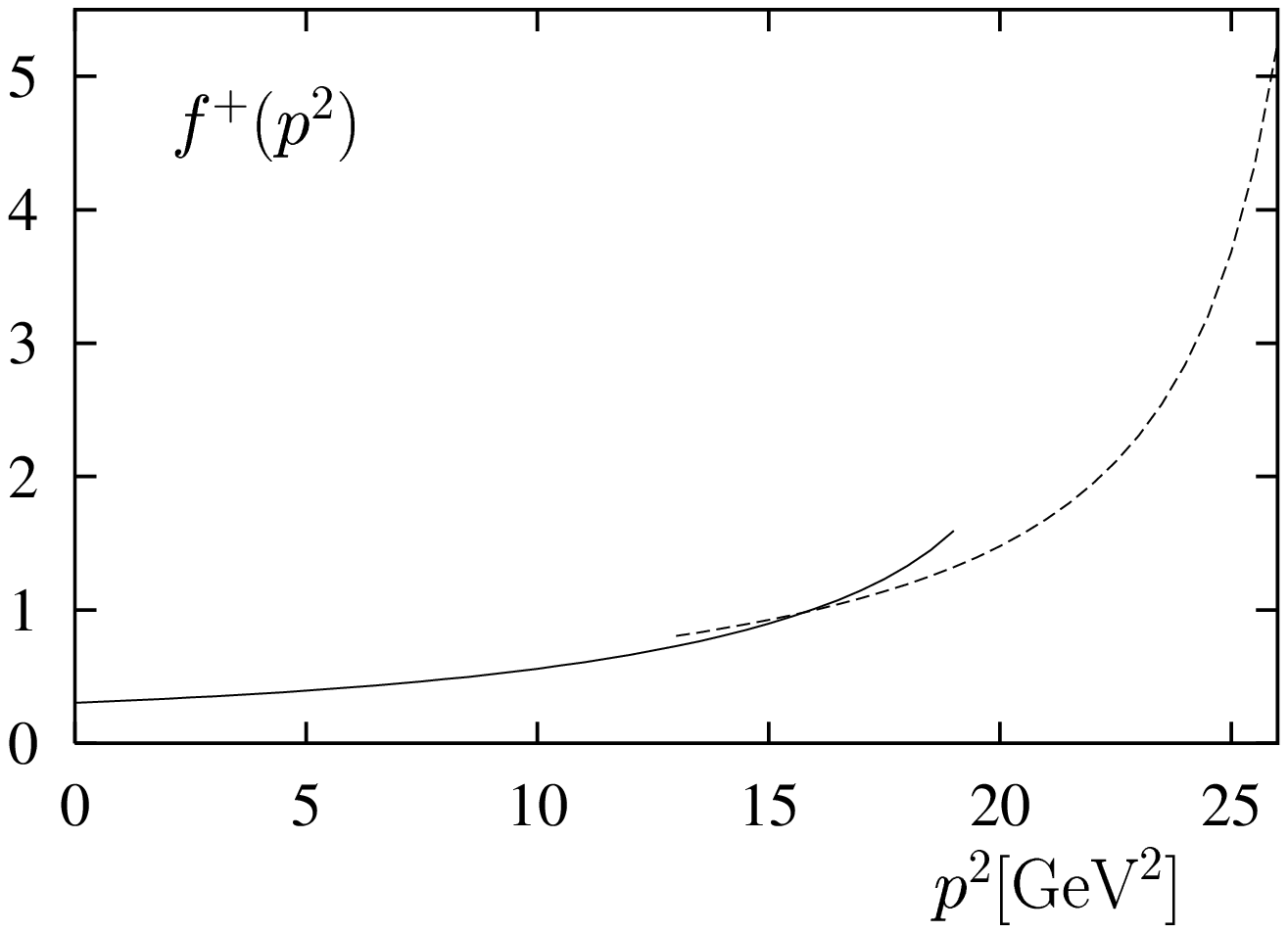,scale=0.9,%
clip=}
}
\caption{\it The $B\to\pi$ form factor $f^+$: direct 
sum rule prediction (solid), and 
single-pole approximation normalized by 
the sum rule estimate for $g_{B^*B\pi}$ (dashed).}
\end{figure}

\begin{figure}[htb]
\centerline{
\epsfig{bbllx=100pt,bblly=209pt,bburx=507pt,%
bbury=490pt,file=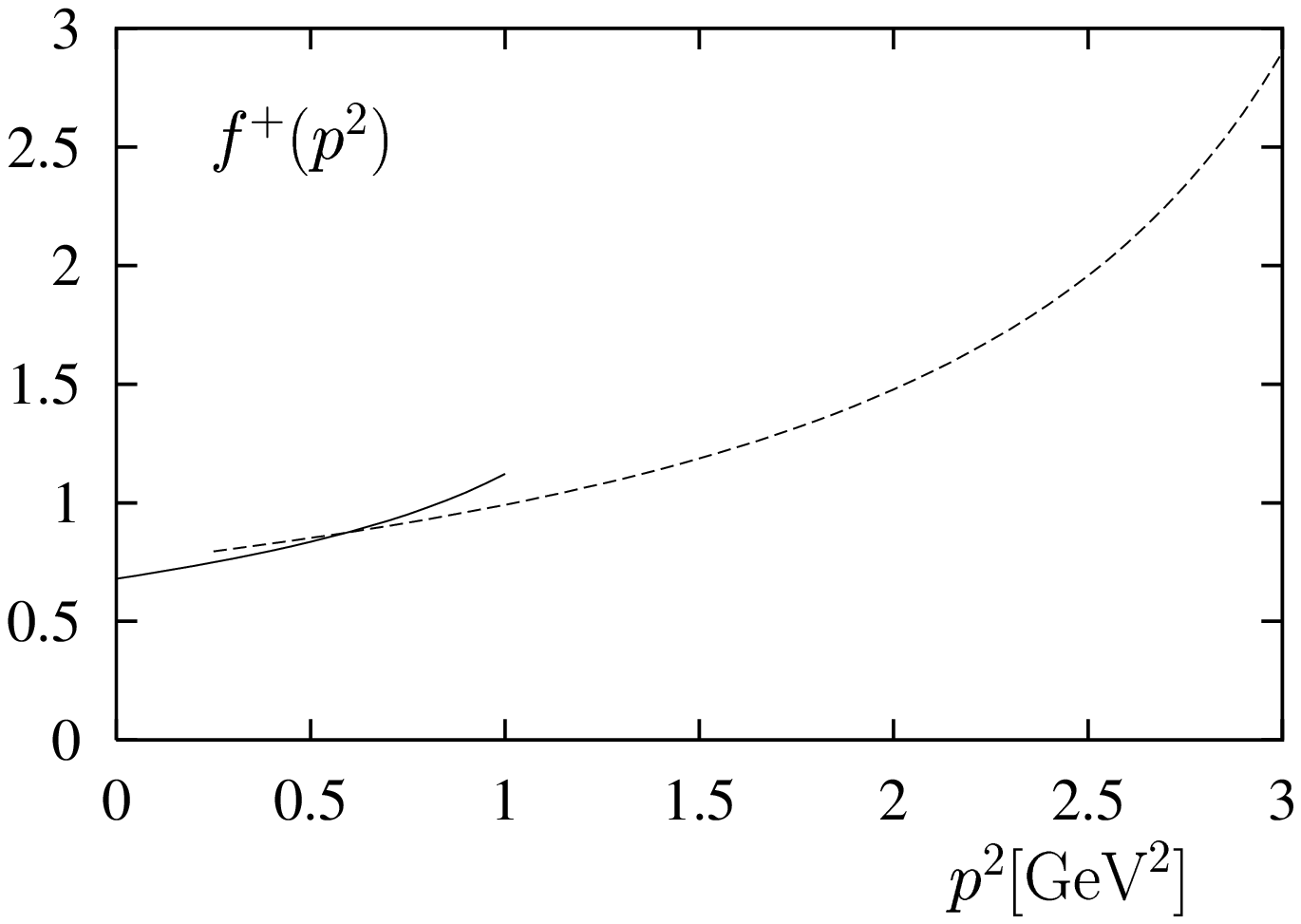,scale=0.9,%
clip=}
}
\caption{\it The $D\to\pi$ form factor $f^+$: direct 
sum rule prediction (solid), and 
single-pole approximation normalized by the sum rule estimate 
for $g_{D^*D\pi}$ (dashed).}
\end{figure}

In contrast to $g_{D^*D\pi}$, the coupling constant $g_{B^*B\pi}$
cannot be measured directly,
since the corresponding decay $B^* \rightarrow B \pi $ is kinematically
forbidden. However, the $B^*B\pi$ on-shell vertex is of great
importance for the understanding of the behaviour of the
$B \to \pi$ 
form factors at large momentum transfer.
Near the kinematic limit the form factor $f^+$ is expected
to be dominated by the $B^*$  pole. The single-pole approximation
given by
\be
f^+(p^2)= \frac{f_{B^*}g_{B^*B\pi}}{2m_{B^*}(1-p^2/m_{B^*}^2)}
\label{onepole}
\ee
is illustrated in Fig. 12 taking $g_{B^*B\pi}$ from (\ref{41})
and $f_{B^*}$ from (\ref{fBstar0}).
Extrapolation of the single-pole model to smaller $p^2$ matches
quite well with the direct estimate from
the light-cone sum rule (\ref{fplus})
at intermediate momentum transfer
$p^2=15$ to $20$ GeV$^2$. This provides us with a consistent and 
complete theoretical description of $f^+$. 
The extrapolation for the $D\to \pi$ form factor using
the analogous  single-pole formula
with $g_{D^*D\pi}$ from (\ref{constD*Dpi}) and $f_{D^*}$ from 
(\ref{fDstar0}) is shown in Fig. 13. Also in this case
we find the direct sum rule result and the pole
model to match nicely at $p^2 \simeq 0.7$ GeV$^2$.

Finally, we refer the reader to \cite{COL95}
for a similar calculation of scalar and axial $B$ meson couplings 
yielding
\be
\Gamma( B(0^{++}) \to B \pi)\simeq \Gamma( B(1^{++})
\to B^* \pi)\simeq 360 ~\mbox{MeV}~,
\ee
and to \cite{AlievBrho} for an investigation of the $B^*B\rho$ coupling. 

\begin{figure}[htb]
\centerline{
\epsfig{bbllx=100pt,bblly=209pt,bburx=507pt,%
bbury=490pt,file=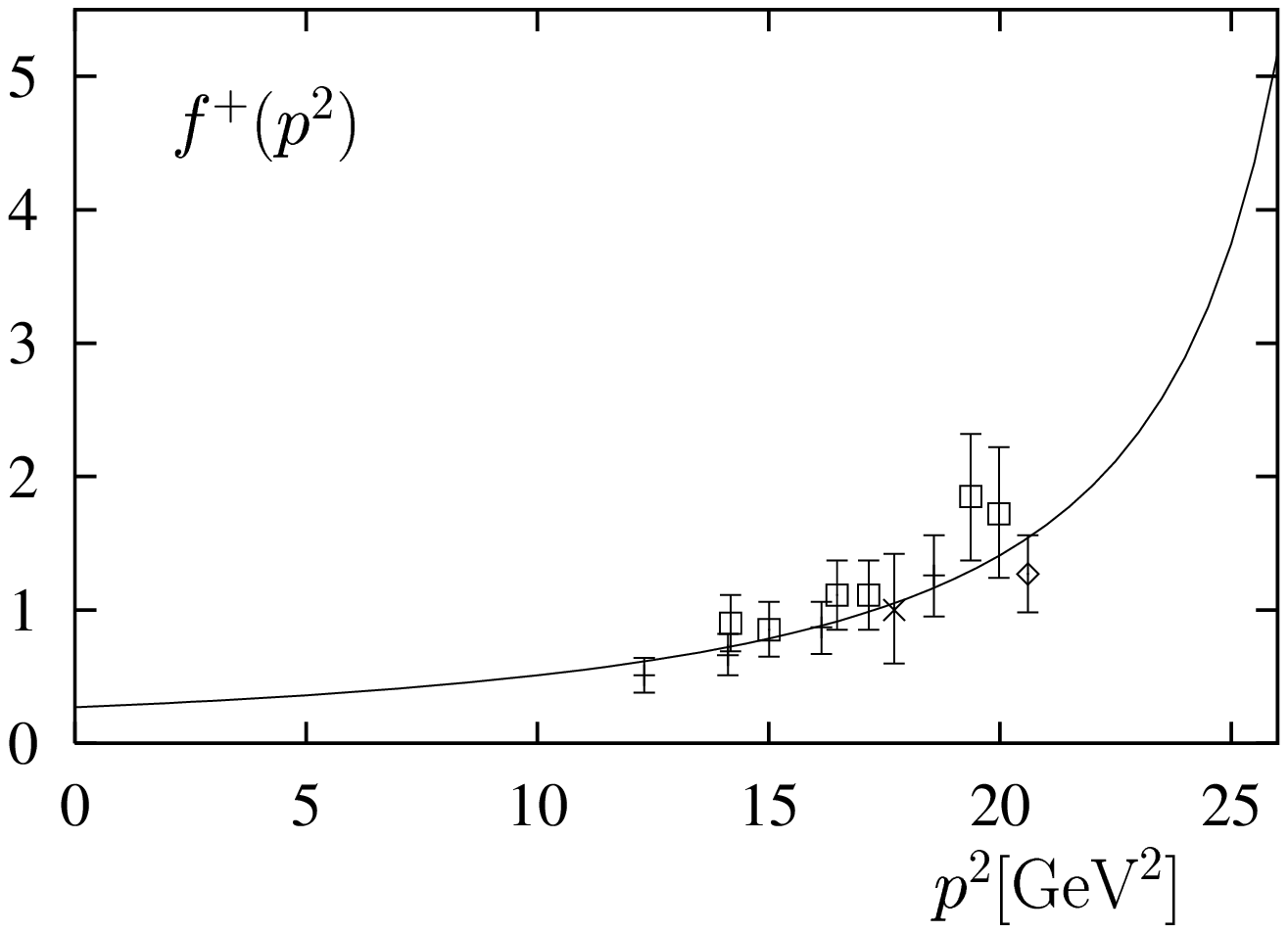,scale=0.9,%
clip=}
}
\caption{\it The sum rule prediction for the $B\to\pi$ form factor $f^+$
in comparison to lattice results \cite{Flynn}.}
\end{figure}

\section{Exclusive semileptonic decays, $V_{ub}$, and all that}

With the $B\to\pi$, $D\to\pi$, and also $B\to\rho$
form factors at hand, we are now in the position to
predict the widths and differential distributions for exclusive
semileptonic decays. In the case of $f^+(p^2)$, 
the sum rules (\ref{fplus}) and (\ref{fin}) together with the single-pole
approximation (\ref{onepole}) provide a complete description
as illustrated in Fig. 12 and 13.
For convenience, we have fitted parametrizations of the form
\be
f^+(p^2) = \frac{f^+(0)}{1-ap^2/m_P^2+b p^4/m_P^4}~,
\label{param}
\ee
$m_P$ denoting the $B$ and $D$ meson mass,
to the theoretical results plotted in these figures.
For the $B \to \pi$ form factor we get
\be
f^+(0) = 0.27,~ a = 1.50,~ b = 0.52.
\label{param1}
\ee
Here, the NLO correction to the twist 2 contribution has been included,
and $f^+(0)$ has been fixed at the value given in (\ref{Bfpnlo}).
At  $p^2< 17$ GeV$^2$  the fit
reproduces the prediction of the 
light-cone sum rule (\ref{fplus}), while  
at $p^2 > 20$ GeV$^2$ it coincides with 
the single-pole approximation (\ref{onepole}). In Fig. 14
the interpolation (\ref{param}) is shown in comparison 
with recent lattice results
\footnote{For a comprehensive review see the article by J.M. Flynn
and C.T. Sachraida in this volume, ref. \cite{flynnsachr}.}.
The agreement is very encouraging. 
The analogous fit of (\ref{param}) to the LO $D \to \pi$ 
form factor plotted in Fig. 13 yields
\be
f^+(0) = 0.68,~ a = 1.16, ~b = 0.32~,
\label{param1D}
\ee  
where $f^+(0)$ has been kept fixed at the value given in (\ref{Dfplo}).

The distribution of the momentum transfer squared 
in $B \to \pi \bar{l} \nu_l$
is given by
$$
\frac{d\Gamma}{dp^2} =
\frac{G^2|V_{ub}|^2}{24\pi^3}
\frac{(p^2-m_l^2)^2 \sqrt{E_\pi^2-m_\pi^2}}{p^4m_B^2}
\Bigg\{ \left(1+\frac{m_l^2}{2p^2}\right)
m_B^2(E_\pi^2-m_\pi^2)\left[f^+(p^2)\right]^2
$$
\be
+ \frac{3m_l^2}{8p^2}(m_B^2-m_\pi^2)^2
\left[f^0(p^2)\right]^2\Bigg\}
\label{dG}
\ee
with $E_\pi= (m_B^2+m_\pi^2 -p^2)/2m_B$ being
the pion energy in the $B$ rest frame, and the form factors being as defined
in (\ref{form3}) and (\ref{f0}). For $l = e$ or $\mu$,
the form factor $f^0$ plays a negligible role 
because of the smallness of the electron and muon masses.
Another interesting observable
is the distribution of the charged lepton energy $E_l$ 
in the $B$ rest frame:
\bq
\frac{d\Gamma}{dE_l}&=&
\frac{G^2|V_{ub}|^2}{64\pi^3m_B}\int^{p_{max}^2}_{p_{min}^2} dp^2
\Bigg\{ \Bigg[ 8E_l(m_B^2-m_\pi^2+p^2)
\nonumber
\\
&&-4m_B(p^2+4E_l^2) +\frac{m_l^2}{m_B}\left(8m_BE_l-3p^2+4m_\pi^2
\right ) -\frac{m_l^4}{m_B}\Bigg] \left[f^+(p^2) \right]^2
\nonumber
\\
&&+\frac{2m_l^2}{m_B}\Bigg[2m_B^2+p^2-2m_\pi^2-4m_BE_l+m_l^2 \Bigg]
f^+(p^2)f^-(p^2)
\nonumber
\\
&&+\frac{m_l^2}{m_B}(p^2-m_l^2)\left[f^-(p^2) \right]^2 \Bigg\}
\label{spectrum}
\eq
with the integration limits 
$p^2_{\stackrel{max}{min}} = m_B(E_l \pm \sqrt{E_l^2-m_l^2})+ O(m_\pi^2)$.
The terms proportional to the pion mass squared not shown explicitly
are taken into account in the numerical calculations.
Here, terms involving the form factor
$f^-$ are suppressed by the lepton mass if $l=e,\mu$.

\begin{figure}[htb]
\centerline{
\epsfig{bbllx=100pt,bblly=209pt,bburx=507pt,%
bbury=490pt,file=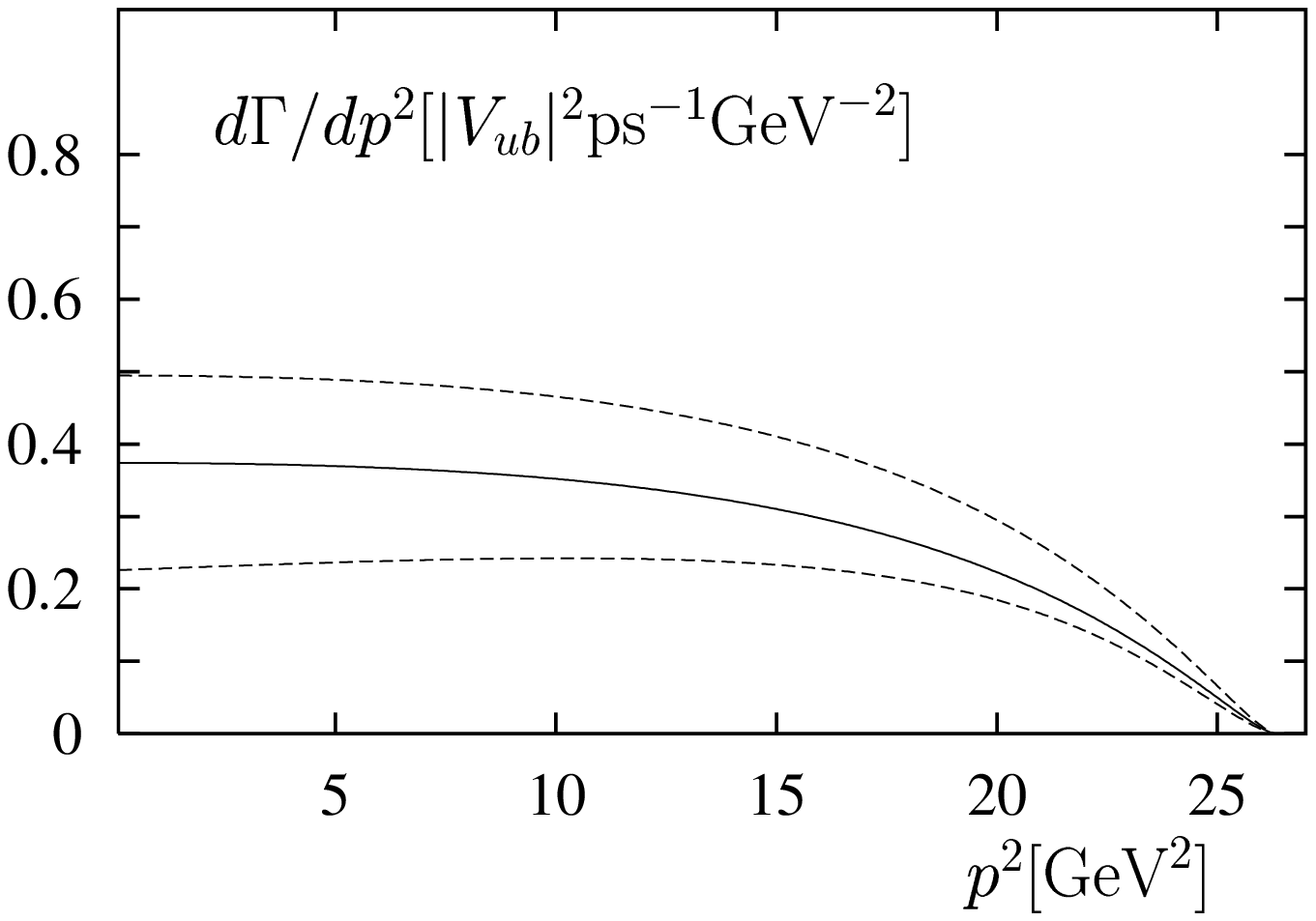,scale=0.9,%
clip=}
}
\caption{\it Distribution of the momentum transfer squared in
$B \to \pi \bar l \nu_l$ ($ l=e, \mu$). 
The dashed curves indicate the theoretical uncertainty
discussed in detail in the text.}
\end{figure}

The above decay distributions  
are shown in Fig. 15 and 
16 for $l=e,\mu$. For the integrated 
width one obtains
\be
\Gamma(B\rightarrow\pi \bar{l} \nu_l) =
 (7.5\pm 2.5) ~|V_{ub}|^2\ \mbox{ps}^{-1}, 
\label{elmu}
\ee
where the theoretical uncertainty reflects 
the uncertainties in $f^+$ (added in quadrature) which have been discussed 
in detail in section 5..

Contrary to the semileptonic decays into $e$ and $\mu$,  
the decay $B \rightarrow \pi \bar{\tau} \nu_\tau$
is quite sensitive to the scalar form factor $f^0$.
Unfortunately, the single-pole approximation used 
to extrapolate the sum rule result on $f^+$
to maximum $p^2$ cannot be applied to $f^0$. 
This is because the scalar $B$ ground state  
which should be about 500 MeV heavier than the pseudoscalar $B$
lies too far above the kinematical endpoint 
$p^2=(m_B-m_\pi)^2$ of the $B\to \pi$ transition in order
to dominate the form factor.
Nearby excited resonances and nonresonant states are expected to give 
comparable contributions.   
For illustrative purposes \cite{KRW}, 
we extrapolate the form factor $f^0$  
linearly from the maximal value $ p^2 = 15$ GeV$^2$ at which 
the sum rules (\ref{fplus}) and (\ref{fplusminus}) still hold  
to the value at $p^2 \simeq m_B^2$ dictated by the
Callan-Treiman limit \cite{Vol}:
\be
\lim_{p^2\to m^2_B} f^0(p^2) = \frac{f_B}{f_\pi}= 1.1 ~\mbox{to}~ 1.6~,
\label{CT}
\ee
where we have used the conservative estimate
$f_B = 150$ to 210 MeV.
This rough extrapolation of the sum rule result is plotted in Fig. 17.
Also shown are lattice data. They are systematically lower than our 
expectation. It should however be noted that in contrast to $f^+$,
NLO effects are still
missing in the sum rule for $f^0$.

Fig. 16 shows the resulting distribution of $E_\tau$.
As anticipated, the spectrum is quite sensitive  
to the large $p^2$ behaviour of $f^0$, and may thus allow 
to determine, or at least constrain the scalar form factor experimentally.
The integrated partial width is 
\be
\Gamma(B \to \pi \bar{\tau} \nu_\tau) = (6.1 \pm 0.4) ~
|V_{ub}|^2\ \mbox{ps}^{-1},
\label{tau}
\ee
yielding, together with (\ref{elmu}), the ratio 
\be
\frac{\Gamma(B\rightarrow\pi \bar{\tau} \nu_\tau)}
{\Gamma(B\rightarrow\pi \bar{e} \nu_e)}= 0.75 ~\mbox{to} ~0.85~.
\label{r}
\ee
This ratio is independent of 
$V_{ub}$, and less sensitive to uncertainties in the sum rule parameters
than the widths themselves. The range quoted above 
corresponds to the variation of $f^0$ within the two extrapolations
considered in Fig. 17.

\begin{figure}[htb]
\centerline{
\epsfig{bbllx=110pt,bblly=209pt,bburx=507pt,%
bbury=490pt,file=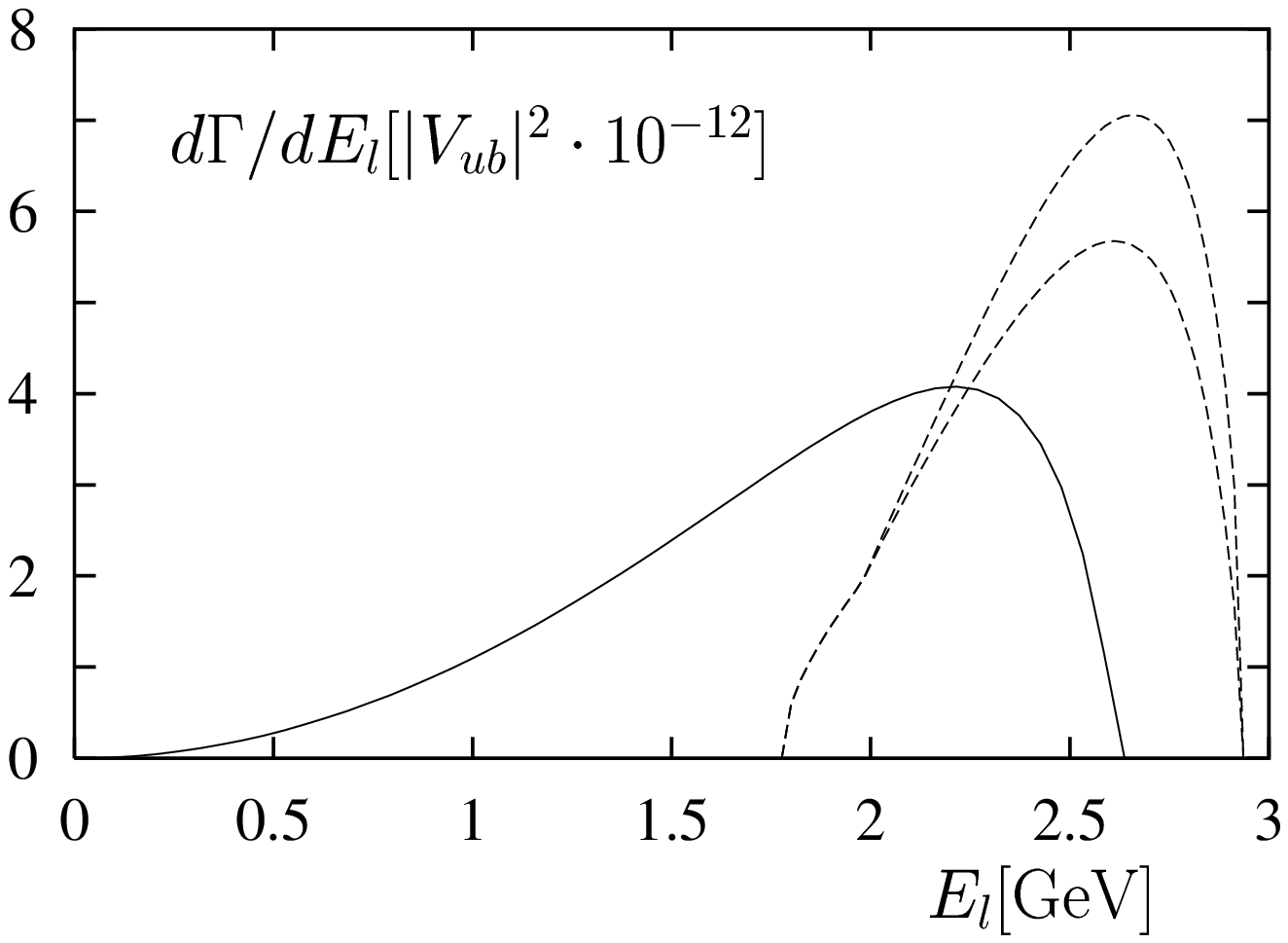,scale=0.9,%
clip=}
}
\caption{\it Distribution of the lepton energy in
$B \to \pi \bar{l} \nu_l$ for $l=e,\mu$ (solid) 
and $l=\tau$ (dashed). In the latter case 
the two curves correspond to the two extrapolations 
of $f^0$ shown in Fig. 17.}
\end{figure}
\begin{figure}[htb]
\centerline{
\epsfig{bbllx=100pt,bblly=209pt,bburx=507pt,%
bbury=490pt,file=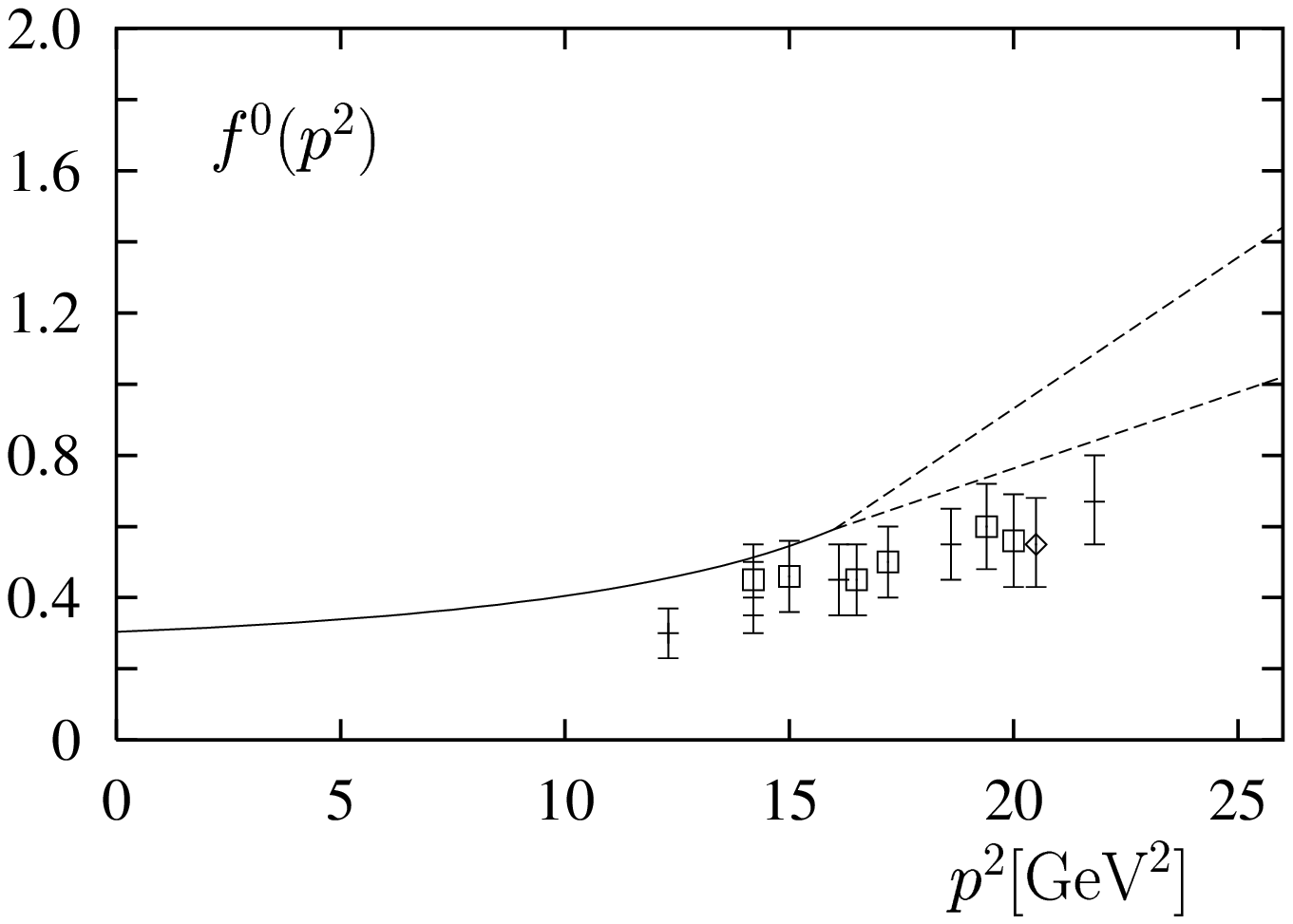,scale=0.9,%
clip=}
}
\caption{\it The $B\to\pi$ form factor $f^0$: direct 
sum rule estimate (solid) and linear 
extrapolations to the limit (\ref{CT}) (dashed).  
The lattice results are from \cite{Flynn}.}
\end{figure}

Recently, the CLEO collaboration \cite{CLEO}
has reported the first observation of  the semileptonic decays
$B\rightarrow \pi \bar{l}\nu$ and $B\rightarrow \rho \bar{l}\nu$
($l = e, \mu $). From the measured branching fraction
$BR(B^0\to\pi^-l^+\nu_l) = (1.8 \pm 0.4\pm 0.3 \pm 0.2)\cdot 10 ^{-4}$~,
and the world average of the $B^0$ lifetime
\cite{PDG},  $\tau_{B^0}=1.56 \pm 0.06 $ ps,
one derives
\be
\Gamma(B^0\rightarrow \pi ^- l^+ \nu_l) =
(1.15 \pm 0.35)\cdot10^{-4}~\mbox{ps}^{-1}~,
\label{CLEOpi7}
\ee
where the errors have been added in quadrature.
Comparison of (\ref{CLEOpi7}) with (\ref{elmu}) yields
\be
|V_{ub}| = 0.0039 \pm 0.0006 \pm 0.0006.
\label{vubpi}
\ee
Here, the first (second) error corresponds to the current
experimental (theoretical) uncertainty.

A similar analysis can be performed for 
$B \rightarrow \rho \bar{l}\nu_l$. Fig. 18 shows the 
distribution of the momentum transfer squared
obtained from the form factors plotted in Fig. 11 \cite{BallBraun}.
The integrated width predicted in \cite{BallBraun} is
\be
\Gamma(B\rightarrow \rho \bar{l} \nu_l)= (13.5\pm 4)
~|V_{ub}|^2~\mbox{ps}^{-1}~.
\label{rhoenu}
\ee
Comparison with the CLEO result,
\be
\Gamma(B^0\rightarrow \rho ^- l^+ \nu_l) =
(1.60 \pm 0.6)\cdot10^{-4}~\mbox{ps}^{-1}~,
\label{CLEOrho7}
\ee
derived from the measured branching ratio
$BR(B^0\ra\rho^-l^+\nu_l) = (2.5 \pm 0.4^{+ 0.5}_{-0.7}\pm 0.5)\cdot 
10 ^{-4}$~
and the $B^0$ lifetime also used in (\ref{CLEOpi7}), gives
\be
|V_{ub}| = 0.0034 \pm 0.0006 \pm 0.0005.
\label{vubrho}
\ee
Within errors, the values of $|V_{ub}|$ extracted from the
two exclusive semileptonic decays are nicely 
consistent with each other, and also coincide with the inclusive 
determination of $V_{ub}$.

A further test of the consistency between theory and 
experiment is provided by comparing
the ratio ($l=e,\nu$)
\be
\frac {\Gamma (B^0 \rightarrow \rho^-l^+\nu_l)}
{\Gamma (B^0 \rightarrow \pi^-l^+\nu_l)} = 1.8 \pm 0.7
\label{neuGamma}
\ee
calculated from (\ref{rhoenu}) and (\ref{elmu})
with the ratio
\be
\frac {BR (B^0 \rightarrow \rho^-l^+\nu_e)}
{BR (B^0 \rightarrow \pi^-l^+\nu_e)} = 1.4^{+0.6}_{-0.4}\pm 0.3 \pm 0.4 
\label{neuBR}
\ee
observed by CLEO.  Although the test is passed by the present estimate, 
the uncertainties on both sides
are still too big to draw any firm conclusion.

\begin{figure}[htb]
\centerline{
\epsfig{file=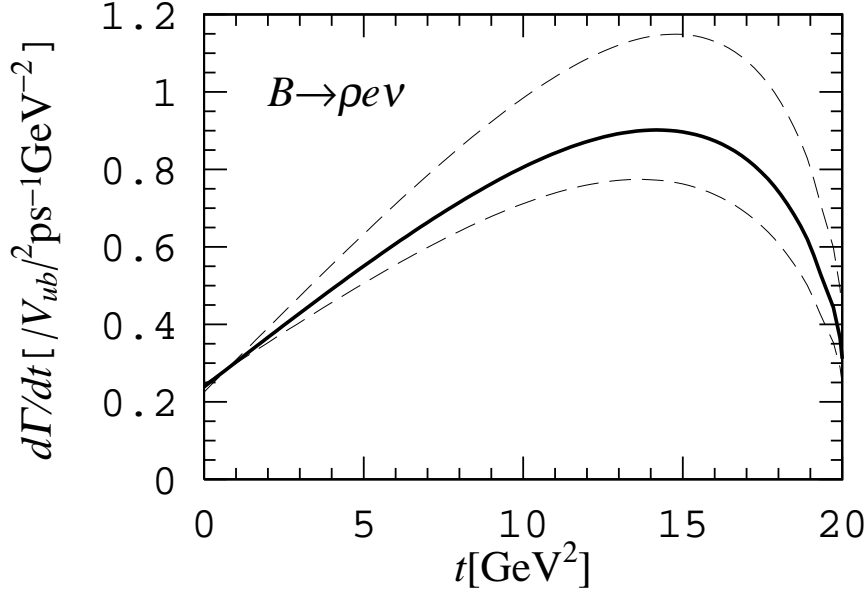,width=12cm,bbllx=50pt,bblly=50pt,bburx=300pt,%
bbury=220pt,clip=}
}
\caption{\it Distribution of the momentum transfer squared in
$B\to\rho \bar{e} \nu_e$.
The dashed curves show the theoretical uncertainty 
(from \cite{BallBraun}).}
\end{figure}

Turning to semileptonic $D$ decays
we show, in Fig. 19,
the distribution of the momentum transfer squared in 
$ D \to \pi \bar{l} \nu_l$.
For the integrated width we get 
\be
\Gamma(D\rightarrow\pi \bar{l} \nu_l) =
0.16 ~|V_{cd}|^2 ~\mbox{ps}^{-1} =
8.0 \cdot 10^{-3} ~\mbox{ps}^{-1}
\label{Gamma}
\ee
using $|V_{cd}|= 0.224 \pm 0.016 $ \cite{PDG}.
This prediction should be compared with
the experimental result
\be
\Gamma(D^0\rightarrow\pi^- e^+ \nu_e) =
(9.2^{+2.9}_{-2.4})\cdot10^{-3} ~\mbox{ps}^{-1}
\label{brD}
\ee
derived from
the branching ratio
$BR(D^0\rightarrow\pi^- e^+ \nu) =
(3.8^{+1.2}_{-1.0})\cdot 10^{-3}$
and the lifetime
$\tau_{D^0}= 0.415 \pm 0.004  ~\mbox{ps}$ ~\cite{PDG}.
The present theoretical uncertainty is estimated to be 
of the order of the experimental error. 
In other words, the CKM-suppressed exclusive semileptonic
$D$ decays are not yet  measured
precisely enough to really challenge theory.
In addition, there is a further demand for better data. One may
use a precise measurement of $D \to \pi \bar{l} \nu_l$ to constrain the
light-cone sum rule 
for the $D \rightarrow \pi$ form factor. Since the evolution 
of the pion wave function and other scale-dependent
input quantities from the charm to the bottom scale
is well under control, this is a promising way to considerably improve the 
sum rule predictions on $B \to \pi$ transitions.

\begin{figure}[ht]
\centerline{
\epsfig{bbllx=89pt,bblly=209pt,bburx=507pt,%
bbury=490pt,file=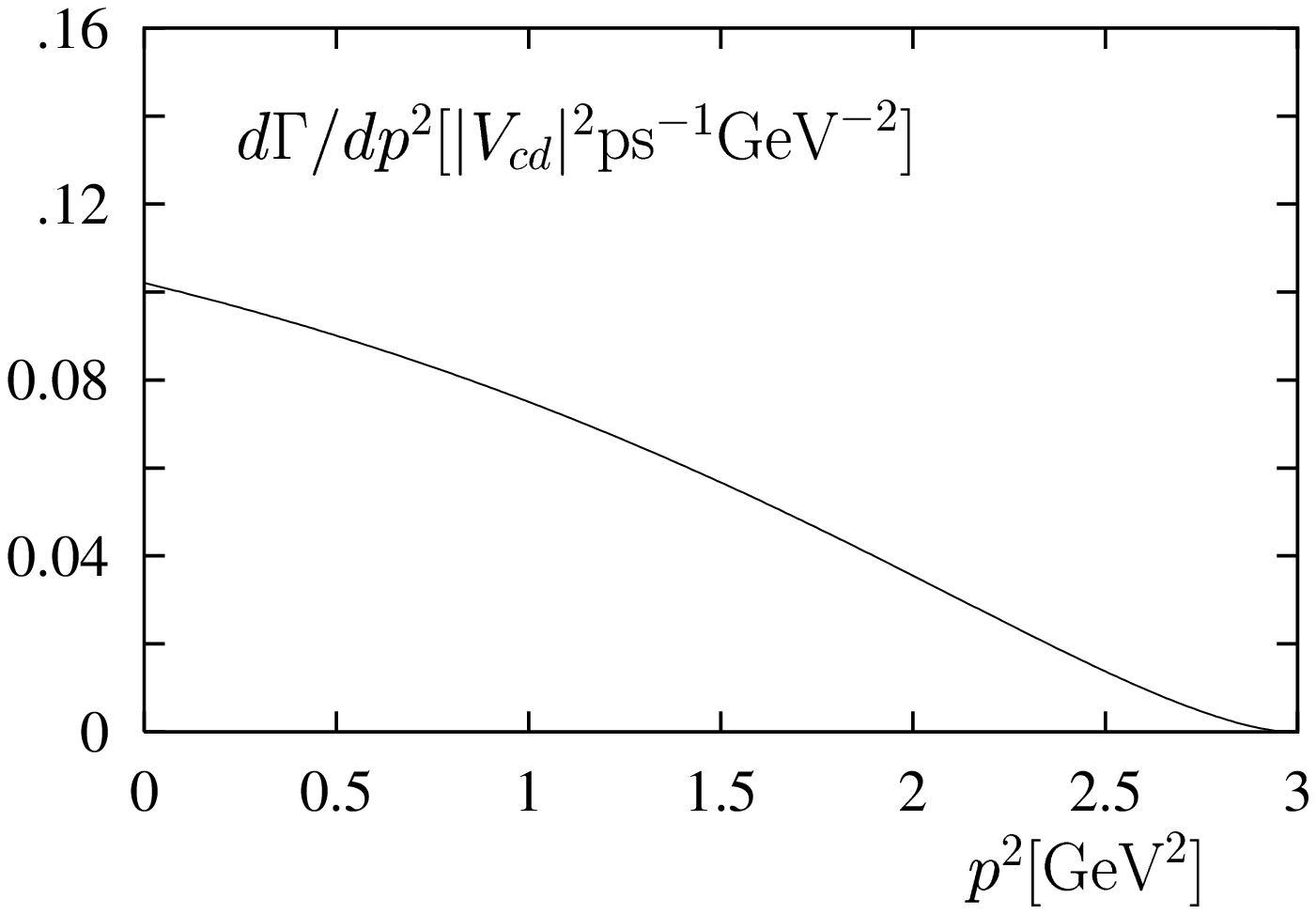,scale=0.9,%
clip=}
}
\caption{\it Distributions 
of the momentum transfer squared in 
$D \rightarrow \pi \bar l \nu_l$ ($l=e, \mu$).}
\end{figure}


\section{Heavy quark limit} 

The light-cone sum rules described in section 5 provide a unique
possibility
to investigate the heavy-mass dependence of the
$B\rightarrow \pi$ form factors.
To this end, one employs the following scaling
relations for mass parameters and decay constants:
\be
m_B = m_b+\bar{\Lambda}~,~~~ s_0^B = m_b^2 + 2m_b\omega_0 ~,
~~~M^2= 2m_b\tau~,
\label{hqet}
\ee
\be
f_B = \hat{f}_B/\sqrt{m_b}, ~~~~ f_{B^*} = \hat{f}_{B^*}/\sqrt{m_b}~,
\label{fBhat}
\ee
where in the heavy quark limit 
$\bar{\Lambda}$, $\omega_0$, $\tau$, $\hat{f}_B$ and $\hat{f}_{B^*}$ are
$m_b$-independent quantities.
With these substitutions it is straightforward to expand
the sum rules (\ref{fplus}) and (\ref{fplusminus}) in $m_b$. 
In both cases, the light-cone expansion
in terms of wave functions with increasing twist is 
consistent with the heavy mass expansion, that is
the higher-twist contributions either 
scale with the same power of $m_b$ as the leading-twist term,
or they are suppressed by extra powers of $m_b$.

We find that the asymptotic scaling 
of the form factors
differs sharply at small
\footnote{see also ref. \cite{CZ1}} and large momentum transfer
\cite{KRW}.
At $p^2 = 0$  
\be
f^+(0)= f^0(0)\sim m_b^{-3/2}~,
\label{0limit}
\ee
\be
f^+(0) +f^-(0)\sim m_b^{-3/2}~,
\label{0limit1}
\ee
whereas at $p^2 = m_b^2-2m_b\chi$,
$\chi$ being independent of $m_b$,
\bq
f^+(p^2) \sim m_b^{1/2}~,
\label{scal1}
\\
f^+(p^2) + f^-(p^2) \simeq f^0(p^2) \sim m_b^{-1/2}~.
\label{scal2}
\eq
The sum rules thus nicely reproduce
the asymptotic dependence of the form factors $f^+$
and $f^-$ on the heavy quark mass $m_b$ derived in \cite{IW,Vol}
for small pion momentum in the rest frame of the $B$ meson.
In addition, the sum rules allow to investigate the opposite
region of maximum pion momentum where neither HQET nor the single-pole 
model can be trusted. In particular, from the sum rule
point of view it is expected
that the excited and continuum states become
more and more important as $p^2 \rightarrow 0$. This is reflected 
in the change of the asymptotic mass dependence.
Claims in the literature which
differ from (\ref{0limit}) to (\ref{scal2}) are 
often based on the pole model and therefore 
incorrect in our opinion. A similar
analysis has been carried out in \cite{BallBraun} with
essentially the same conclusions.

Furthermore, the sum rule (\ref{fin}) for the $B^*B\pi$
coupling constant suggests
\be
g_{B^*B\pi} \sim m_b~,
\label{scalg}
\ee
in agreement with the expectation from HQET \cite{NW,IW,Wise}.
At the expense of additional parameters such as $\bar{\Lambda}$,
$\omega_0$, etc. 
light-cone sum rules can also be used to
determine the $1/m_Q$ corrections. For $g_{B^*B\pi}$ 
and $g_{D^*D\pi}$,
a simple quantitative estimate of the latter is obtained
by fitting the numerical results (\ref{41}) and (\ref{constD*Dpi})
to the form 
\be\label{1/m}
g_{B^*B\pi} = \frac{2 m_B}{f_\pi}\cdot \hat g
\Bigg(1+\frac{\Delta}{m_B}\Bigg)
\ee
and the analogous expression for $g_{D^*D\pi}$.
This yields \cite{BBKR}
\be
\hat g =0.32\pm 0.02~,~~
\Delta =(0.7 \pm 0.1)~ \mbox{GeV} ~.
\label{fit}
\ee
In Table 2 , the above  
value of the reduced
coupling constant $\hat{g}$ is 
compared to the results in other approaches.
The next-to-leading term in the heavy mass is sizeable,
increasing from about 15\% for $g_{B^*B\pi}$ to 40\% for $g_{D^*D\pi}$.

Finally, in the heavy quark limit the ratio
\be
r=\frac{g_{B^*B \pi}f_{B^*}\sqrt{m_{D}}}
{g_{D^*D \pi}f_{D^*}\sqrt{m_{B}}}
\label{ratio3}
\ee
is expected to approach unity. Moreover, it has been shown 
\cite{BLN} that $r$ is subject to $1/m_Q$ corrections 
only in next-to-leading order. Indeed,
the ratio $r=0.92$ 
derived from the light-cone sum rules deviates surprisingly
little from unity, in agreement with the HQET expectation.

\section{ The nonfactorizable amplitude for $B \ra J/\psi K$}

As a final example for applications of QCD sum rules to 
exclusive heavy meson decays we consider nonleptonic two-body
decays. This class of processes is theoretically 
much more complicated than the (semi)leptonic decays
discussed so far. As compared to the latter, effects from
(a) hard gluon exchange at short distances,
(b) soft interactions of quarks and gluons including 
nonspectator effects, (c) hadronization, and 
(d) final state interactions among the hadronic 
decay products change things considerably.
Up to now, only the hard-gluon effects can be systematically taken
into account in the framework of improved QCD perturbation
theory. The result is an 
effective weak Hamiltonian at the physical scale 
$\mu \simeq m_Q \ll m_W$, 
given by a sum of local operators 
with renormalized Wilson coefficients
\footnote{See, e.g., ref. \cite{NS}.}. 

We restrict our discussion to
the decay $B \rightarrow J/\psi K$ which will play an
important role at future $B$-factories and which brings
the main theoretical difficulties to light. The piece of the
effective Hamiltonian relevant for this decay mode
may be written in the form 
\be
H_W= \frac{G}{\sqrt{2}}V_{cb}V^*_{cs}\{(c_2
+\frac{c_1}3) O_2+2c_1\tilde{O}_2\}~,  
\label{H}
\ee
where 
\be
O_2(\mu)=(\bar{c}\Gamma^\rho c)(\bar{s}\Gamma_\rho b),\
\tilde{O}_2(\mu)=(\bar{c}\Gamma^\rho \frac{\lambda^a}2c)(\bar{s}\Gamma_\rho
\frac{\lambda^a}2 b)
\label{o}
\ee
with $\Gamma_\rho = \gamma_\rho(1-\gamma_5)$. 
The Wilson coefficients $c_{i}(\mu)$ contain the effects
from QCD interactions at short distances below the scale 
set by the inverse $b$-quark mass. 
The hadronic  matrix elements of the four-quark operators (\ref{o})
are supposed to incorporate the long-distance effects (b) to (d).
The problem of calculating these matrix elements is extremely
demanding and still far from a satisfactory solution. 
 
In a radical first approximation, one may factorize
the matrix elements of $H_W$ for $B \to J/\psi K$ 
into products of 
hadronic matrix elements of the currents that compose $H_W$.
Strong interactions at scales lower than $\mu$ 
between quarks entering different 
currents as well as nonspectator effects are thereby
completely neglected. Moreover, 
the matrix element of the operator
$\tilde{O}_2$ vanishes because of colour conservation so that 
\be
\langle J/\psi K\mid H_W\mid B\rangle 
= \frac{G}{\sqrt{2}}V_{cb}V^*_{cs} \left(c_2(\mu)
+\frac{c_1(\mu)}3 \right) 
\langle J/\psi K \mid O_2(\mu) \mid B\rangle~.
\label{factO}
\ee
The factorized matrix element of the operator $O_2$ is given by
\be
\langle J/\psi K\mid O_2(\mu)\mid B\rangle
= \langle J/\psi\mid \bar{c}\Gamma^\rho c  \mid 0 \rangle
\langle K  \mid \bar{s}\Gamma_\rho b \mid B\rangle 
= 2f_\psi f_{B \ra K}^+m_\psi(\epsilon^\psi  \cdot q) ~,
\label{factoriz}
\ee
where 
\be
f_{\psi} = 405~ \mbox{MeV}
\label{fpsi}
\ee
is the decay constant determined 
by the leptonic width 
$\Gamma( J/\psi$\-$ \ra l^+l^-) = 5.26 \pm 0.37$ keV, and
\be
f_{B \ra K}^+ =  0.55 \pm 0.05 
\label{fBK}
\ee
is the $B\ra K$ form factor at the
momentum transfer $p^2=m_\psi^2$ estimated \cite{BKR} from a 
light-cone sum rule similar 
to the one for the $B \ra \pi$ 
form factor given in (\ref{fplus}).
Obviously, $\epsilon^\psi$ denotes the $J/\psi$ polarization vector,
and $q$ the $K$ four-momentum.  

Already at this point one encounters a principal problem:
since the matrix elements of quark currents in (\ref{factoriz})
are scale-independent, the $\mu$-dependence of  
$\langle J/\psi K \mid O_2(\mu) \mid B\rangle $ which is 
supposed to  
cancel the $\mu$-dependence of the Wilson coefficients 
in (\ref{factO}) in order to give a
physically sensible result, is lost. 
Hence, the above approximation
can at best be 
valid at a particular value of $\mu$ which could be called
the factorization scale $\mu_F$. The conventional assumption $^6$
is $\mu_F = O(m_b)$. 

Using the next-to-leading order coefficients $c_{1,2}(\mu)$ 
in the HV scheme with 
$\Lambda^{(5)}_{\overline{MS}}=225$ MeV from 
\cite{Buras2} and
taking $\mu=m_b \simeq 5$ GeV, 
one has 
\be
c_2(\mu) + \frac{c_1(\mu)}{3} = 0.155~.
\label{c2c1}
\ee
Together with  
(\ref{fpsi}) and (\ref{fBK}), this yields 
\be
BR( B\rightarrow J/\psi K) = 0.025\%~, 
\label{BRfact} 
\ee
a branching ratio which is considerably 
smaller than the measurements
\cite{PDG,CLEOnl}
\be
BR( B^- \rightarrow J/\psi K^- )= (0.101 \pm 0.014 )\%~,
\label{CLEOpsi}
\ee
\be
BR( B^0 \rightarrow J/\psi \bar K^0 )= (0.075 \pm 0.021) \% ~.
\label{CLEO0}
\ee

The quantitative failure and the scale problem pointed out above
imply that naive factorization of matrix elements does not work.
Factorization has to be accompanied by a 
reinterpretation  of the Wilson coefficients. For  
decays such as $B \ra J/\psi K$,  
the short-distance coefficient
$c_2(\mu) + c_1(\mu) / 3 $
is substituted by an effective coefficient $a_2$ which is supposed to
incorporate possible nonfactorizable contributions.
Phenomenologically \cite{BSW}, 
$a_2$ is treated as 
a free parameter to be determined from experiment. From 
(\ref{factO}), (\ref{factoriz}),
and (\ref{CLEOpsi}),
the most precise of the two measurements, one finds
\be
|a_2^{B\psi K}|= 0.31 \pm 0.02~, 
\label{a2cleo}
\ee
where the quoted error is purely experimental.
The sign of $a_2^{B\psi K}$ remains undetermined.
The above value is close to the outcome of a comprehensive
analysis \cite{NS} \footnote{We refer here to the  
so-called new model for the heavy-to-light form factors.}
of nonleptonic two-body $B$-decays.

The big difference between the short-distance and effective
coefficient (\ref{c2c1}) and (\ref{a2cleo}), respectively,
points at the existence of sizeable nonfactorizable contributions.
The latter are also needed to
cancel or at least soften the strong 
$\mu$-dependence of (\ref{c2c1}).
A deeper study shows that the dominant nonfactorizable effects
should arise from the matrix element
of the operator  $\tilde{O}_2$. Writing the latter in the 
convenient parametrization 
\be
\langle J/\psi K\mid \tilde{O}_2(\mu) \mid B\rangle = 
2 f_\psi \tilde{f}_{B \psi K}(\mu) m_\psi (\epsilon^\psi \cdot q)~,
\label{nf}
\ee
and adding it to the factorized
matrix element (\ref{factoriz}) of $O_2$, one gets
for the complete decay amplitude: 
\be
\langle J/\psi K\mid H_W\mid B\rangle 
= \sqrt{2}G~V_{cb}V^*_{cs}
a_2^{B\psi K}f_\psi f_{B \ra K}^+m_\psi(\epsilon^\psi\cdot q) 
\label{ampl}
\ee
with the effective coefficient \cite{KR} 
\be
a_2^{B\psi K}
=c_2(\mu)+\frac{c_1(\mu)}3 + 
2c_1(\mu)\frac{\tilde{f}_{B\psi K}(\mu)}{f_{B\ra K}^+}~.
\label{a2}
\ee
In Fig. 20, we show the partial width for $B\to J/\psi K$ 
as a function of the parameter $\tilde{f}_{B\psi K}$
associated with $\langle \tilde{O}_2 \rangle$. Note that 
$\tilde{f}_{B\psi K}= 0$ corresponds to naive factorization, while 
the fitted value of $a_2^{B\psi K}$ given in (\ref{a2cleo}) 
implies
\be
\tilde{f}_{B\psi K}(\mu=m_b)= +0.04~~\mbox{or}~ -0.12~.
\label{tildemb}
\ee
For a smaller scale the value of $\tilde{f}_{B\psi K}$
is slightly shifted to the right on the real axis, e.g.,
\be
\tilde{f}_{B\psi K}(\mu = \frac{1}{2} m_b)= +0.06 ~~\mbox{or}~ -0.09~.
\label{tildemu}
\ee
We see that a nonfactorizable amplitude of 10 to 20~\% of the size
of the factorizable one is sufficient to conciliate expectation 
with experiment. 

\begin{figure}[htb]
\centerline{
\epsfig{bbllx=100pt,bblly=209pt,bburx=507pt,%
bbury=530pt,file=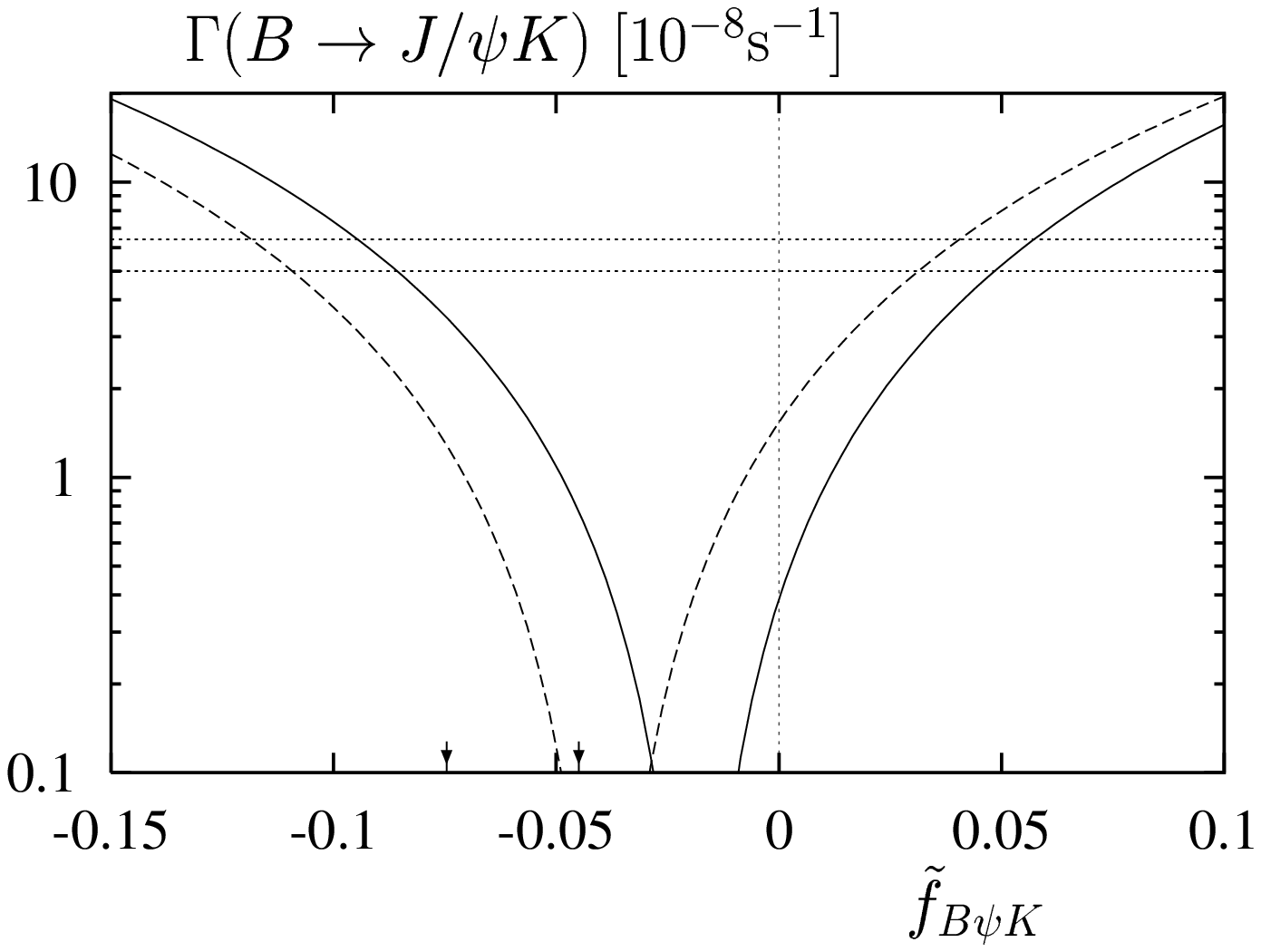,scale=0.9,%
clip=}
}
\caption{\it The partial width for $B \rightarrow J/ \psi K $ 
as a function of $\tilde{f}_{B\psi K}$ parameterizing
the nonfactorizable contribution to the decay amplitude.
The horizontal dotted lines represent
the average of the experimental widths
given in the text. The solid (dashed) curves show the theoretical
expectation for $\mu=m_b/2$ ($\mu=m_b$).
The arrows indicate the QCD sum rule estimate. 
}
\end{figure}

The theoretical calculation of $a_2^{B\psi K}$ and of the analogous 
coefficients for other two-body decays \cite{BSW} is one
of the most important tasks in heavy flavour physics.
As a first step in this direction, 
we have undertaken a rough
estimate of $\tilde{f}_{B \psi K}$ using again QCD sum rule methods
\footnote{This 
work was done in collaboration with B. Lampe. Results were already  
reported in \cite{KR}.}.
Following the general idea put forward in \cite{BS}, 
we choose the four-point correlation function 
\be
\widetilde{\Pi}_{\mu\nu}(p,q)=\int~ d^4x~d^4y~d^4z~e^{iqx+ipy}
\langle 0 \mid T\{j_{\mu5}^K(x)j_\nu^\psi(y)\tilde{O_2}(z)j^B_5(0)\}\mid 0
\rangle 
\label{corr}
\ee
with $j^B_5= \bar{b}i\gamma_5 u$, $ j_\nu^\psi= \bar{c}\gamma_\nu c $, and
$j_{\mu5}^K= \bar{u}\gamma_\mu \gamma_5s$ 
being the generating currents of the mesons 
participating in the decay $B \to J/\psi K$, and 
$p+q$, $p$ and $q$ being the respective four-momenta.

Generalizing the procedure applied in section 2 and 3
to the two- and three-point correlation functions 
(\ref{corr2}) and (\ref{corr3}), respectively,
one writes a dispersion relation for (\ref{corr}) in terms of 
intermediate hadronic states in the $B$, $J/\psi$, and $K$ channel.
The ground state contribution to
\be
\widetilde{\Pi}_{\mu\nu}(p,q)= 
i\frac{\langle
0\mid j_{\mu 5}^K\mid K \rangle
\langle
0\mid j_\nu^\psi \mid J/\psi\rangle
\langle J/\psi K \mid\tilde{O}_2 \mid B\rangle
\langle B\mid j_5^B \mid 0\rangle}
{(m_{K}^2-q^2)(m_{\psi}^2-p^2)(m_{B}^2-(p+q)^2)}+~.....~, 
\label{res}
\ee 
contains the matrix element (\ref{nf}) of interest.
In addition, one has contributions from  
excited resonances and continuum states denoted above by ellipses 
which lead to a very complicated singularity structure. 
On the other hand, in the Euclidean   
region $ Q^2 =-q^2 > O(1~\mbox{GeV}^2)$, $p^2 \leq 0$, 
$(p+q)^2 \leq 0$ 
one can expand the correlation function (\ref{corr})
in terms of local operators:
\be
\widetilde{\Pi}^{\mu\nu}(p,q)
= \sum_d \widetilde{C}^{\mu\nu}_d(p,q,\mu)
\langle  \Omega_d(\mu) \rangle~.
\label{ope9}
\ee
Because of the restriction to short distances it suffices
to keep only the operators with low dimensions given in (\ref{oper2}).
We have included all relevant operators up to $d=6$.
The corresponding coefficients $\tilde{C}^{\mu\nu}_d(p,q,\mu)$ 
have been calculated from the diagrams shown in Fig. 21.
By equating (\ref{res}) and (\ref{ope9})
one can derive a sum rule for the matrix element   
$\langle J/\psi K \mid\tilde{O}_2 \mid B\rangle$.

There are two complications that are 
not present in the two- and three-point sum rules discussed above. 
One problem is the presence of a light continuum in the 
$B$ channel below the pole of the ground state $B$-meson. 
This contribution to (\ref{res}) can be associated with processes 
of the type $B \to$ ``$D^*D_s$'' $\to J/\psi K$, 
where an intermediate state carrying 
$D^*D_s$ quantum numbers rescatters into the final
$J/\psi K$ state. Formally, in (\ref{corr}) it is created from
the vacuum by the combined 
action of the operator product $\tilde{O}_2 j_5^B$. 
As a reasonable solution we suggest to 
cancel this unwanted piece against those terms in the OPE (\ref{ope9})
with the quark content $c\bar{c}s\bar{q}$
which develop a nonzero
imaginary part at $(p+q)^2 \geq 4m_c^2$.
In the approximation considered this is the case for
the four-quark condensate contribution represented by  
Fig. 21c.  

The second problem concerns the 
subtraction of contributions from excited resonances
and continuum states in the remaining sum rule.
A closer look at the $J/\psi$ channel
of the correlation function (\ref{corr})
reveals that the higher charmonium resonances
contribute with alternating signs. 
Therefore, the usual subtraction procedure  
in which the dispersion integral over the excited and continuum 
states is approximated by its perturbative counterpart
is not reliable here. In order to proceed we employ 
explicit, although rough
models for the hadronic spectral functions. 
Since the number of additional parameters has to be manageable, only 
the first excited resonances are included in each channel
besides the $B$, $J/\psi$ and $K$ ground states.
In total, in this approximation the correlator (\ref{res})  
contains three free parameters in addition to $\tilde{f}_{B\psi K}$.

We then Borel transform (\ref{res}) and (\ref{ope9})  
in the $B$-meson channel and take  
moments in the charmonium channel.
In the $K$-meson channel, $q^2$ is kept
spacelike. Fitting the hadronic representation 
(\ref{res}) to the OPE result (\ref{ope9}) for various values of the
Borel mass $M$ and $q^2$, and for several moments, we find 
\be
\tilde{f}_{B\psi K} = -(0.045~to~0.075).
\label{ftilde}
\ee
The scale $\mu$ implicit in this estimate is set by the Borel mass.
The central value in the allowed range turns out to be given by 
$M \simeq \sqrt{m_B^2-m_b^2} \simeq \frac{1}{2}m_b \simeq 2.4$ GeV.
Substituting (\ref{ftilde}) in (\ref{a2}), and evaluating the 
short-distance coefficients $c_{1,2}(\mu)$ also at $\mu= M$, 
one gets
\be
a_2^{B \psi K}= -0.29 +0.38 -(0.19 ~to~ 0.31) =-(0.10 ~to~ 0.22) ~,
\label{a2number}
\ee
where the three terms in the first relation refer to the three terms 
in (\ref{a2}) in the same order. 
Interestingly, the sum rule approach seems to 
favour the negative solution for $\tilde{f}_{B \psi K}$. 
Although in comparison with (\ref{tildemu}) our estimate  
falls somewhat short, the gap between theory 
and experiment is narrowed considerably as can be seen from Fig. 20. 

\begin{figure}[htb]
\centerline{
\epsfig{bbllx=35pt,bblly=235pt,bburx=523pt,%
bbury=580pt,file=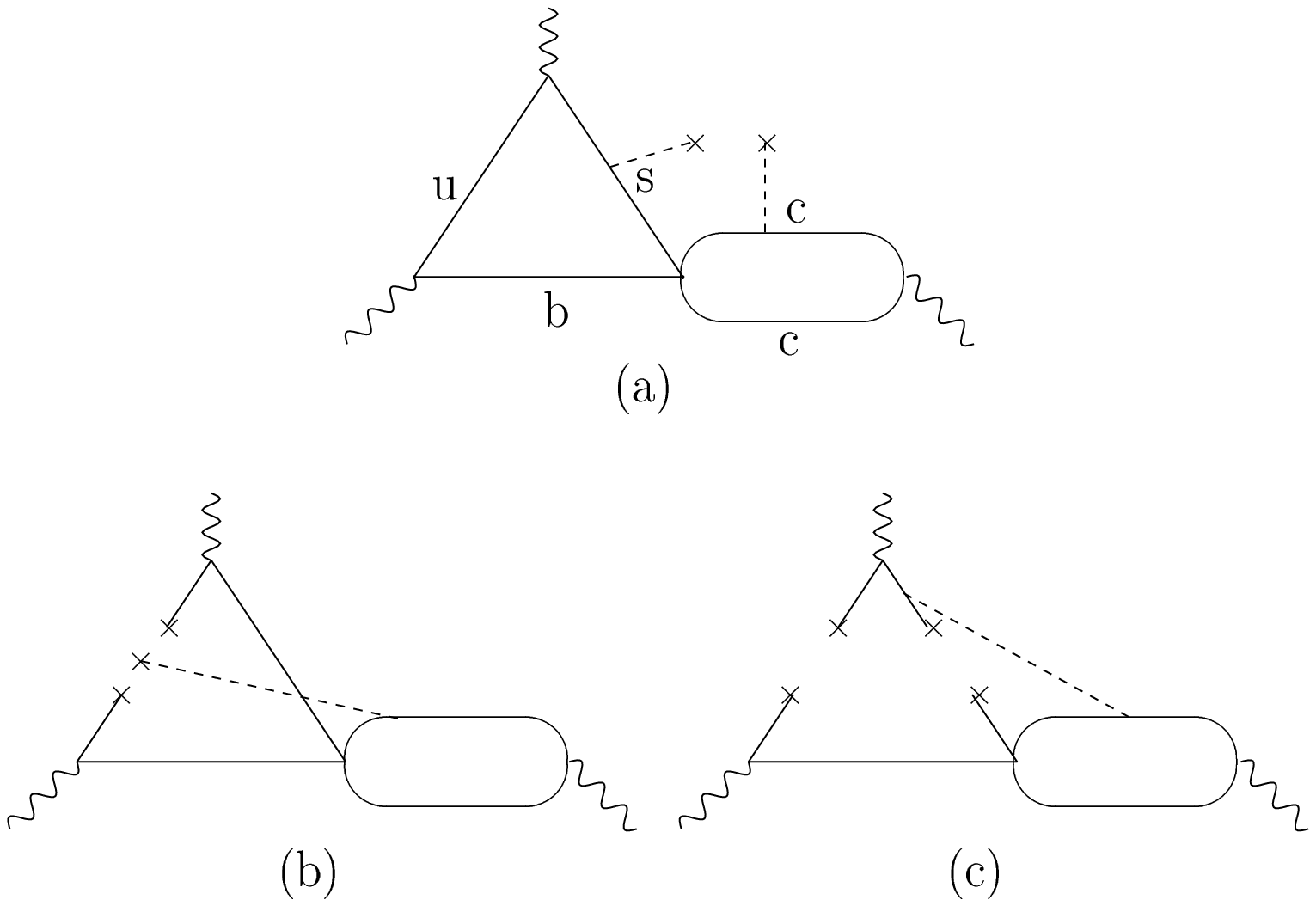,scale=0.9,%
clip=}
}
\caption{\it Diagrams determining the Wilson coefficients of the OPE 
of the correlation function $\tilde{\Pi}_{\mu\nu}$.
The symbols are as in Fig. 1.}
\end{figure}

Several comments are in order. Firstly, the nonfactorizable matrix 
element (\ref{nf}) is small as compared to the factorizable one 
given in (\ref{factoriz}),
numerically, $|\tilde{f}_{B\psi K}/f^+_{B\ra K}(m_\psi^2)|\simeq 0.1 $. 
Nevertheless, it has a strong quantitative impact on $a_2^{B\psi K}$ 
because of the large coefficient
$|2c_1/(c_2+c_1/3)| \simeq 20 ~to~ 30 $. 
Secondly, the factorizable amplitude proportional to $c_1/3$  
and the nonfactorizable one proportional to $\tilde{f}_{B\psi K}$
are opposite in sign and hence tend to cancel. 
In fact, if 
$|\tilde{f}_{B\psi K}|$ is taken at the upper end of the estimated 
range, the cancellation is almost complete resulting in a considerable 
enhancement of the branching ratio (\ref{BRfact}).
Note that both terms are nonleading in $1/N_c$.
This is 
exactly the scenario anticipated by the $1/N_c$ -- rule 
for $D$ decays \cite{BGR} which has found theoretical
support by the global sum rule analysis of \cite{BS}. 
In \cite{BS93,Halperin}, a similar trend was claimed for 
$B\to D \pi$.
Thirdly, our estimate 
yields a negative overall sign for $a_2^{B\psi K}$ in contradiction to 
naive factorization (first two terms in (\ref{a2})),
and also to a global fit of the factorized decay amplitudes  
with two universal coefficients $a_1$ and $a_2$ to the data \cite{NS}.
It should be stressed, however, that in this fit 
the positive sign of $a_2$ actually results from 
the channels $B^- \ra D^0\pi^-, D^0 \rho^-$, $D^{*0} \pi^-$, 
and $D^{*0}\rho^-$,
and is then assigned also to the $J/\psi K$ channel.
This assignment may not be correct. Certainly, 
the sum rule approach described above 
provides no justification for such an assumption. 
On the contrary, diagrams of the kind shown
in Fig. 21 suggest some channel-dependence of the nonfactorizable
matrix elements. For class II processes involving 
$a_2 = c_2 + c_1/3 + 2c_1 \tilde{f}/f^+$ (see (\ref{a2})),
the channel-dependence is enhanced by the large coefficient $2c_1$,
while the factorized matrix elements come with the small 
coefficient  $c_2 + c_1/3$. A concrete numerical example is provided by 
(\ref{a2number}). The opposite is the case for class I processes 
involving $a_1 = c_1 + c_2/3 + 2c_2 \tilde{f}/f^+$:
the nonfactorizable contributions are damped
by the small coefficient $c_2$, while
the factorized term has a large coefficient.
Therefore, $a_1$ is indeed expected to be universal to a good 
approximation, but $a_2$ should exhibit some channel-dependence,
in particular when comparing decays with very different final states
such $D \pi$ and $J/\psi K$.

As a last remark, from the sum rule point of view
there is also no simple relation between $B$ and $D$ decays. 
In \cite{NS}, arguments are presented
which support a change of sign in $a_2$ when going from $D$ to $B$
decays. However, it is not obvious that
these arguments hold independently of the particle composition of
the final state. In the sum rule approach, for example, 
one has significant differences in the OPE of the correlation 
functions such as (\ref{corr}) for   
$B \to D \pi$, $B \to J/\psi K$, and $D \to K \pi$ as can be
imagined from Fig. 21.
From this point of view at least, 
the relation of $B$ and $D$ decays is expected to
involve more than just a change of mass scales.

\section{Conclusion}

QCD sum rule techniques have proved to be very useful in calculating
hadronic matrix elements for exclusive decays of $B$ and $D$ mesons.
In this review, applications are discussed to decay constants, 
form factors, and amplitudes of nonleptonic two-body decays,
as well as to the $B^*B\pi$ and $D^*D\pi$ couplings.
We have not considered exclusive radiative decays such as  
$B\to K^* \gamma$ and $B\to \rho \gamma$,
but we at least want to mention that here sum rules have 
been employed to determine the matrix elements of the 
leading magnetic penguin operator 
\cite{penguin3point,Ballpenguin,COL96,ABS} and to estimate 
long-distance effects \cite{KSW,AB,KRSW}. The above examples are by no
means exhaustive.

Using the sum rule results we have presented predictions 
on decay distributions and integrated
widths for $B \to \pi \bar{l} \nu_l$, $B \to \rho \bar{l} \nu_l$,
and $D \to \pi \bar{l} \nu_l$.
Comparison with the CLEO measurements \cite{CLEO}
of exclusive semileptonic $B^0$ decays 
yields values for $V_{ub}$ in good agreement with each other and
with the determination from inclusive data \cite{vubincl}.
Agreement between expectation and measurement is also found
for the Cabibbo-suppressed semileptonic $D^0$ decay \cite{PDG}.
 
From the sum rule estimate of the $D^*D\pi$ coupling 
we have calculated the decay width for $D^* \to D \pi$ and
compared our result with other estimates. The experimental 
upper limit \cite{PDG} is still about three times larger than the 
expected width.

Furthermore, we have described an attempt to estimate nonfactorizable
matrix elements of nonleptonic decay amplitudes with the help of sum
rule techniques. Using $B\to J/\psi K$ as a prototype example,
it is argued that nonfactorizable effects play an essential role.  
With the sum rule estimate of the nonfactorizable contributions included
the effective coefficient $a_2^{B\psi K}$  is found to be consistent 
with the value extracted from the experimental
branching ratio \cite{PDG}. However, the sign of $a_2^{B\psi K}$ predicted
by the sum rule analysis is opposite to the sign determined from data,
if channel-independent, universal coefficients
$a_{1,2}$ are assumed \cite{NS}. We have explained why
in the sum rule approach universality can be expected for $a_1$, but
not for $a_2$. 
This issue certainly requires further clarification.
As long as one is not able to actually calculate the effective
coefficients, one cannot claim a complete
theoretical understanding of the exclusive nonleptonic decays.

Among the different variants of sum rules we have put particular
emphasis on the light-cone sum rules which provide a powerful tool
in problems involving a pion, kaon or $\rho$ meson such as
heavy-to-light form factors and couplings.
In this approach, the light hadrons are described on mass-shell by a set of
wave functions with different twist and quark-gluon
multiplicity, representing distribution amplitudes in the fraction 
of the hadron light-cone momentum carried by the constituents. 
This avoids the notorious model-dependence of extrapolations
from Euclidean to physical momenta in light channels.
The light-cone sum rules also seem to be fully consistent with
the heavy mass expansion in contrast to some of the sum
rules based on short-distance expansion.
Moreover, the derivation of light-cone sum rules 
is often technically easier than conventional
sum rule calculations. 

We have addressed in some detail the present theoretical
uncertainties, and the prospects for improvement.
On the theoretical side, we see room for it by determining
the nonasymptotic features of the light-cone wave functions
more accurately, and by including higher-order
perturbative effects in the sum rules. Work is under way in both
directions: the re-analysis of the twist 2 pion wave function
in \cite{BJ}, the determination of $\rho$ wave functions in \cite{rho},
and the calculations of the $O(\alpha_s)$
correction to the $B \to \pi$ form factor $f^+$ in \cite{KRWY,Bagan}
are recent examples. On the experimental side,
the advent of new and more precise measurements at
future $B$ and tau-charm factories should allow to tightly
constrain the input parameters and to test the reliability
of the sum rule approach in a very challenging way. 
Because of the universality of the nonperturbative input it
appears conceivable to decrease the uncertainties from presently
20 to 30 ~\% to about 10 \%.  

Finally, we have pointed out the
encouraging agreement of lattice and sum rule calculations in 
the case of $f_B$, $f_D$, and the $B\to\pi$ and $B\to\rho$ 
form factors. This agreement should be enough motivation to join efforts. 
We believe that it would be very fruitful to combine the flexibility 
of the sum rule method with the rigorous nature of the lattice
approach.

{\bf Acknowledgements}

We are thankful to V.M. Belyaev, V.M. Braun, B. Lampe, 
Ch. Winhart, S. Weinzierl, and O. Yakovlev for collaboration
on various subjects of this review and for useful discussions.
This work was supported by the Bundesministerium f\"ur Bildung,
Wissenschaft, Forschung und Technologie, Bonn, Germany,
Contract 05 7WZ91P (0).

\section*{Appendix 1}
\app
Here, we collect the formulae for the light-cone wave functions
of the pion and specify the parameters. It is important to note
that the asymptotic form of these functions and the scale
dependence are given by perturbative QCD \cite{CZ,BF2}.

The twist 2 wave function $\varphi_\pi$ is
expressed as an expansion in Gegenbauer polynomials:
 \begin{equation}
\varphi_\pi(u,\mu) = 6 u(1-u)\Big[1+a_2(\mu)C^{3/2}_2(2u-1)+
 a_4(\mu)C^{3/2}_4(2u-1)+\ldots\Big]\,,
\label{expansion}
\end{equation}
where
\bq
C_2^{3/2}(2u-1)=\frac{3}2[5(2u-1)^2-1]~,
\nonumber
\\
C_4^{3/2}(2u-1)=\frac{15}8[21(2u-1)^4-14(2u-1)^2+1]~.
\label{G2}
\eq
The  normalization is such that
\be
\int^1_0 du \varphi_\pi(u,\mu)=1 ~.
\ee
The nonperturbative effects are contained in the coefficients
$a_n$.
In LO, they are multiplicatively renormalizable and have
the following scale dependence:
\be 
a_n(\mu)=a_n(\mu_0)\left( 
\frac{\alpha_s(\mu)}{\alpha_s(\mu_0)}\right)^{\gamma_n/b}~,
\label{anom}
\ee 
where $b=11- 2n_f/3$ is the LO coefficient of the QCD beta function,
$n_f$ being the number of active flavours, and
\be
\gamma_{n}=C_F\left[-3 -\frac{2}{(n+1)(n+2)}+4\left(\sum_{k=1}^{n+1}
\frac1k\right)\right]
\label{gamman}
\ee
are the anomalous dimensions \cite{CZ}.
We see that $a_n(\mu)$, $n \geq 2$ vanishes for
$\mu \rightarrow \infty $. Therefore,
these terms describe nonasymptotic features of the
wave function (\ref{expansion}).

The initial values of the nonasymptotic coefficients 
can be estimated from two-point sum rules \cite{CZ} for the moments
$\int u^n \varphi_\pi(u,\mu) du$ at low $n$. 
The nonperturbative information encoded 
in the quark and gluon condensates is 
thereby transmuted into the long-distance properties of the wave function.
Alternatively, one can determine the coefficients directly
from light-cone sum rules for known hadronic quantities
such as the $\pi NN$ and $\omega\rho\pi$ couplings.
For the numerical results shown in this review we have used the
following estimate at $\mu_0=0.5$ GeV \cite{BF}:
\be
a_2(\mu_0)=\frac23, ~a_4(\mu_0)=0.43~.
\label{aa}
\ee 
On the basis of the approximate conformal symmetry 
of QCD it has been shown \cite{BF2} that the expansion (\ref{expansion})
converges sufficiently fast so that the terms with $n>4$ are
negligible.

The scale $\mu$ to be used in the 
light-cone sum rules (\ref{fplus}) and (\ref{fplusminus})
for the $B \to \pi$ form factors, and in the sum rule (\ref{fin})
for the $B^*B\pi$ coupling
is somewhat ambiguous, in
particular in LO approximation. As a reasonable choice, we take    
\be
\mu_b =\sqrt{m_B^2-m_b^2} \simeq 2.4\, \mbox{\rm GeV}~.
\label{bscale}
\ee
The analogous choice for the $D$ meson form factors and coupling is 
\be
\mu_c = \sqrt{m_D^2-m_c^2} \simeq 1.3\,\mbox{\rm GeV}~.
\label{cscale}
\ee
These scales characterize the typical virtuality of the $b$,
respectively $c$ quark, and coincide within a factor of two also with the
respective value of the Borel mass $M$.
With (\ref{anom}) and (\ref{aa}), one gets for the evoluted coefficients
$$
a_2(\mu_b) = 0.35, ~~a_4(\mu_b) = 0.18~,
$$
\be
a_2(\mu_c) = 0.41, ~~a_4(\mu_c) = 0.23~.
\ee
The evolution of $a_n(\mu)$ is also known in NLO,
where mixing effects occur \cite{kad}.
Using the same input values (\ref{aa}) and $\mu_b$ from (\ref{bscale}), 
one finds \cite{KRWY}
\be  
a_2(\mu_b) = 0.22,~~ a_4(\mu_b)= 0.084~.
\ee

The twist 3 two-particle wave functions $\varphi_p$ and
$\varphi_{\sigma}$ are related to
the three-particle wave function $\varphi_{3\pi}$ by equation of motion.
It is therefore sufficient to specify the latter
\cite{CZ,Gorsky,BF2}.
Including the first three nonasymptotic terms, 
which is consistent with retaining $a_{2,4}$ in $\varphi_\pi$, one has
$$
\varphi_{3\pi}(\alpha_i,\mu)=360 \alpha_1\alpha_2\alpha_3^2
\Big(1+\omega_{1,0}(\mu)\frac12(7\alpha_3-3)
$$
\be
+\omega_{2,0}(\mu)(2-4\alpha_1\alpha_2-8\alpha_3+8\alpha_3^2)
+\omega_{1,1}(\mu)(3\alpha_1\alpha_2-2\alpha_3+3\alpha_3^2)\Big] 
\label{3pi}
\ee
yielding
\bq
\varphi_p(u,\mu)=1+B_2(\mu)\frac12(3(u-\bar{u})^2-1)+B_4(\mu)
\frac18(35(u-\bar{u})^4
-30(u-\bar{u})^2+3)
\label{bbb}
\eq
and
\be
\varphi_\sigma (u,\mu)=6u\bar{u}\Big[ 1+C_2(\mu)\frac{3}2(5(u-\bar{u})^2-1)
+C_4(\mu)\frac{15}8(21(u-\bar{u})^4-14(u-\bar{u})^2+1)\Big]
\label{tw3}
\ee
with $\bar{u}=1-u$ and
\begin{eqnarray}\label{wf3}
B_2=30\frac{f_{3\pi}}{\mu_\pi f_\pi}\,, ~~
B_4=\frac32\frac{f_{3\pi}}{\mu_\pi f_\pi}
(4\omega_{2,0}-\omega_{1,1}-2\omega_{1,0})\,,
\nonumber\\
C_2=\frac{f_{3\pi}}{\mu_\pi f_\pi}(5-\frac12\omega_{1,0})\,,~~
C_4=\frac1{10}\frac{f_{3\pi}}{\mu_\pi f_\pi}(4\omega_{2,0}-\omega_{1,1})~.
\label{ccc}
\end{eqnarray}
The parameter $f_{3\pi}(\mu)$ and the coefficients
$\omega_{i,k}(\mu)$ have again been estimated from sum rules \cite{CZ}:
$$
f_{3\pi}(1\mbox{GeV}) 
= 0.0035~\mbox{\rm GeV}^2~, $$
\be
\omega_{1,0}(1\mbox{GeV}) = -2.88\,
,~~ \omega_{2,0}(1\mbox{GeV})= 10.5\,,~~ 
\omega_{1,1}(1\mbox{GeV}) = 0~.
\ee
After renormalization \cite{CZ,BF2} to the relevant scales $\mu_b$ and $\mu_c$ 
one has
\bq
f_{3\pi}(\mu_b) = 0.0026~ \mbox{\rm GeV}^2,~~
 \omega_{1,0}(\mu_b)= -2.18,~~
 \omega_{2,0}(\mu_b)= 8.12,~~
 \omega_{1,1}(\mu_b)= -2.59\,~,
\nonumber\\
f_{3\pi}(\mu_c) = 0.0032~ \mbox{\rm GeV}^2,~~
\omega_{1,0}(\mu_c)= -2.63 ,~~
\omega_{2,0}(\mu_c)= 9.62 ,~~
\omega_{1,1}(\mu_c)= -1.05 ~.
\label{3pinorm}
\eq
The parameter $\mu_\pi = m_\pi^2/(m_u + m_d)$ can be inferred from
the PCAC relation (\ref{condens}):
\be
\mu_\pi(1\, \mbox{\rm GeV}) = 1.65\,\mbox{\rm GeV},~~
\mu_\pi(\mu_c) = 1.76\,\mbox{\rm GeV},~~
\mu_\pi(\mu_b) = 2.02\,\mbox{\rm GeV}~.
\ee

For convenience, we also list the complete set of twist 4 wave functions 
given in ref. \cite{BF2}. It includes
four three-particle wave functions specified by two parameters:
\bq
\varphi_\perp (\alpha_i,\mu)&=&30\delta^2(\mu) 
(\alpha_1-\alpha_2)\alpha_3^2[\frac13+2
\varepsilon(\mu) (1-2\alpha_3)] ~,
\nonumber
\\
\varphi_\parallel (\alpha_i,\mu)&=&120\delta^2(\mu)\varepsilon(\mu) 
(\alpha_1-\alpha_2)\alpha_1\alpha_2\alpha_3~,
\nonumber
\\
\tilde{\varphi}_\perp (\alpha_i,\mu)&=&30\delta^2(\mu)\alpha_3^2(1-\alpha_3)[\frac13+2
\varepsilon(\mu) (1-2\alpha_3)] ~,
\nonumber
\\
\tilde{\varphi}_\parallel (\alpha_i,\mu)&=&-120\delta^2\alpha_1\alpha_2\alpha_3[\frac13+
\varepsilon(\mu) (1-3\alpha_3)] ~,
\label{tw4gluon}
\eq
and two two-particle wave functions related to the former
by equations of motion:
\begin{eqnarray}
g_1(u,\mu)&=&\frac{5}2\delta^2(\mu)\bar{u}^2u^2+\frac{1}{2}\varepsilon(\mu)
\delta^2(\mu)[\bar{u}u(2+13\bar{u}u)+10u^3\ln u(2-3u+\frac65u^2)
\nonumber
\\
&&{}+10\bar{u}^3\ln \bar{u}(2-3\bar{u}+\frac65\bar{u}^2)]\,,
\nonumber
\\
g_2(u,\mu)&=&\frac{10}3\delta^2(\mu)\bar{u}u(u-\bar{u})\,.
\end{eqnarray}
The parameter $\delta^2$ is actually defined by the matrix element
\be
\langle \pi |g_s\bar{d}\tilde{G}_{\alpha\mu}\gamma^\alpha u|0 \rangle=
i\delta^2f_\pi q_\mu ~.
\label{delta}
\ee
Renormalizing the values \cite{nov}
\be
\delta^2(1 \mbox{GeV})
=0.2 ~ \mbox{GeV}^2 
\label{deltares}
\ee
and \cite{BF} 
\be
\varepsilon(1 \mbox{GeV}) =0.5  
\label{varepsil}
\ee
obtained from sum rule estimates
to the relevant scales $\mu_c$ and $\mu_b$, one finds
\bq
\delta^2(\mu_c)=0.19 ~\mbox{GeV}^2
,~~ \varepsilon(\mu_c) =0.45\,,
\nonumber\\
\delta^2(\mu_b)=0.17 ~\mbox{GeV}^2,~~ 
\varepsilon(\mu_b) =0.36\,.
\eq

\appende

\section*{Appendix 2}
\app
Here, we give the explicit expressions for various subdominant contributions
to the light-cone sum rule (\ref{fplus}):
\\
\\
\noindent the surface term $t^+$ 
$$
t^+(s_0^B,p^2,M^2)=
\exp\left(-\frac{s_0^B}{M^2}\right)
\Bigg\{
\frac{\mu_\pi(m_b^2+p^2)}{6m_b(m_b^2-p^2)}
\varphi_\sigma(\Delta)
$$
$$
-\frac{4m_b^2}{(m_b^2-p^2)^2}\left(
1+\frac{s_0^B-p^2}{M^2}\right) g_1(\Delta)
+
\frac{4m_b^2 }{(s_0^B-p^2)(m_b^2-p^2)}
\frac{dg_1(\Delta)}{du}
$$
\be
+\frac2{m_b^2-p^2}\Bigg[1+\frac{m_b^2+p^2}{m_b^2-p^2}\left(
1+\frac{s_0^B-p^2}{M^2}\right) 
\Bigg]\int^\Delta_0g_2(v)dv-
\frac{2(m_b^2+p^2)}{(m_b^2-p^2)(s_0^B-p^2)}g_2(\Delta)\Bigg\}
\label{t+}~,
\ee
\\
\\
\noindent the contribution of three-particle wave functions $f_G^+$
$$
f_G^+(p^2,M^2)=
-\int_0^1\!\!u du\!\int \frac{{\cal D}\alpha_i
\Theta( \alpha_1+u\alpha_3-\Delta)}{(\alpha_1+u\alpha_3)^2}
$$
\be
\times \exp\!\left[-\frac{m_b^2-p^2
(1-\alpha_1-u\alpha_3)}{(\alpha_1+
u\alpha_3)M^2}\right]\!\Phi_3(u,\alpha_i,M^2,p^2) \nonumber~,
\label{formSR}
\ee
where
\be
\Phi_3 = \frac{2f_{3\pi}}{f_{\pi}m_b} 
\varphi_{3\pi}(\alpha_i)
\left[1-\frac{ m^2_b -p^2 }{(\alpha_1+u\alpha_3)M^2}\right]    
-\frac1{uM^2} \Bigg[2\vp_\perp (\alpha_i)-\vp_\parallel (\alpha_i)+
2\tilde{\varphi}_\perp (\alpha_i)-\tilde{\varphi}_\parallel (\alpha_i)\Bigg]~,
\label{xxz}
\ee
\\
\\
\noindent and the perturbative correction $\delta^+(p^2,M^2)$ 
\begin{eqnarray}
\delta^+(p^2,M^2)
= \frac{1}{\pi}\int\limits_{0}^{1} du ~\varphi_\pi (u,\mu) 
\int\limits_{m_b^2}^{s_0^B}\frac{ds}{m_b^2}
\mbox{Im}T_1\Big(\frac{p^2}{m_b^{2}},\frac{s}{m_b^{2}},u,\mu\Big)
\exp\left(\frac{-s}{M^2}\right)~,
\label{sumrules}
\end{eqnarray}
where
$$\frac{1}{\pi} \mbox{Im}T_1(r_1,r_2,u,\mu) = 
  \delta(1-\rho) \left[ \pi^2 -6 + 3\ln \frac{m_b^{2}}{\mu^2}   
 -2 \mbox{Li}_2(r_1) \right. 
$$
$$ + 2 \mbox{Li}_2(1-r_2)
\left.  -2 \left( \ln \frac{r_2-1}{1-r_1} \right)^2 + 2 
\left( \ln r_2 +\frac{1-r_2}{r_2} \right)
                     \left( 2 \ln(r_2-1) - \ln(1-r_1) \right)
 \right]$$
$$+ \theta(\rho-1)\left[ 8 \left. \frac{\ln(\rho-1)}{\rho-1}
\right|_+
 +2 \left( \ln r_2 + \frac{1}{r_2} -2 
-2 \ln(r_2-1) + \ln 
\frac{m_b^{2}}{\mu^2} \right)
             \left. \frac{1}{\rho-1} \right|_+ \right.$$
$$
-2 \frac{r_2-1}{(r_1-r_2)(\rho-r_1)} 
   \left( \ln \rho -2 \ln(\rho-1) + 1 -
\ln \frac{m_b^{2}}{\mu^2} 
\right)
$$
$$
+2 \frac{1-r_1}{(r_1-r_2)(r_2-\rho)} \left( \ln \frac{\rho}
{r_2} -2 \ln \frac{\rho-1}{r_2-1} \right) 
\left. -4 \frac{\ln \rho}{\rho-1} + 2 \frac{1}{r_2-\rho}
 \left( \frac{1}{\rho} - \frac{1}{r_2} \right)
   +\frac{1}{\rho^2} -\frac{1}{\rho} \right]$$  
$$ 
+\theta(1-\rho) \left[ 2 \left( \ln r_2 + \frac{1}{r_2} 
-2 \ln(r_2-1) 
-\ln \frac{m_b^{2}}{\mu^2}
\right) \left. \frac{1}{\rho-1} \right|_+
 \right.
 $$
\begin{eqnarray}
\label{spden}
 -2 \frac{1-r_1}{(r_1-r_2)(r_2-\rho)} 
 \left( \ln r_2 + 1 -2 \ln(r_2-1) -\ln \frac{m_b^{2}}{\mu^2}\right) \left.
 -2 \frac{1}{r_2-\rho} \frac{1-r_2}{r_2} \right] 
\end{eqnarray} 
with
\be
\rho=r_1+u(r_2-r_1),~~ 
\int d\rho f(\rho) \left. \frac{1}{1-\rho} \right|_+ =
  \int d\rho \left( f(\rho) - f(1) \right) \frac{1}{1-\rho}.
\ee

The surface term $t^{\pm}$ appearing in the sum rule (\ref{fplusminus}) 
is given by 
\be
t^{\pm}(s_0^B,p^2,M^2)=
\exp\left(- \frac{s_0^B}{M^2}\right)
\left[- \frac{\varphi_\sigma(\Delta)}{6\Delta}  
+ \frac{2m_bg_2(\Delta)}{\mu_\pi(m_b^2-p^2)}\right] ~.
\label{tt}
\ee
\appende

\end{document}